\documentclass[structabstract]{aa}  
\usepackage{graphicx}
\usepackage{amssymb}
\usepackage{latexsym}
\usepackage{amsmath}
\def\lsim{\mathrel{\rlap{\lower4pt\hbox{\hskip1pt$\sim$}}
    \raise1pt\hbox{$<$}}}                % less than or approx. symbol
\def\gsim{\mathrel{\rlap{\lower4pt\hbox{\hskip1pt$\sim$}}
    \raise1pt\hbox{$>$}}}                % greater than or approx. symbol
\usepackage{amsfonts}
\usepackage{color}
\usepackage{ulem}
\usepackage{natbib}
\bibpunct{(}{)}{;}{a}{}{,}
\newcommand{\ds}{\displaystyle}

\newcommand{\expo}{\mathrm{e}}

\newcommand{\promille}{%
  \relax\ifmmode\promillezeichen
        \else\leavevmode\(\mathsurround=0pt\promillezeichen\)\fi}
\newcommand{\promillezeichen}{%
  \kern-.05em%
  \raise.5ex\hbox{\the\scriptfont0 0}%
  \kern-.15em/\kern-.15em%
  \lower.25ex\hbox{\the\scriptfont0 00}}
\begin{document}

\title{Estimating the phase in ground-based interferometry: performance
  comparison between single-mode and multimode schemes}

\author{E. Tatulli \and N. Blind \and J.P. Berger \and A. Chelli \and F. Malbet}
 \institute{Laboratoire d'Astrophysique, Observatoire de Grenoble, 38041
  Grenoble cedex France \\ \email{etatulli@obs.ujf-grenoble.fr}}

\date{Received xxx; accepted xxx}

%\pagerange{\pageref{firstpage}--\pageref{lastpage}} \pubyear{2008}

\abstract{}{In this paper we compare the performance of multi and single-mode
  interferometry for the estimation of the phase of the complex
  visibility.}{We provide a theoretical description of the
  interferometric signal which enables to derive the phase error in
  presence of detector, photon and atmospheric noises, for both multi
  and single-mode cases.}{We show that, despite the loss of flux
  occurring when injecting the light in the single-mode component
  (i.e. single-mode fibers, integrated optics), the spatial
    filtering properties of such single-mode devices often enable 
    higher performance than multimode concepts. In
    the high flux regime speckle noise dominated, 
    single-mode interferometry is always more efficient, and its performance is significantly better when the correction provided
    by adaptive optics becomes poor, by a factor of $2$ and more when the Strehl ratio is lower than $10\%$. In low light
    level cases (detector noise regime), multimode interferometry reaches better performance, yet the gain never exceeds $\sim 20\%$, which corresponds to the percentage of photon loss due to the injection in the guides. 
Besides, we demonstrate that
  single-mode interferometry is also more robust to the turbulence
  in both cases of fringe tracking and phase referencing, at the exception of narrow field of views ($\la 1^{''}$).}{Our
    conclusion is therefore that, from a theoretical point of view and
    contrarily to a widespread opinion, fringe trackers built using
    single-mode optics should be considered as a solution both practical and competitive.}


\titlerunning{Multimode vs. single-mode interferometric phase}

\maketitle

\section{Introduction}
Performance of ground-based optical interferometers is severely limited by the atmospheric turbulent piston which introduces a random optical path
difference between the beams that are combined to produce fringes. 
These fringes are thus randomly moving on the detector,
blurring the signal and preventing to integrate on time longer than
the coherence time of the atmosphere, typically a few tens of
milliseconds in the infrared. As a consequence, the limiting-magnitude
and ultimate precision of ground-based interferometers are
dramatically reduced.\\
By estimating and compensating in real-time
the interferometric phase  -- in other words by ``locking'' the fringes
on the detector, fringe tracking and phase referencing devices are
powerful instruments to circumvent this problem, allowing to
integrate the interferograms on much longer time frames. 
Such instruments, which noticeably improve the sensitivity and the
accuracy of interferometers are since a few years undergoing major
developments, as can attest the amount of concepts which
are currently studied for various interferometers (\citealp{lebouquin_1,
  sahlmann_1}: VLTI; \citealp{berger_1}: CHARA; \citealp{jurgenson_1}: MROI).\\
In all the different concepts proposed, one key issue still in
debate among the instrumental community is the relevance of using
spatial filtering devices such as single-mode fibers or integrated
optics to carry/combine the beams. It is often argued that
spatial filtering is not suited for that type of instruments since 
only a fraction of the total flux is injected in the spatial filter
component. This coupling efficiency, which is typically of the order
of the Strehl ratio \citep{foresto_1}, can indeed be low in presence
of strong turbulence and/or poor adaptive optics (AO)
correction. However, the issue is not that simple. Single-mode filters
only keeps the part of the incoming flux which is related to the
coherent part of the corrugated wavefront. In other words single-mode 
devices only propagate the first mode of the
electro-magnetic field leaving at its output a plane
wavefront -- that is a deterministic signal, the price to pay being a  loss of flux correlated to the strength of the turbulence. 
At the contrary, multimode
schemes such as bulk optics are preserving the total flux but keeps at
the output the incoherent part of the wavefront (producing randomly
moving speckle patterns in the image) which makes the interferogram
sensitive to turbulence. Hence choosing between single-mode and
multimode schemes corresponds to decide whether losing flux or losing
coherence in the signal will be the best strategy to optimize the performance of the interferometer.\\
As a matter of fact, the efficiency of single-mode devices in terms of
precision and robustness of the estimation of the amplitude of the
visibility has already been demonstrated experimentally \citep{foresto_2} and
theoretically as well \citep{tatulli_1}. However concerning the
estimation of phase -- which is the quantity of interest for fringe
tracking and phase referencing instruments, the situation is not clear
yet. If simulations have been initiated in some specific cases to estimate the effects of spatial filtering (\citealp{buscher_1}: phase
jumps; \citealp{tubbs_1}: computation of the coherence time), no
theoretical formalism regarding this issue has, to our knowledge, been
presented so far.\\
In this paper, we derive the error of the interferometric phase in
presence of detector, photon and atmospheric noises, both for the
single-mode and multimode cases in presence of partial AO correction, following a formalism analogue to the
one  previously developed for the squared visibility in 
\citet{tatulli_1}. We then compare the performance of single-mode and
multimode schemes for the estimation of the phase of the
interferograms, and apply our analysis in the framework of fringe-tracking and phase referencing methods.

\section[]{The phase in ground-based interferometry} \label{sec_formalism}
\begin{figure}
\begin{center}
\includegraphics[width=0.5\textwidth]{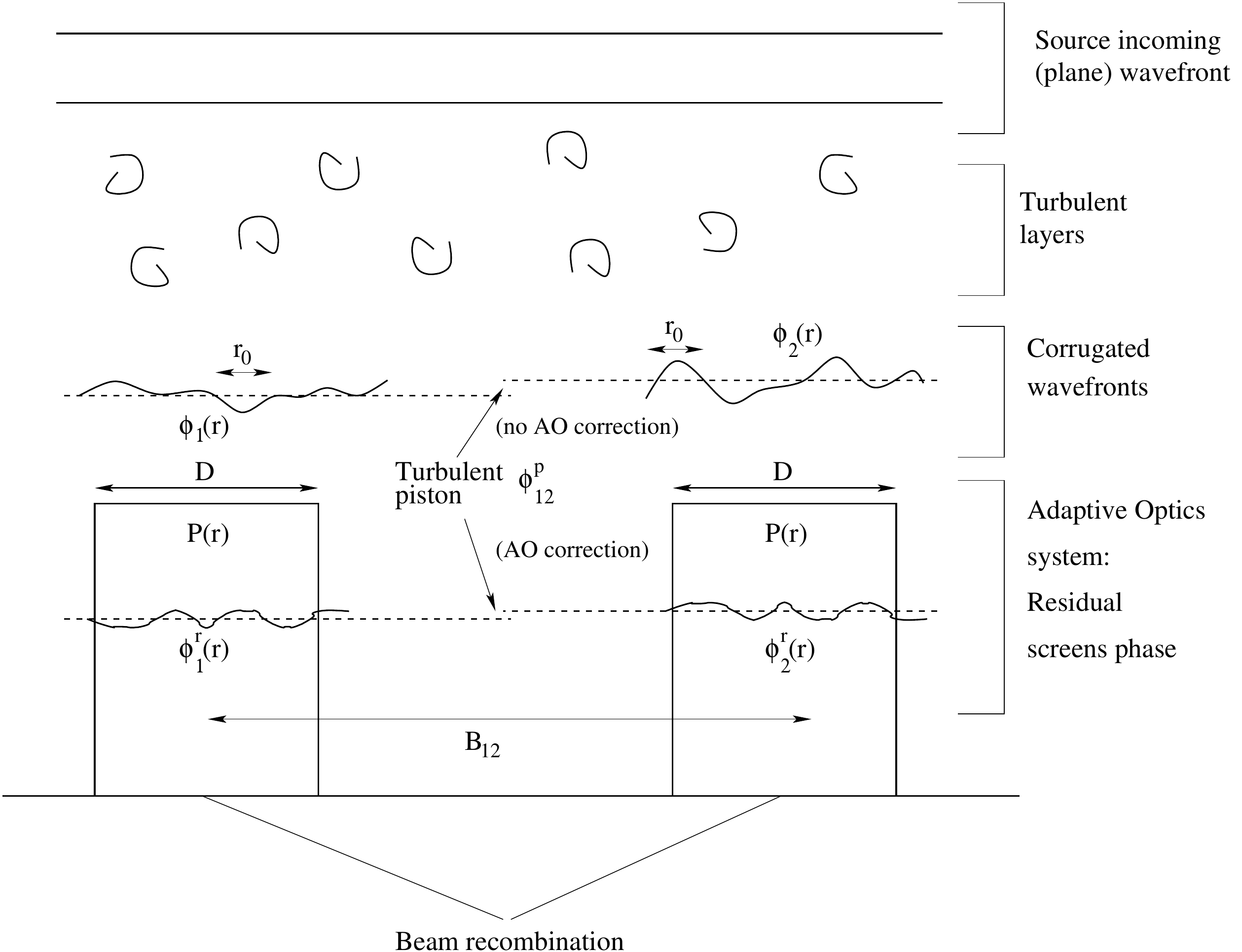}
\caption{\label{fig_turb} Sketch of the effect of the turbulence
in interferometry. On each telescope pupil $P({\vec{r}})$
separated by the baseline distance $B_{12}$, the wavefront is affected
by a random  phase screen, respectively $\phi_1({\vec r})$ and $\phi_2({\vec r})$.  The
strength of the turbulence depends on the parameters $D/r_0$ where $D$
is the diameter of the telescope and $r_0$ is the Fried parameter
\citep{fried_1}, describing the typical size of a coherent cell in the
wavefront. Furthermore, the optical path difference between the two
beams is randomly shifted by the turbulent
piston $\phi^p_{12}$ which is the phase difference between the two
telescopes, averaged on the pupils. If an AO system is present on each telescope, then the
wavefront is partially corrected in real time, and the quantity of
interest are the residual phases $\phi^r_1({\vec r})$ and $\phi^r_2({\vec r})$ and
the associated residual turbulent piston.}
\end{center}
\end{figure}
\begin{table*}
\caption{Expression of the variance of the phase for the three noise
  regimes, respectively detector, photon and atmospheric
  regimes, both in the multimode and single-mode cases. See text and
  Appendix \ref{app_sigma} for a description of the parameters involved. \label{tab_multi}}
\begin{tabular}{lll}
\multicolumn{3}{l}{Multimode case} \\
\hline
$\ds \sigma^2_{det_{\phi}}$&=&$\ds  \frac{1}{2}  \frac{N_{pix}\sigma^2_{det}}{N^2t_1t_2|V_{12}|^2\mathrm{e}^{-2\sigma^2_{\phi_r}}}$\\ \\*[1em]
$\ds \sigma^2_{phot_{\phi}}$&=&$\ds \frac{1}{2} 
\frac{t_1+t_2}{Nt_1t_2|V_{12}|^2\mathrm{e}^{-2\sigma^2_{\phi_r}}}$\\ \\*[1em]
$\ds \sigma^2_{atm_{\phi}}$&=&$\ds \frac{1}{2}\frac{\int
  \left[P({\vec r})P({\vec r^{\prime}})\mathrm{e}^{-\frac{1}{2}\mathcal{D}_{\phi_r}({\vec r},{\vec r^{\prime}})}\right]^2\mathrm{d}{\vec r}\mathrm{d}{{\vec r^{\prime}}}-\mathrm{e}^{-4\sigma^2_{\phi_r}}\int
  \left[P({\vec r})P({\vec r^{\prime}})\mathrm{e}^{\frac{1}{2}\mathcal{D}_{\phi_r}({\vec r},{\vec r^{\prime}})}\right]^2\mathrm{d}{\vec r}\mathrm{d}{{\vec r^{\prime}}}}{\left[\int
    P^2({\vec r})\mathrm{d}{\vec r}\right]^2\mathrm{e}^{-2\sigma^2_{\phi_r}}}$\\ 
&&\\
\multicolumn{3}{l}{Single-mode case} \\
\hline
$\ds \sigma^2_{det_{\phi}}$&=&$\ds \frac{1}{2} \frac{N_{pix}\sigma^2_{det}}{\rho_0^2N^2t_1t_2|V_{12}|^2\mathrm{e}^{-2\sigma^2_{\phi_r}}}$\\ \\*[1em]
$\ds \sigma^2_{phot_{\phi}}$&=&$\ds  \frac{1}{2} 
\frac{(t_1+t_2)\overline{\mathcal{S}}}{\rho_0Nt_1t_2|V_{12}|^2\mathrm{e}^{-2\sigma^2_{\phi_r}}}$\\ \\*[1em]
$\ds \sigma^2_{atm_{\phi}}$&=&$\ds \frac{1}{2}\frac{\left[\int P({\vec r})P({\vec r^{\prime}})\mathrm{e}^{-\frac{1}{2}\mathcal{D}_{\phi_r}({\vec r},{\vec r^{\prime}})}\mathrm{d}{\vec r}\mathrm{d}{{\vec r^{\prime}}}\right]^2-\mathrm{e}^{-4\sigma^2_{\phi_r}}\left[\int P({\vec r})P({\vec r^{\prime}})\mathrm{e}^{\frac{1}{2}\mathcal{D}_{\phi_r}({\vec r},{\vec r^{\prime}})}\mathrm{d}{\vec r}\mathrm{d}{{\vec r^{\prime}}}\right]^2}{\left[\int
      P({\vec r})\mathrm{d}{\vec r}\right]^4\mathrm{e}^{-2\sigma^2_{\phi_r}}}$
\end{tabular}
\end{table*}

\subsection[]{General description and underlying assumptions}
{\bf The ground-based interferometer:} As sketched in Fig. \ref{fig_turb}, we consider an interferometer with two telescopes, separated by the baseline distance $B_{12}$. We assume that both telescope are identical, that is they are described by the same pupil function $P(r)$ of diameter $D$, therefore by the same  collecting area $\Sigma_P$  defined as $\Sigma_P = \int [P({\vec r})]^2\mathrm{d}{\vec r}$. Their transmission are however different, namely  $t_1$ and $t_2$ respectively. The wavefront over each telescope is affected by randomly moving phase screens, respectively $\phi_1(r)$ and  $\phi_2(r)$. The typical cell size of these turbulent phases is $r_0$ \citep{fried_1} and the strength of the turbulence is characterized by the quantity $D/r_0$. When Adaptive Optic systems are present on each telescope, they partially compensate in real-time the wavefront corrugations of the atmosphere. The correction is however not perfect and it remains residual phases $\phi^r_1(r)$ and  $\phi^r_2(r)$ affecting the two telescopes. Because of this turbulence, the optical path difference (opd) between the two light paths to the star through the two telescopes is randomly shifted by the so-called turbulent piston $\phi^p_{12}$, i.e. the  phase difference between the two telescopes averaged over the pupils:
\begin{equation}
\phi^p_{12} = \int
P({\vec r})\phi^r_1({\vec r}) \mathrm{d}{\vec r} - \int P({\vec r})\phi^r_2({\vec r}) \mathrm{d}{\vec r} \label{eq_phi12p}
\end{equation}
 {\bf The interferogram:} A 2-telescope interferogram consists in an incoherent part, which is
the sum of the photometric fluxes coming from both
telescopes, and a modulated coherent part -- namely the fringes,
proportional to the so-called complex coherent flux  $F^c_{12}$ which
depends on the  flux $N$ (photons per surface unit and per time unit) and  on the complex visibility $V_{12}$ of the
source corresponding to the baseline frequency $f_{12}=B_{12}/\lambda$ of the
interferometer. $\lambda$ is the effective wavelength of the interferogram. 
The modulation of the fringes can be whether
temporal when the opd is scanned with moving
piezo-mirrors (temporal coding), whether spatial when the opd is
scanned with dedicated output pupils which separation defines the
frequency of the modulation (spatial coding). \\
Finally, the phase of the interferogram $\Phi_{12}$, that is the phase shift with respect to the zero opd, originates from the source phase $\phi^{obj}_{12}$, a potential instrumental phase $\phi^{ins}_{12}$, that we will assume equal to zero in the following\footnote{considering a non zero
  instrumental phase $\phi^{ins}_{12}$ would only introduce an extra
  shift of the interferogram but would  not affect the
  performance of the interferometer, as long as this instrumental
  phase is stable or calibratable.}, and the turbulence piston phase $\phi^{p}_{12}$ .

\subsection[]{Estimating the interferometric phase and its associated error}
Estimating the interferometric phase $\Phi_{12}$ requires to define and compute from each
interferogram an appropriate complex estimator $\widetilde{F^c_{12}}$ of the complex coherent flux $F^c_{12}$. Several methods are possible to build this estimator
such as ``ABCD'' techniques in the image plane \citep{colavita_1,
  tatulli_2} or analysis in the Fourier plane \citep{roddier_1,
  foresto_2}. In any case, the measured phase 
is then the argument of the complex estimator $\widetilde{F^c_{12}}$:  
\begin{equation}
\Phi_{12} = \mathrm{atan}\left[\frac{\mathrm{Im}\left(\widetilde{F^c_{12}}\right)}{\mathrm{Re}\left(\widetilde{F^c_{12}}\right)}\right]\end{equation}  
In this framework, one can show that in first approximation, i.e. when the error on $\widetilde{F^c_{12}}$ is small compared to its amplitude, the
variance of the instantaneous phase (i.e. measured over one
interferogram) can be expressed as following \citep[see Eq. (5), case $j=k$]{chelli_1}:
\begin{equation}
\sigma^2_{\phi} = \frac{1}{2}\frac{\mathrm{E}\left(\left|\widetilde{F^c_{12}}\right|^2\right)-\mathrm{Re}\left[\mathrm{E}\left(\widetilde{F^c_{12}}^2\right)\right]}{\left[\mathrm{E}\left(\widetilde{F^c_{12}}\right)\right]^2} \label{eq_errphi}
\end{equation}
where E denotes the expected value. \citet{chelli_1} has also demonstrated that the variance of the interferometric phase is independent of the object phase. For sake of simplicity, we will therefore consider that the source of interest is centro-symmetric, namely $\phi^{obj}_{12} = 0$. \\
Regardless of the method used to
build the estimator from the interferogram, the statistics of the
phase will depends on whether the interferometer is using
a multimode or a single-mode design. In the following, we explore how
the atmospheric spatial fluctuations of the turbulent wavefront 
affects the fringe pattern, in both multimode and single-mode cases.

\subsection[]{The coherent flux in multimode interferometry}
In multimode interferometry, the total flux on the detector
remains constant (neglecting scintillation) and the interferograms
are created in speckle patterns randomly moving with the
fluctuations of the turbulent wavefronts over the two telescopes. \\
There are two different ways to combine the beams in multimode interferometry \citep{chelli_2}: whether in the image plane, a technique which is suited to perform spatial coding of the fringes on the detector, as known as Fizeau \citep{beckers_1} and Michelson \citep{mourard_1}  mountings\footnote{in Fizeau combination, the output pupil is homothetic to the entrance one, whereas in Michelson scheme this property is not conserved.}, or in the pupil plane which is commonly used to temporally sample the interferogram through dedicated moving piezo-electric mirrors \citep{dyck_1}. In both cases however, the expression of the coherent flux can be written as following (as shown in the demonstration of  Appendices \ref{app_image_plane} and \ref{app_pupil_plane} of this paper and in Appendices A1 and A2 of \citealt{buscher_1}):
\begin{equation}
F^c_{12} = \Sigma_PN\sqrt{t_1t_2}V_{12}T_{12} \label{eq_fc_multi}
\end{equation}
$T_{12}$ is the normalized interferometric transfer
function resulting of the cross-correlation of the pupil of the two
telescopes, corrugated by the residual atmospheric screen phases $\phi^r_1({\vec r})$ and  $\phi^r_2({\vec r})$, (see Fig. \ref{fig_turb}). It writes:
\begin{equation}
 T_{12} = \frac{\int [P({\vec r})]^2
   \mathrm{e}^{i(\phi^r_1({\vec r})-\phi^r_2({\vec r}))}\mathrm{d}{\vec r}}{\int
   [P({\vec r})]^2\mathrm{d}{\vec r}} \label{eq_coupling_multi}
\end{equation}
As an analogy with classical AO systems on single-pupil telescopes (see
e.g. \citealt{fusco_1}), $|T_{12}|^2$ can be seen as the instantaneous
{\it interferometric} Strehl ratio, in the case of multimode interferometry.

\begin{figure*}
\begin{center}
\begin{tabular}{cc}
\includegraphics[width=0.4\textwidth]{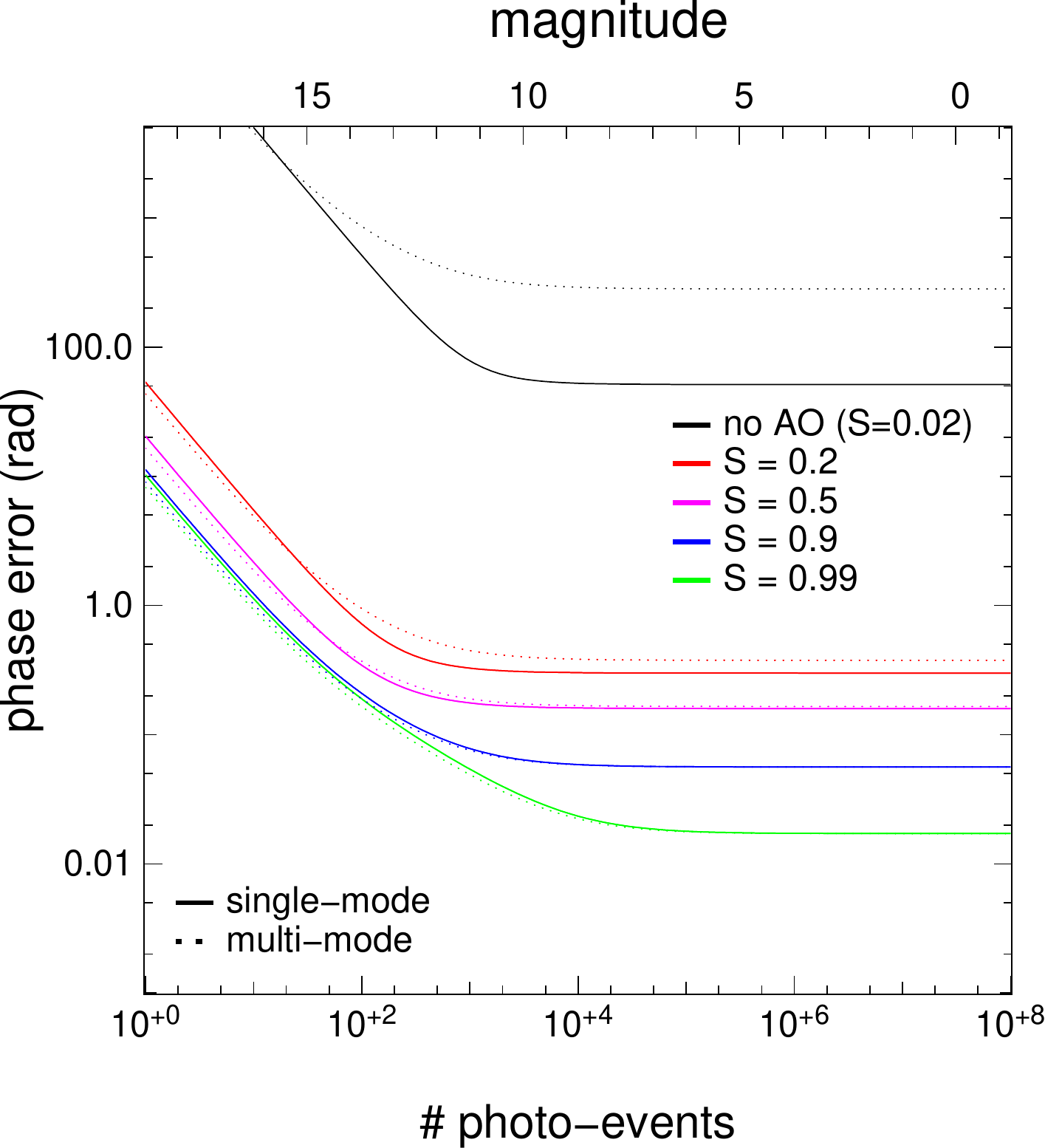}
&  \includegraphics[width=0.4\textwidth]{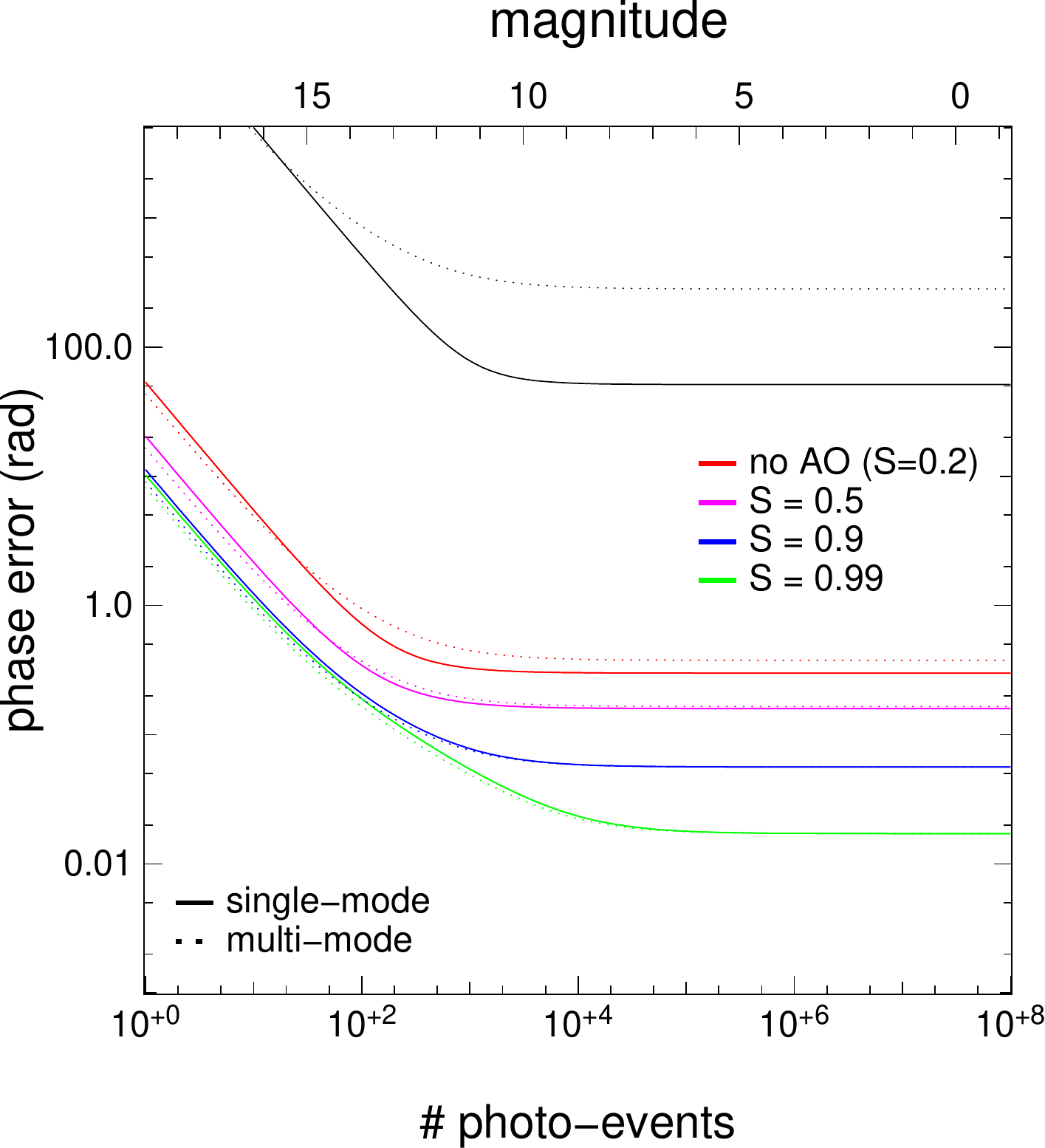} \\
\includegraphics[width=0.4\textwidth]{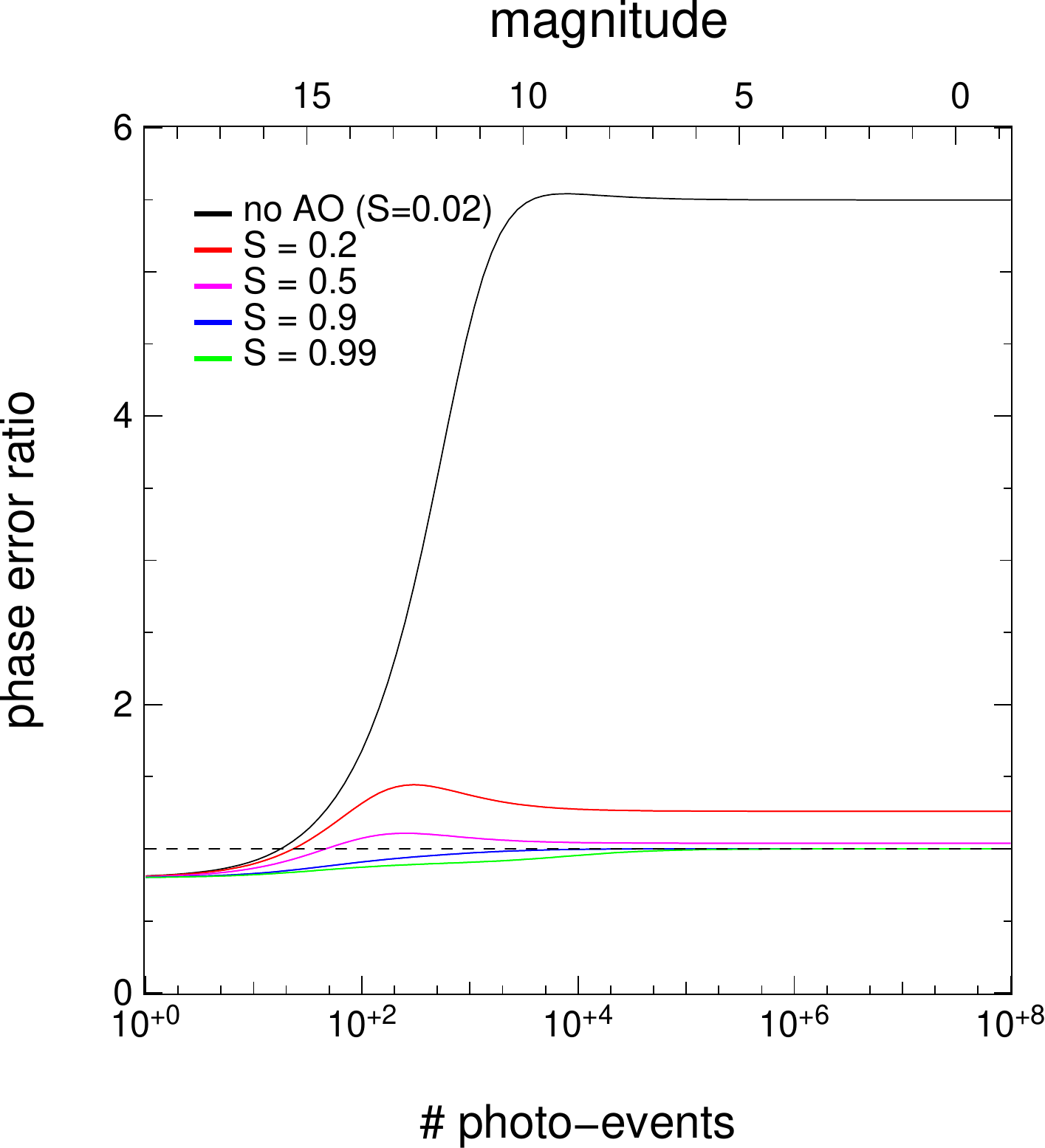}
&  \includegraphics[width=0.4\textwidth]{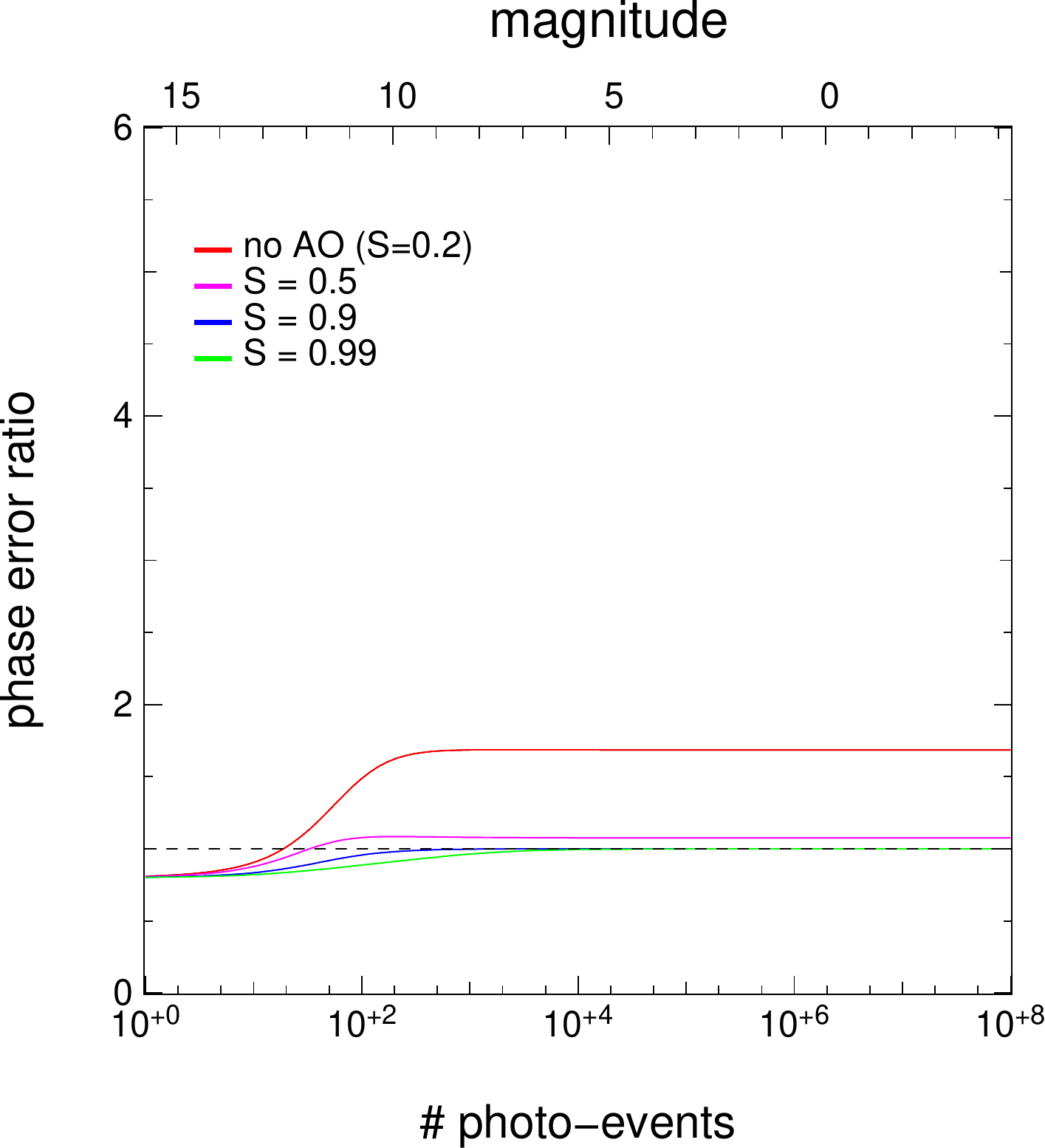}
\end{tabular}
\caption{\label{fig_perf} Global performance comparison
    between multimode and single-mode interferometry, all sources of
    noise being considered. Top:
  Error of the phase in multimode (dotted lines) and single-mode
  (solid lines) as the function of the number of detected photoevents  (per time unit)
  and K-band magnitude. For both cases, we can see the three regimes:
  the detector noise (in $1/N$), then (shortly) the photon noise (in
  $1/\sqrt{N}$) and finally the speckle noise (saturation). Bottom:
  Ratio of multimode vs. single-mode phase error. Plots are shown for
  several levels of correction (Strehl ratios). The fixed parameters are
  $\sigma_{det}=4e^{-}/$pix, $|Vij|=1$, $N_{pix}=4$. Two turbulence strengths are considered $D/r_0 = 8$ (left) and $D/r_0 =2$ (right) which correspond to the average turbulence conditions of the VLTI with UTs ($D=8$m) and ATs ($D=1.8$m) respectively. For the conversion between detected photoevents and
  magnitude scale, we have considered a total transmission coefficient
  of $15\%$, a detector quantum efficiency of $50\%$, an integration
  time of 30ms and a spectral resolution of 35 with 2 pixels by
  spectral channel.}
\end{center}
\end{figure*}

\subsection[]{The coherent flux in single-mode interferometry}
The main property of single-mode devices such as fibers or integrated
optic chips is to perform a spatial filtering of the input wavefront
so that only its Gaussian part is transmitted. As a consequence, a
single-mode device turns the  input spatial wavefront fluctuations 
into intensity fluctuations at the output. Doing so, each
outgoing wavefront is flat, hence the shape of the interferogram  is
deterministic, only depending on the instrumental configuration. 
The trade-off is that only a fraction of the flux is
transmitted, corresponding to the coherent energy of the turbulent
wavefront. The single-mode instantaneous coherent flux thus takes the form (see Appendix \ref{app_singlemode_noise}, Eq. (\ref{eq_singlemode_fc})):
\begin{equation}
F^c_{12} = \Sigma_P N\sqrt{t_1t_2}\rho^{12}(V) \label{eq_fc_mono}
\end{equation}
where $\rho^{12}(V)$ is the interferometric coupling coefficient that
depends both on the source extension and on the level of AO correction (\citealt[see Eq. (3)]{tatulli_1}, and Eq. (\ref{app_eq_rhoij}), Appendix \ref{app_singlemode_noise} of this paper).  Focusing on compact sources, that is astrophysical objects unresolved by a single telescope, $\rho^{12}(V)$ takes a simple expression of the form $\rho^{12}(V) = \rho_0 V_{12} \rho_{12}$ \citep[Eqs. (4,\,5,\,6)]{tatulli_3}. The  coherent flux thus rewrites:
\begin{equation}
F^c_{12} = \Sigma_P N\sqrt{t_1t_2} \rho_0 V_{12} \rho_{12} \label{eq_fc_mono_compact}
\end{equation}
where $\rho_{12}$ is independent of the source properties:
\begin{equation}
\rho_{12} =  \frac{\int P({\vec r})\mathrm{e} ^{i(\phi^r_1({\vec r}))}\mathrm{d}{\vec r}} {\int
  P({\vec r})\mathrm{d}{\vec r}}\frac{\int P({\vec r}) \mathrm{e}^{-i(\phi^r_2({\vec r}))}\mathrm{d}{\vec r}}{\int
  P({\vec r})\mathrm{d}{\vec r}} \label{eq_coupling_mono}
\end{equation}
and where $\rho_0$ is the maximum achievable coupling efficiency, shown to be $\sim 80\%$ \citep{shaklan_1}, due to geometrical
mismatch between the telescope Airy disk profile and the Gaussian
profile of the propagated mode. \\
Eq. (\ref{eq_coupling_mono}) has to be compared with
Eq. (\ref{eq_coupling_multi}). We can see that they are almost similar,
but with one major difference: in Eq. (\ref{eq_coupling_multi}), the
product of the phasors is integrated over the pupil
whereas in Eq. (\ref{eq_coupling_mono}), each phasor is {\it first} integrated
over the pupil,  then the product is performed. As noticed by \citet{buscher_1}, such difference is
the mathematical expression of the property of spatial filtering.\\
Again, $|\rho_{12}|^2$ -- which is equal to $1$ in the case of perfect AO correction/absence of atmospheric turbulence -- can be seen by as the instantaneous {\it interferometric} Strehl ratio, but for the single-mode case.

\begin{figure*}
\begin{center}
\begin{tabular}{cc}
\includegraphics[width=0.4\textwidth]{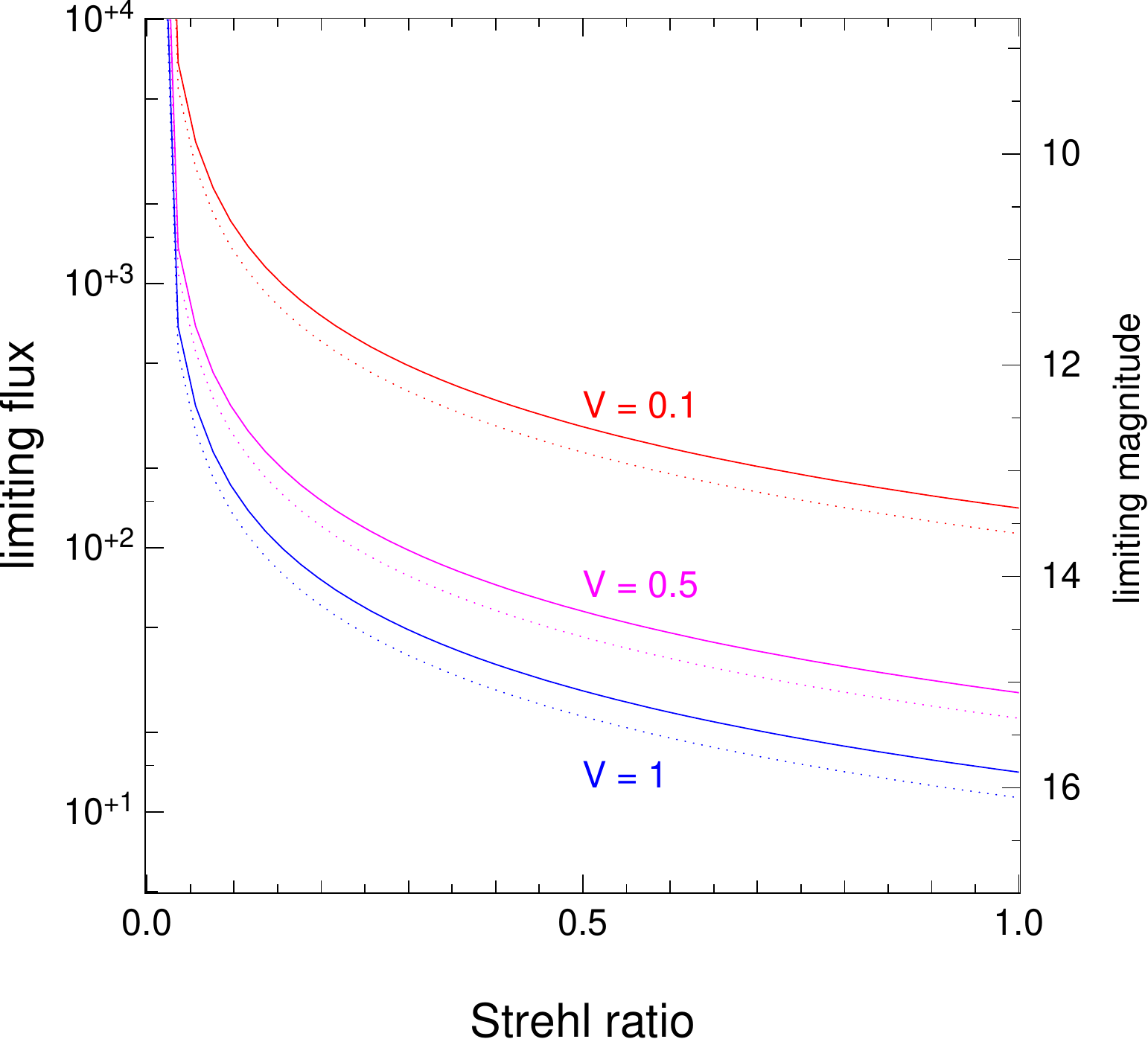} & \includegraphics[width=0.4\textwidth]{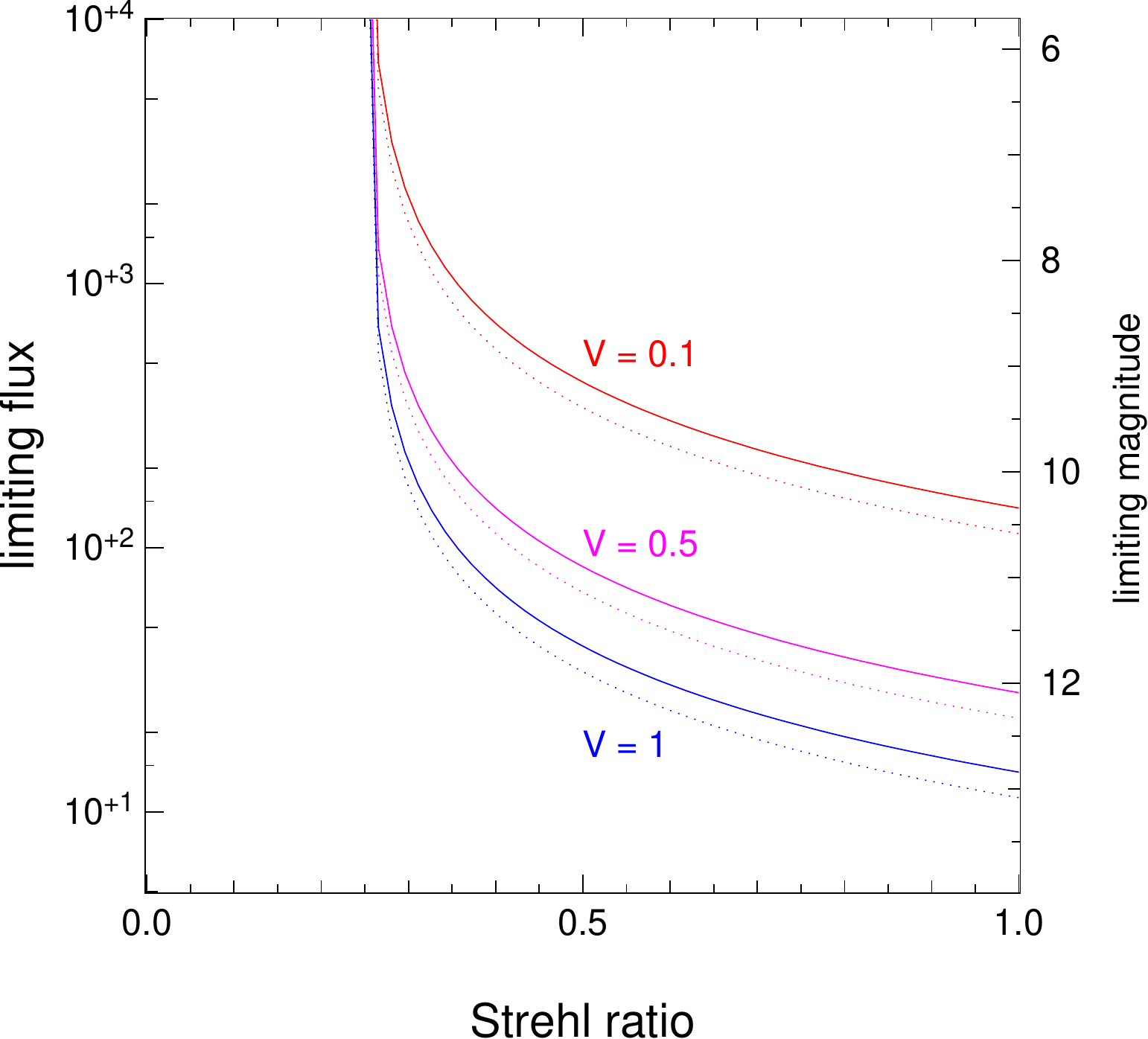}
\end{tabular}
\caption{\label{fig_limitmag} Limiting flux as a function of the
  Strehl ratio (detector noise regime case), for turbulence strengths of $D/r_0=8$ (left) and $D/r_0=2$ (right). Plots
  are shown for an unresolved (V=1), fairly resolved (V=0.5) and fully
  resolved (V=0.1) source, respectively.  Curves are going by pair,
  for single-mode (solid lines) and multimode (dotted lines) cases. The fixed
  parameters have the same values than previous figures.}
\end{center}
\end{figure*}

\section[]{Performance comparison}
\subsection{Phase noise}\label{subsec_noise}
From Eqs. (\ref{eq_errphi},\,\ref{eq_fc_multi},\,\ref{eq_fc_mono_compact}), we derive the expression of the
error of the phase for both single-mode and multimode cases, as
developed in Appendix \ref{app_sigma}. We show that the variance of
the interferometric phase can be decomposed as the quadratic sum of
three terms, corresponding to  the three regimes respectively detector, photon and atmospheric regimes:
\begin{equation}
\sigma^2_{\phi} = \sigma^2_{det_{\phi}} + \sigma^2_{phot_{\phi}} +  \sigma^2_{atm_{\phi}}
\end{equation}
where the detail of each term is given in Table
\ref{tab_multi}. The phase error depends on parameters coming
from the source and the instrument which we recall here: $N$ is
the number of incoming photons per surface unit  and per time unit, $|V_{12}|$
is the amplitude of the visibility, $t_1$ and $t_2$ are the
transmissions of both telescopes, $N_{pix}$ is the number of pixels to
sample the interferogram, $\sigma_{det}$ is the detector noise. The
phase error also depends on atmospheric terms: the long exposure Strehl ratio
$\overline{\mathcal{S}}$ and its associated coherent energy
$\mathrm{e}^{-\sigma^2_{\phi_r}}$ where $\sigma^2_{\phi_r}$ is the
variance of the residual phase of a single telescope \citep{noll_1},
and of the phase structure function in presence of partial AO
correction $\mathcal{D}_{\phi_r}({\vec r})$ as modelled in Appendix \ref{app_ao}.\\
Figure \ref{fig_perf} (top) shows the multimode and single-mode phase
error as a function of the number of incoming photoevents
(alternatively, source $K$-band magnitude), for different levels
of AO correction (i.e. different Strehl ratios).  In a general way,
the behavior of the error is in both cases similar: at low fluxes, the detector noise dominates with a
slope of the error in $1/N$, then it goes by a (short)
photon noise regime in  $1/\sqrt{N}$ and eventually reaches
for the brightest sources  a plateau due to speckle noise -- analog to the one
known in the case of the visibility \citep{goodman_1, tatulli_1} -- which limits
the ultimate precision on the phase. Note already that
in order to keep the phase noise below reasonable levels, that is 
smaller than 1 radian, it is mandatory to equip interferometers with
AO systems, to at least insure a low-order correction of the turbulent
phase providing Strehl ratios greater than $\sim 0.1$ for $D/r_0=8$ and $\sim 0.5$ for $D/r_0=2$.\\
Going in further details, Fig. \ref{fig_perf} (bottom) shows the ratio
of the error of the multimode phase by the error of the single-mode
phase as a function of the number of photoevents. Clearly, two
different behaviors occur whether we are considering photon-starved or
photon rich regime, as discussed in the following.

\subsection{Low light level regime - limiting magnitude}\label{sec_limag}
In the case of faint sources, the phase error is dominated by the detector noise $\sigma_{det_{\phi}}$. The comparison between both multimode and single-mode cases is however not straightforward because the number of pixels required to sample the fringes is specific to each technique,  and depends on the chosen combination scheme. Let us first review here the possible technical solutions:\\
{\it Multimode - (image plane) multiaxial combination:} 
In this case,  each speckle in which the interferograms are formed must be sampled correctly, that is it must be crossed by at least two pixels \citep{chelli_2}. Then, if we take the whole image of size $\lambda/r_0$, the total number of pixels required becomes $N_{pix} = 2D/r_0$. As a consequence, in such a combination mode, $N_{pix}$ is dependent of the turbulence.  In the case of the VLTI at Cerro Paranal, the average $r_0$ is $\simeq 1$m in the K-band, which gives turbulence strengths of $D/r_0 \simeq 8$ and $D/r_0 \simeq 1.8$ for the UTs ($D=8$m) and ATs ($D=1.8$m). This corresponds to a number of pixels that can vary quite a lot with the strength of the turbulence, respectively $N_{pix} = 16$ (UTs) and $N_{pix} = 4$ (ATs).\\
{\it Multimode - (pupil plane) coaxial combination:} Providing  that the interferogram is scanned faster than the coherent time of the atmosphere in order to ``freeze'' the fringes, the minimum number
of pixels to obtain full information on the temporal interferogram is 3, which corresponds to the 3 degrees of freedom: incoherent flux, fringe amplitude, and phase. However, instead of this ``ABC'' scheme with 120 degrees between the three channels, it is frequently more practical to implement an ``ABCD'' scheme with 90 degree phase shifts, hence requiring 4 pixels. Note that in the framework of phase tracking, it is also possible to only measure the sine component of the fringe with ``AC'' scheme  with 180 degrees between the two channels. In any case, the number of pixels needed in a coaxial pupil plane combiner is between 2 and 4. \\
{\it Single-mode - multiaxial combination:}  On the contrary of the multimode/multiaxial combination, the number of pixels is here independent of the turbulence as the shape and frequency of the interferogram are fixed by the design of the beam combiner. Typically, the interferogram consists in a sinusoidal signal with a frequency defined by the separation of the output pupils (so-called the coding frequency), and where its amplitude is modulated by the Gaussian envelope of the single-mode device. \citet{tatulli_4} have shown that the optimum number of pixels which respects the Shannon criterion ($>2$ pixels per fringe) and prevents from an overlapping of the photometric and interferometric peaks in the Fourier plane is $N_{pix}=10$.\\
{\it Single-mode - coaxial combination:} As in the coaxial multimode case, one just need here to sample the interferogram with respect to the 3 degrees of liberty. As a result, the number of pixels is again between 2 and 4. Note that instead of the usual temporal coding, fringes can be scanned simultaneously thanks to ``ABCD-like'' integrated chip devices \citep{benisty_1}. \\
As a consequence, coaxial schemes -- both in multimode and single-mode cases -- appear more appropriate since they are using substantially less pixels than multiaxial ones. We remark however that these conclusions apply in the framework of fringe tracking where pair-wise combinations are favored. They may differ in the context of  interferometric imaging instrument where an important number of baselines is involved. In the following we will consider coaxial combiners with $N_{pix}=4$, corresponding to the standard ``ABCD'' sampling. 
\\ \\
As multimode and single-mode fringe trackers finally require the same number of pixels, a straightforward comparison of the expressions of the detector noise shows that multimode combiners will achieve in this regime slightly better performance by a factor of $1/\rho_0$, because single-mode spatial filters can not transmit $100\%$ of the flux for the geometrical reasons mentioned in previous sections. If we define the limiting magnitude\footnote{note that the definition of the  limiting magnitude depends on the estimator chosen to measure the phase. In this paper, the interferometric phase is merely estimated at the baseline frequency $f_{12}$, that is at the top of the interferometric peak (see Appendices \ref{subapp_multimode_noise} and \ref{app_singlemode_noise}). If a different estimator is used, like integrating the high frequency peak over the frequency range [$f_{12}-D/\lambda, f_{12}+D/\lambda$] \citep[see e.g.][in the case of the squared visibility estimators]{roddier_1,mourard_1} the expression, and therefore the value of the limiting magnitude will change accordingly.} such as the error of the phase is
equal to $1$ rad, which -- apart from very low AO correction levels --
occurs in this detector noise regime, the corresponding limiting flux is given by:
\begin{equation}
K_{lim}^{[\mathrm{multi,single}]} = \frac{\sqrt{2}\sqrt{N_{pix}}\sigma_{det}}{[1,\rho_0]V_{12}\mathrm{e}^{-\sigma^2_{\phi_r}}}
\end{equation}
with the factor $1$ or $\rho_0$ depending whether we are using
multimode or single-mode interferometers. As illustrated by Fig. \ref{fig_limitmag}, the gain in limiting magnitude for multimode combiners is $\sim 0.25$ magnitude.

\subsection{High light level regime: the speckle noise}\label{sec_specknoise}
\begin{figure}
\begin{center}
\includegraphics[width=0.44\textwidth]{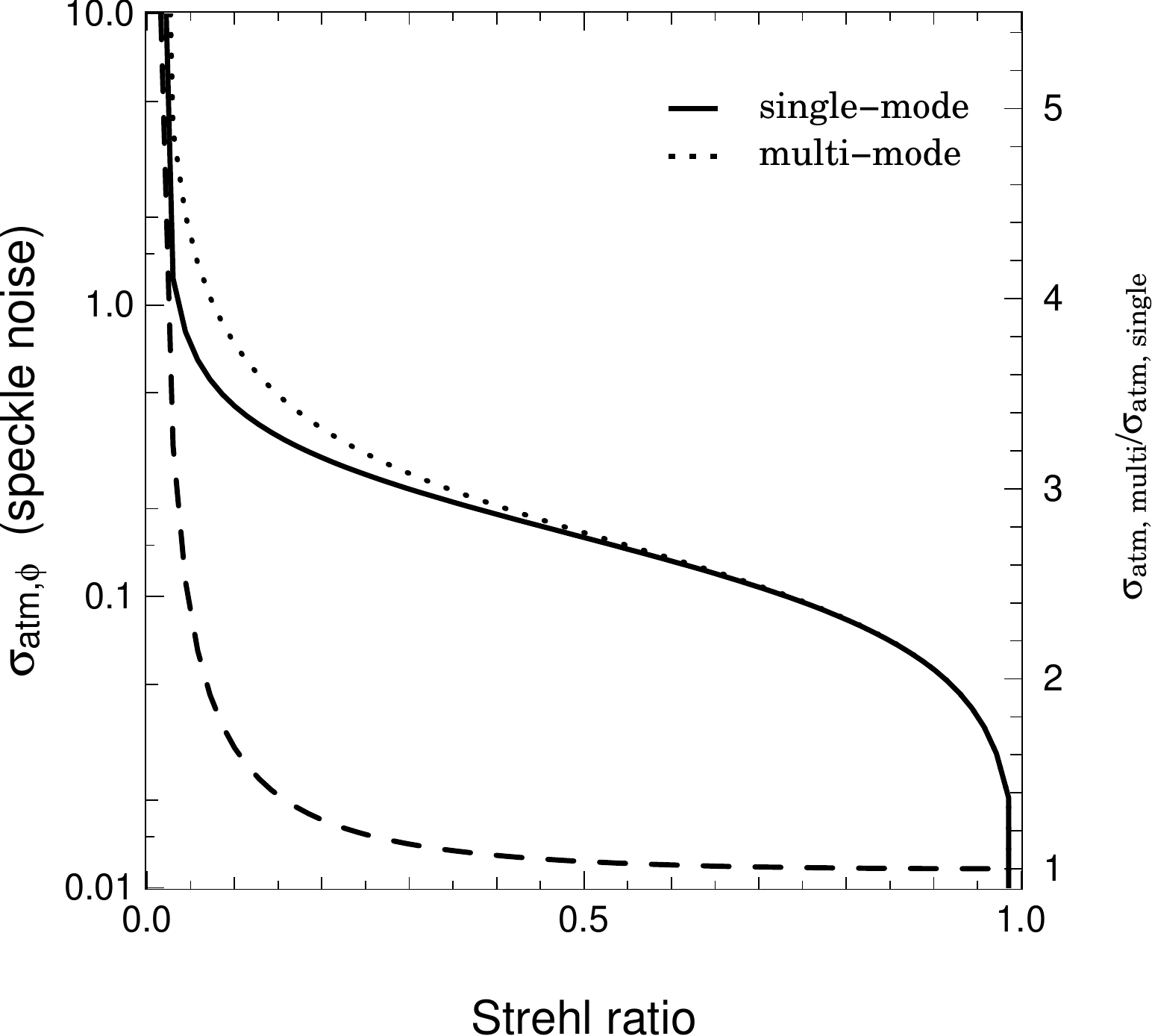}
\caption{\label{fig_specknoise} Phase speckle noise as a function of
  the Strehl ratio for both single-mode (solid line) and multimode
  (dotted line) cases, with $D/r_0 = 8$. The ratio between the multimode and the
  single-mode phase errors is plotted in dashed line (y-axis on the
  right side).}
\end{center}
\end{figure}In the photon-rich regime, the speckle noise dominates and the error on the interferometric phase reaches a plateau, that is, is not dependent on the flux of the source anymore. As illustrated on Fig. \ref{fig_specknoise}, single-mode interferometers always provides in such a case lower phase error than multimode ones, emphasizing the remarkable properties of spatial filtering of the turbulent wavefront by single-mode devices. Such behavior has been already noticed  by \citet{tatulli_1} for the estimation of the squared visibility,
showing that the so-called {\it modal} speckle noise of the
visibility was, for a given AO correction, always smaller than
classical speckle noise of multimode interferometers. 
The concept of modal speckle noise can also be applied for the
single-mode interferometric phase. However at the difference of the
squared visibility for which the modal speckle noise is 0 for a point
source, the phase modal speckle noise always exists, independently of
the size of the source.\\
Note that the gain of using single-mode schemes is
all the more important than the the level of correction is low. If for
fairly good correction with Strehl ratio above $20\%$, the difference
remains marginal with a factor $\sim 1-1.5$ between the single-mode
and multimode phase error, the situations where bright sources are
observed with low/none AO correction will highly benefit of
single-mode interferometers. In such cases the precision of the phase
can increase by at least a factor of $2$ and much more when using
spatial filtering of the corrugated wavefront with a typical Strehl
  ratio below $10\%$. This is a typical
counter-intuitive example where it is more
profitable to loose a substantial part of the flux and keep only the
coherent part of the perturbed wavefronts than conserving the whole
flux at the price of introducing additional atmospheric noise.

\section{Application to fringe tracking}
In this section, we apply the formalism developed previously in the
context of on-axis and off-axis fringe tracking, that is when the phase is
estimated and compensated in real time to stabilize the fringes on
the detector. We focus on the relative performance of fringe tracking
systems using whether multimode or single-mode schemes.

\subsection{Coherent integration}

\begin{figure}
\begin{center}
\includegraphics[width=0.4\textwidth]{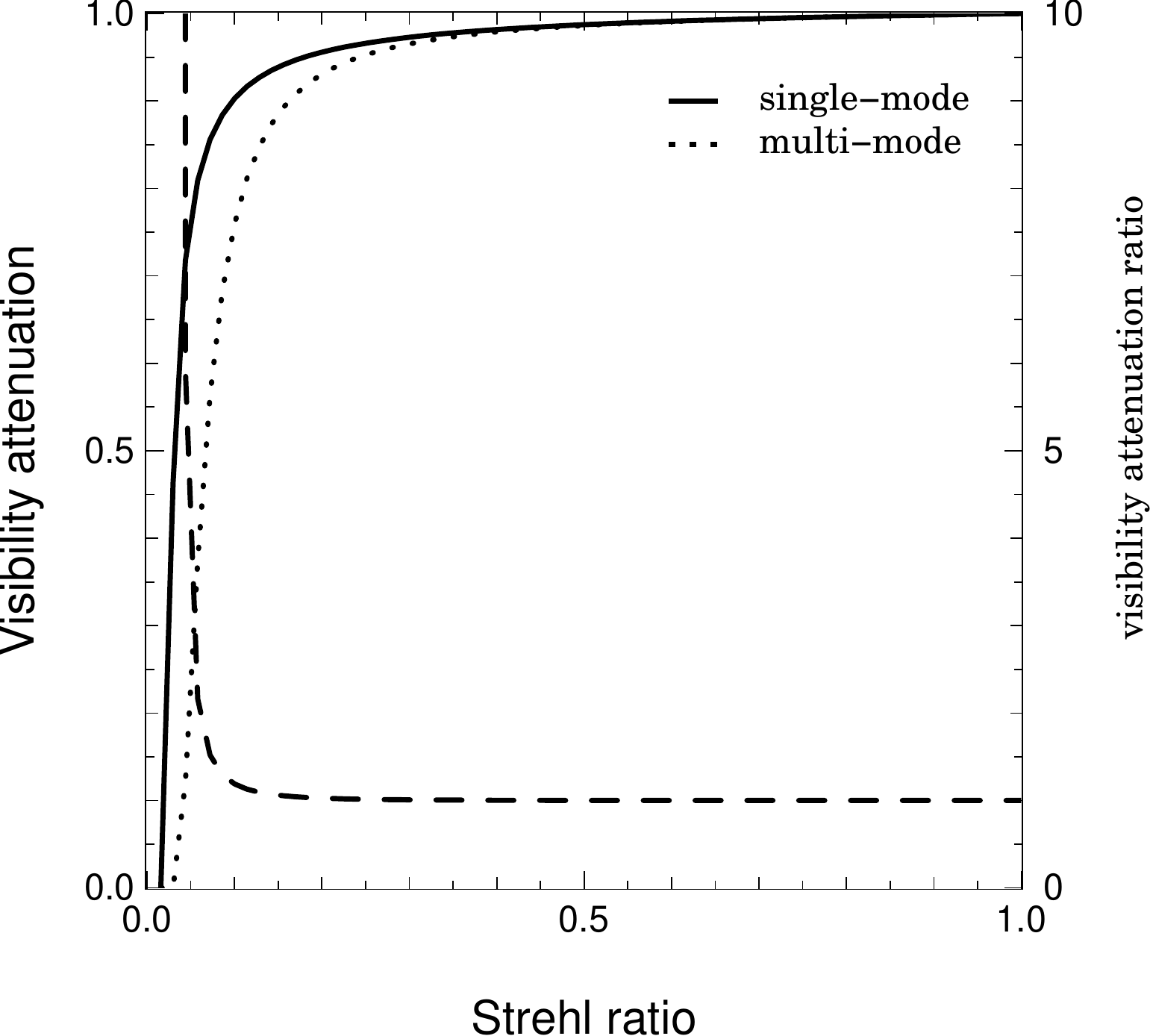}
\caption{\label{fig_visatt} Long exposure visibility attenuation due
  to the atmospheric noise of the measurement of the phase, as a function of
  the Strehl ratio  for multimode (dotted line) and single-mode (solid
  line) systems, with $D/r_0 = 8$. The ratio of the attenuation
between single-mode and multimode case is plotted in dashed line (y-axis on the
  right side).}
\end{center}
\end{figure}
For each interferogram, fringes are shifted from the zero optical
path difference by the turbulent piston. This piston is corrected
in real time by the fringe tracker which estimates the phase of the
fringes and compensates the optical path difference in real time by
moving dedicated mirrors. 
\\
Centering the fringes in real time allows to perform coherent
integration of the signal, that is to integrate on time scales much longer
than the coherence time of the atmosphere. However, the estimated
piston used for the opd correction is affected by a random
measurement error $\epsilon(t)$. As a result, the interferogram is not
perfectly centered and is still moving with an excursion depending on the statistics of the noise. If the interferograms are integrated 
over a time long enough for the realizations of the random error
$\epsilon(t)$ to span all the range of its probability law, then, by
rules of ergodicity, the visibility will be affected by a loss of contrast\footnote{For sake of simplicity, we assume here that the phase is compensated instantaneously, hence not taking into account
the delay of the fringe tracking loop  between the measurement of the phase and its correction. The problem of time delay, which is independent of the multimode or
single-mode nature of the fringe-tracking system, is treated in \citet{conan_1}.}  which writes:
\begin{equation}
<V>_t = <V>_{\epsilon} = V e^{-\frac{\sigma^2_{\phi}}{2}}
\end{equation}
where $<>_{\epsilon}$ is the ensemble average over the realizations of
the noise, and $\sigma^2_{\phi}$ is the variance of this noise
as computed in Sect. \ref{subsec_noise}. 
Hence, for long integration times, the
fact that the instantaneous turbulent phase is corrected only to the
precision of its estimation, introduces an attenuation of the coherent
flux (equivalently, of the visibility) by a factor $e^{-\frac{\sigma^2_{\phi}}{2}}$. Such bias therefore depends on
whether single-mode or multimode fringe tracking is used.\\
For high-light level regime in which fringe tracking is mostly
expected to work, Fig. \ref{fig_visatt} shows the attenuation of the visibility
as a function of the Strehl ratio for both single-mode and multimode
systems. We can see that acceptable loss of contrast, typically $\ga 0.8$,
is achieved as soon as moderate AO corrections with Strehl
$\ga 0.1$ (for $D/r_0=8$) is provided. On the contrary, for lower performance of the AO system, the attenuation
coefficient drops rapidly and the advantage of tracking the fringes is
lost due to the uncompensated turbulent fluctuations of the phase over
the pupils.\\
One important conclusion to draw is that, for a given AO correction, the visibility attenuation induced by the noise of the phase is always smaller when using single-mode fringe tracking systems instead of multimode ones. In other words, the maximum achievable atmospheric contrast is always higher using single-mode
fringe tracker than multimode systems.  This is especially true for
low AO correction cases where the attenuation can becomes twice larger and more, as shown in Fig. \ref{fig_visatt}.

\subsection{Phase jumps}
\begin{figure*}
\begin{center}
\begin{tabular}{ccc}
\includegraphics[width=0.3\textwidth]{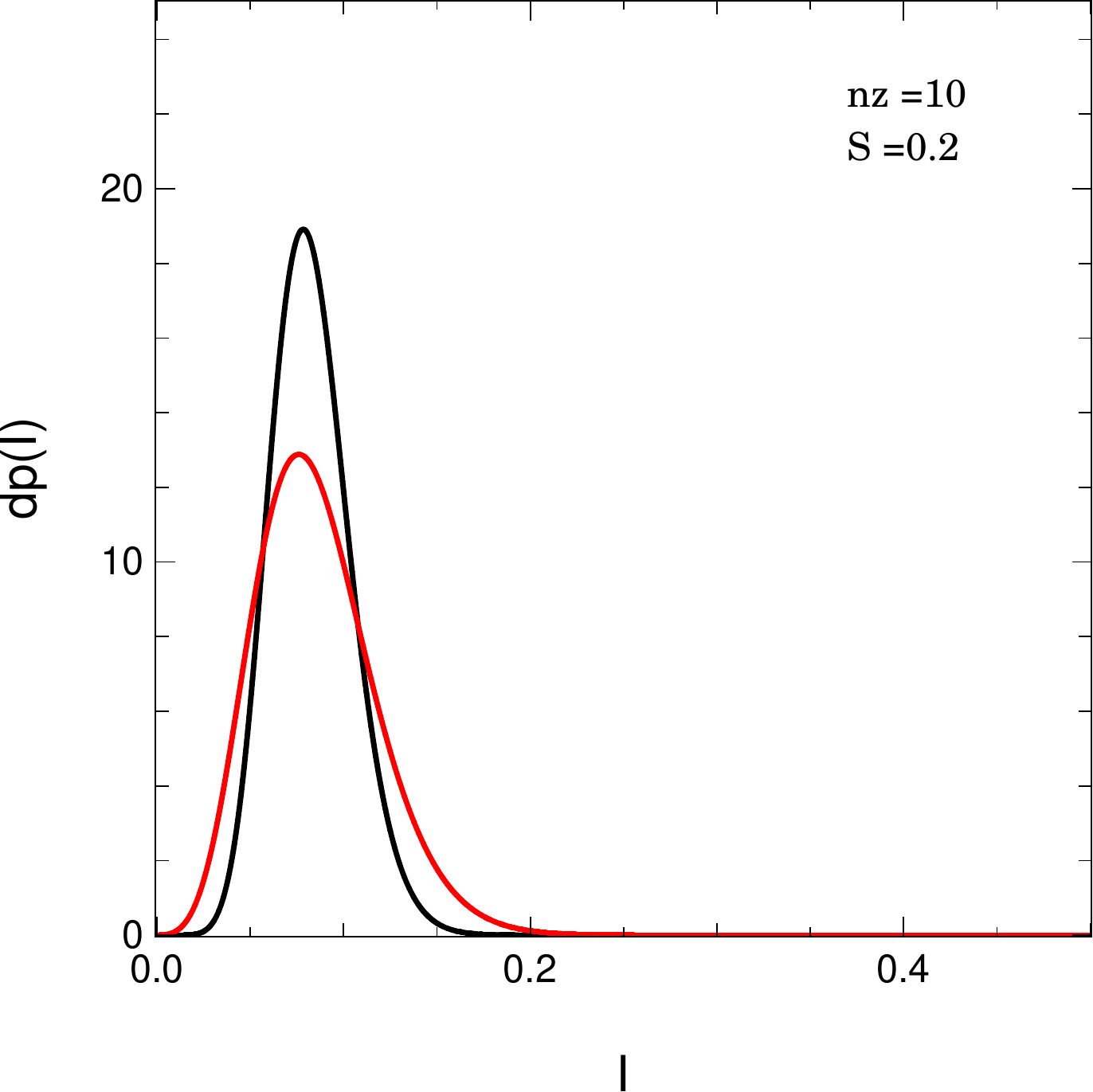}
&  \includegraphics[width=0.3\textwidth]{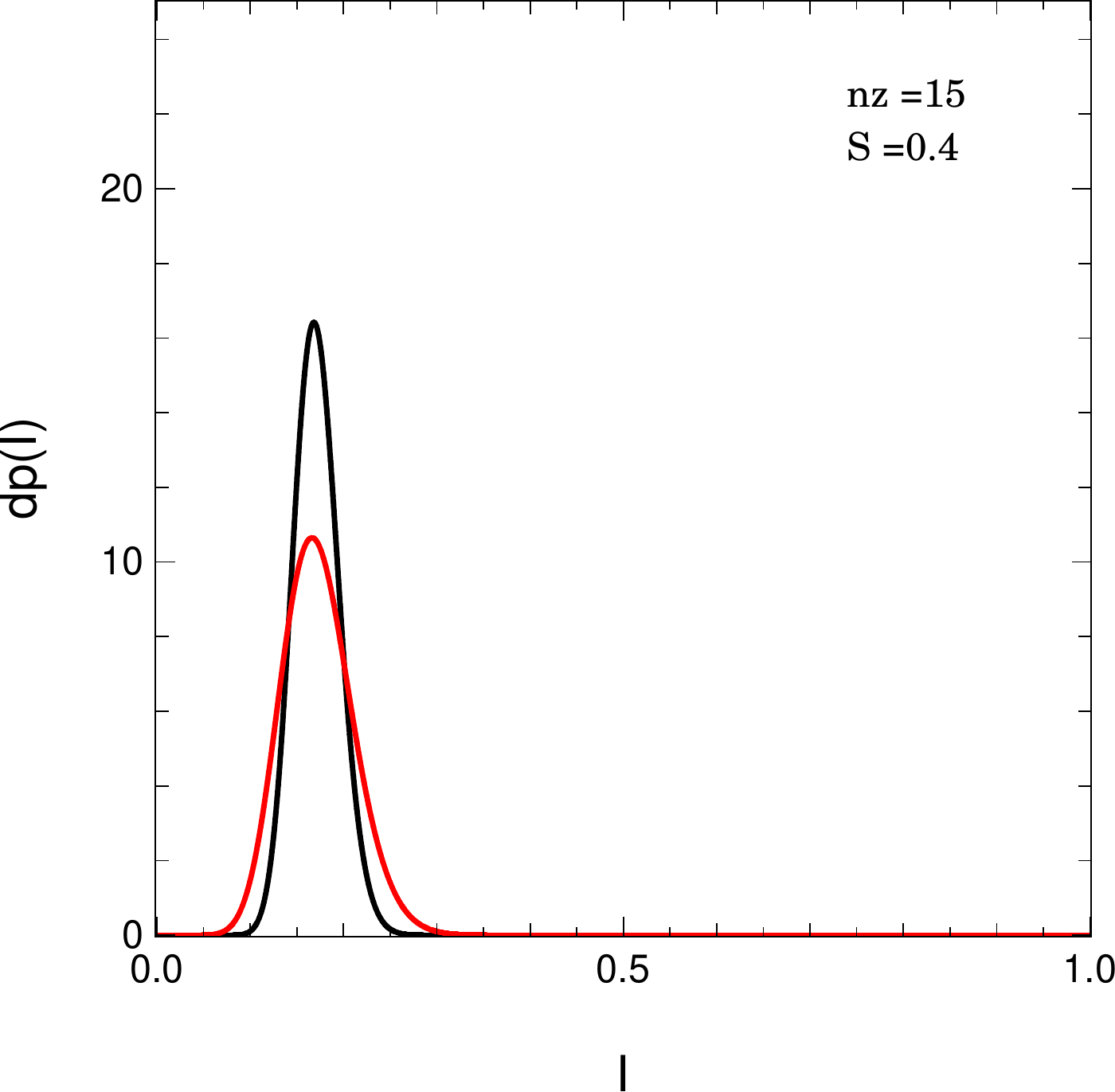}
&  \includegraphics[width=0.3\textwidth]{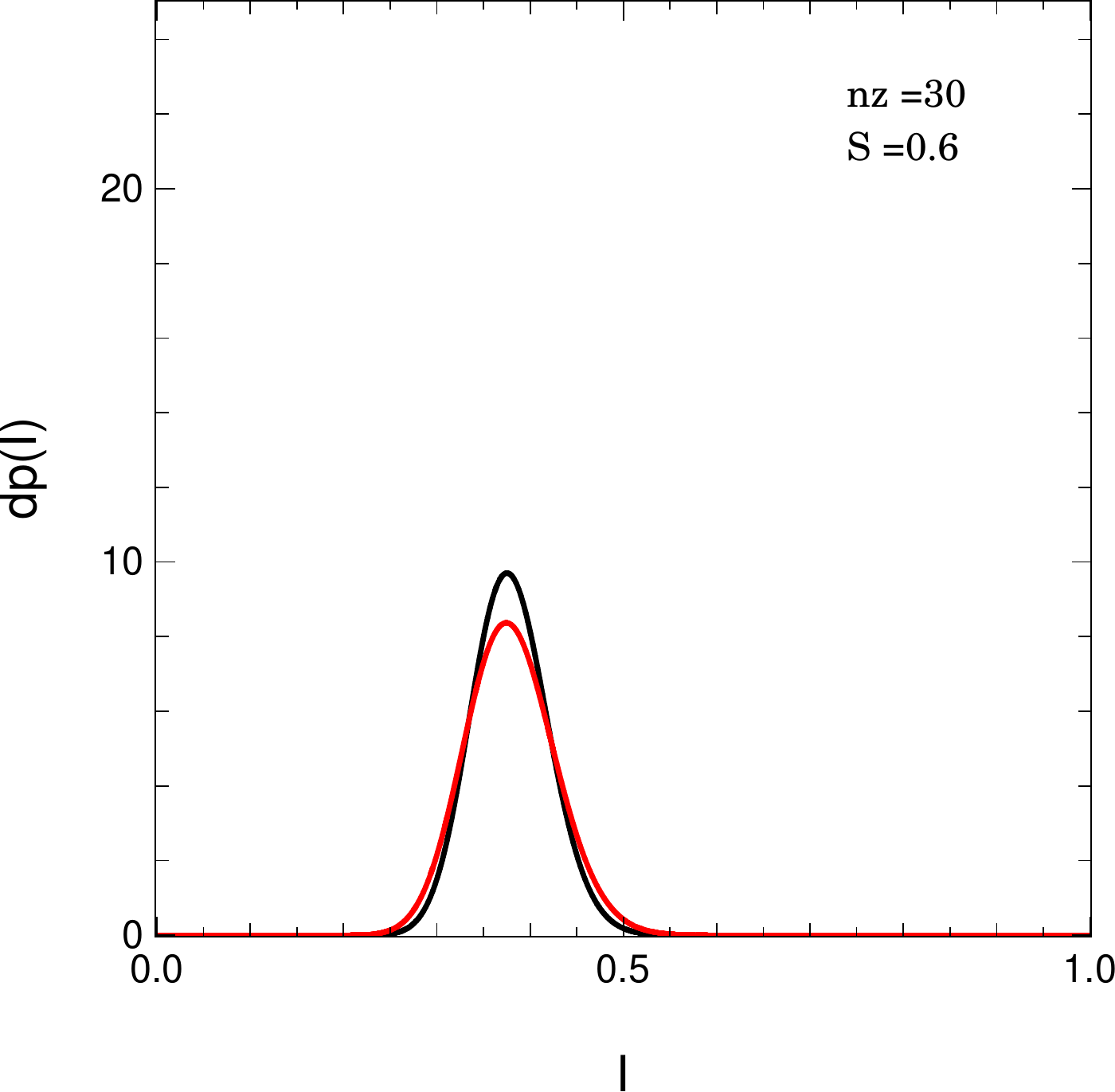}\\
\includegraphics[width=0.3\textwidth]{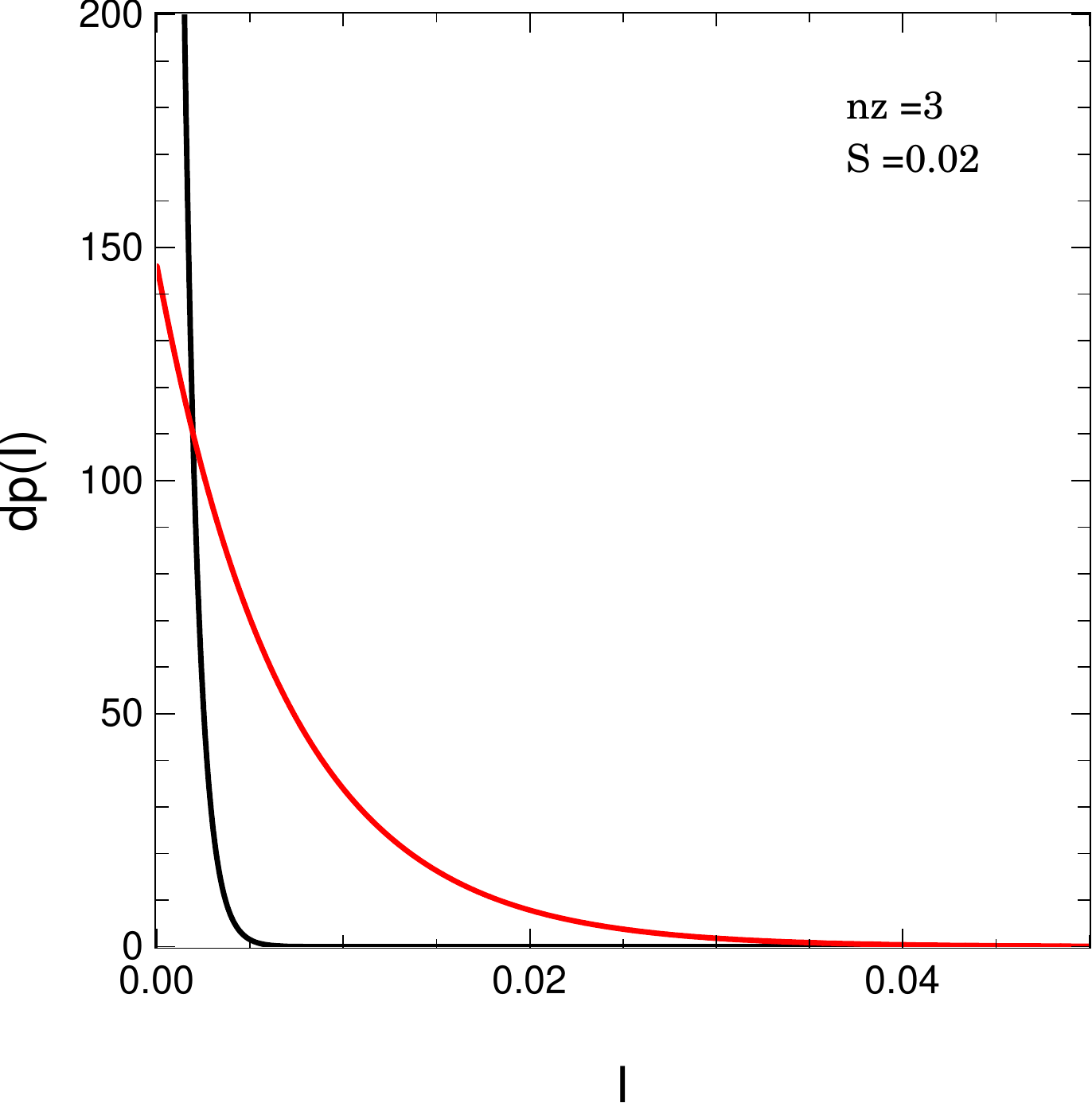}
&  \includegraphics[width=0.3\textwidth]{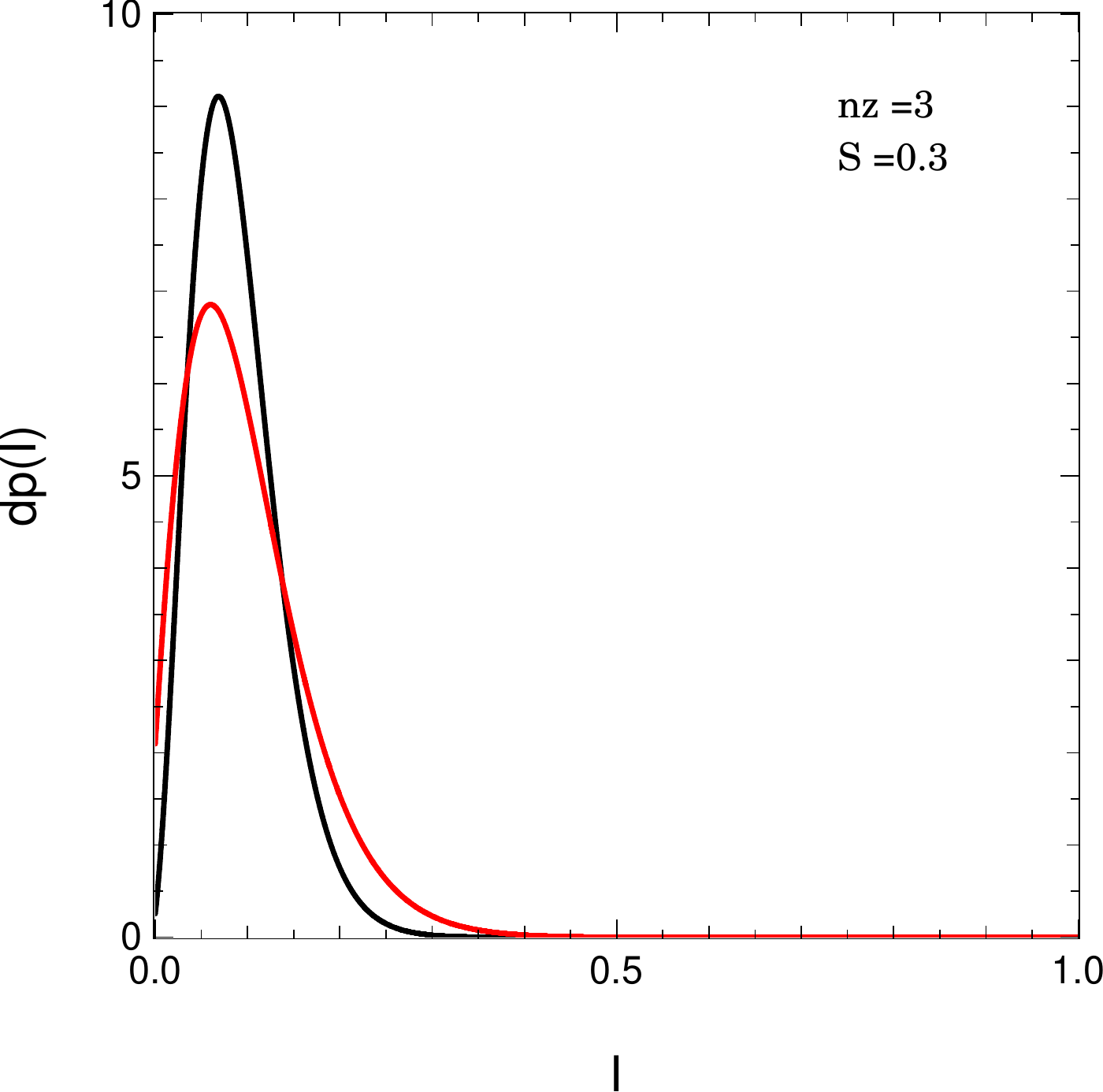}
&  \includegraphics[width=0.3\textwidth]{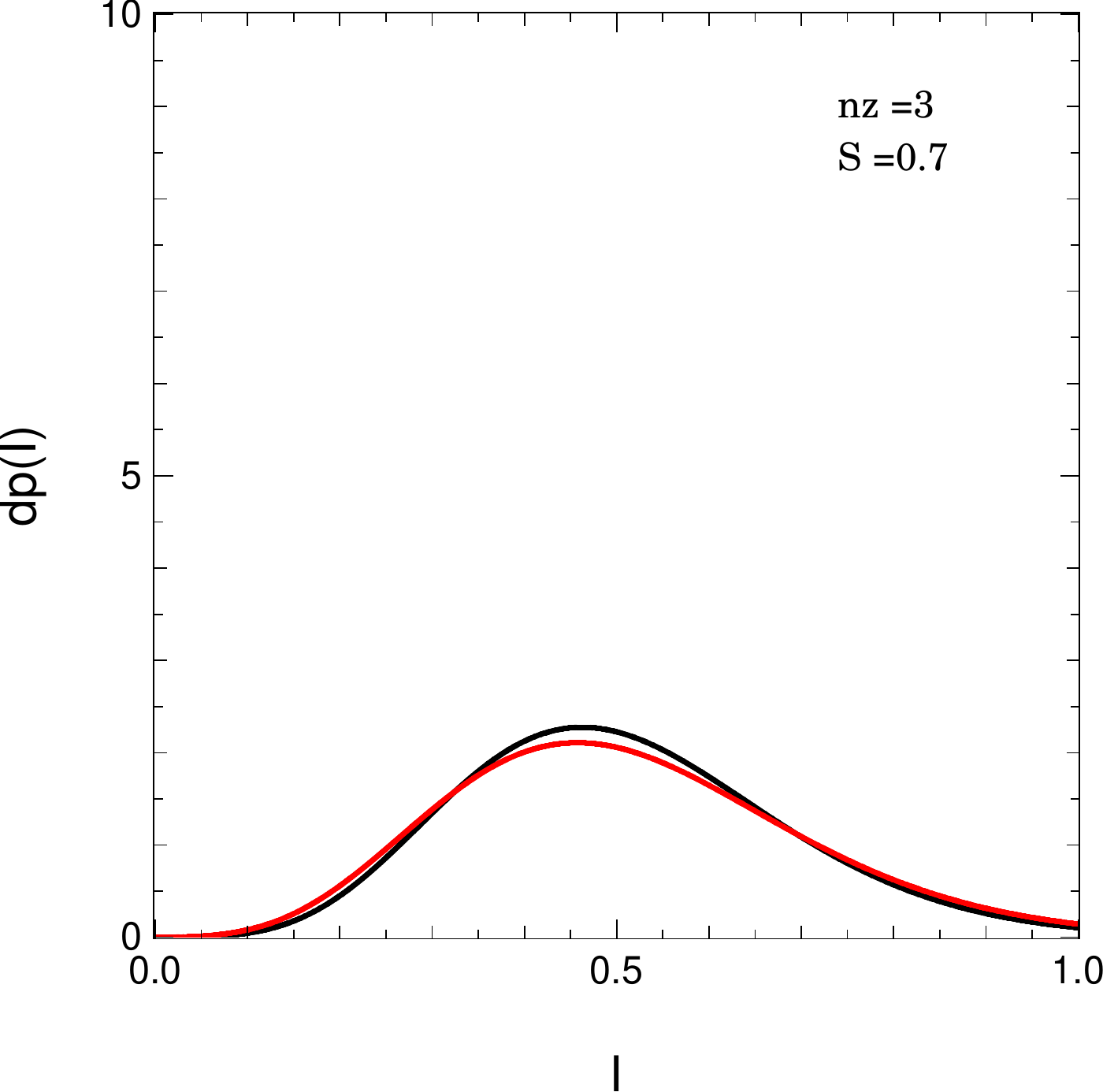}
\end{tabular}
\caption{\label{fig_rician} Density probability of the multimode
  instantaneous Strehl ratio ($I=|T_{12}|^2$, red lines) and
  single-mode one ($I=|\rho_{12}|^2$, black lines) for various levels of AO
correction and turbulence strengths. Top: constant turbulence strength
$D/r0 = 8$ and the number of
corrected Zernike are, from left to right, $n_z = 10, 15, 30$. Bottom:
constant AO correction $n_z = 3$ (tip-tilt only), with 
various turbulence strengths, from left to right $D/r_0 = 8, 4, 2$.}
\end{center}
\end{figure*}
It is usually implicitly assumed that the phase estimated in
interferometry is effectively the shift of the fringes with respect to
the zero opd, namely the piston phase. By definition, the piston phase shift is the average differential phase across the apertures, as illustrated on Fig. \ref{fig_turb} and described in Eq. (\ref{eq_phi12p}) that can be rewritten:
\begin{equation}
\phi^p_{12}(t) = \int_{\Sigma} [\phi^r_1({\vec r},t) - \phi^r_2({\vec r},t)] \mathrm{d}{\Sigma} \label{eq_pistonphase}
\end{equation}
following the formalism of \citet[see Eq. (6)]{buscher_1},  where $\mathrm{d}{\Sigma}$ is an elemental within the aperture $\Sigma$. \\
As first brought up by this author, the practical estimation $\widetilde{\phi^p_{12}}(t)$  of such piston phase is performed by taking the argument of the complex phasor averaged over the pupils, that is: 
\begin{equation}
\widetilde{\phi^p_{12}}(t) = \mathrm{arg}  \left(\int_{\Sigma}
  e^{i(\phi^r_1({\vec r},t) - \phi^r_2({\vec r},t))} \mathrm{d}{\Sigma} \right) \label{eq_phaseestim_multi}
\end{equation}
in the multimode case, and:
\begin{equation}
\widetilde{\phi^p_{12}}(t) = \mathrm{arg} \left(\int_{\Sigma} e^{i \phi^r_1({\vec r},t)}\mathrm{d}{\Sigma}  \int_{\Sigma} e^{-i \phi^r_2({\vec r},t)} \mathrm{d}{\Sigma} \right) \label{eq_phaseestim_mono}
\end{equation}
in the single-mode one. \\
By comparing Eq. (\ref{eq_pistonphase}) to Eqs. (\ref{eq_phaseestim_multi},\,\ref{eq_phaseestim_mono})  we can see that the true piston phase $\phi^p_{12}$ and the estimated one $\widetilde{\phi^p_{12}}$ are literally not the same, unless when there is no phase fluctuation (i.e. no turbulence or a perfect correction of the wavefronts) across the apertures. \\
When considering
a small amount of fluctuations $\delta\phi_{12}({\vec r},t)$ such as
$\phi^r_1({\vec r},t) - \phi^r_2({\vec r},t) = \delta\phi_{12}({\vec r},t)$, a Taylor expansion of Eqs. (\ref{eq_phaseestim_multi},\ref{eq_phaseestim_mono}) can show that $\phi^p_{12}$ and
$\widetilde{\phi^p_{12}}$ will then deviate roughly as $\int_{\Sigma}
[\delta\phi_{12}({\vec r},t)]^3 \mathrm{d}\Sigma$ \citep{buscher_1}.  
But if these fluctuations are large (typically above 1 radian), there can be strong
differences between the piston phase and the argument of
the complex phasor. More precisely, strong divergences are occurring when the complex phasor has a very small amplitude and eventually crosses
the origin of the complex plane. In such cases, rapid phase jumps of
the argument of the phasor can be experienced whereas these jumps are not
seen in the piston phase. As a consequence the correction
performed by the fringe tracker may be highly wrong and fringes potentially lost, especially if the science camera is working at a different wavelength
than that of the fringe tracker \citep{buscher_1}, as it is the case
e.g. for the FINITO instrument (fringe tracking in H-band and
correction in K-band, \citet{gai_1, lebouquin_1}). \\
It is therefore important to know the probability of phase jumps to
occur in order to ensure an observational/instrumental context in which these events are avoided as much as possible. So far, the conclusions about this point are not clear : if \citet{tubbs_1} implied that spatial filtering could be the source of such anomalies, \citet{buscher_1} on the contrary argues that AO correction and most of all spatial filtering help to reduce these singularities. All the previous analysis were however based only on simulations. We propose here a theoretical analysis of this phenomenon. We emphasize that our analysis focuses on the rate of occurrence of phase jumps, as defined above. We do not study the consequences of these phase jumps in terms of effective loss of fringe tracking. Such a causality will depend on the system (e.g. single wavelength vs. multiple spectral channels methods) used to practically estimate and correct the opd. Considering that a phase jump would systematically induce a failure in the fringe tracking system therefore corresponds to the worst case scenario.\\
The fact that the coherent flux drops to zero because of turbulent
phase fluctuation depends on  whether the interferometric transfer
function $|T_{12}|$ for multimode systems or the
interferometric coupling coefficient  $|\rho_{12}|$ in the single-mode
case, drops to zero. We thus want here to establish the probability
density of these quantities, and study how likely they can take very
small values. \citet{canales_1} have studied the distribution of speckle statistics
in presence of partial AO correction for mono-pupil telescopes.
 They have shown that the density probability $dp(I)$ of the intensity
 I at the center of the image -- that is by definition, the instantaneous Strehl ratio --
 follows a Rician statistics of the form:
\begin{equation}
dp(I) = \frac{1}{2\sigma^2}\exp\left(-\frac{I+a^2}{2\sigma^2}\right)I_0\left(-\frac{a\sqrt{I}}{\sigma^2}\right)
\end{equation}
where $a^2$ and $\sigma^2$ depends on the first and second order
moments of the real and imaginary part of the complex phasor
describing the AO corrected turbulent wavefront. Note that such analytical definition of the density probability implicitly assumes a time-constant $r_0$ to characterize the turbulence. This hypothesis is the limitation of our model as $r_0$ may actually vary on time scales of minutes and shorter, driving to brief episodes of very small $r_0$ that would trip up AO systems and lead
to very small Strehl ratio for a brief period of time. However, modelling this effect requires more complex and heavy simulations of partially AO-corrected turbulence which are out out the scope of this paper.\\
We have emphasized in Sect. \ref{sec_formalism} that $|T_{12}|^2$ and
$|\rho_{12}|^2$ represent the instantaneous interferometric Strehl ratio, for the
multimode and single-mode cases, respectively. Therefore, by
straightforward analogy with an interferometric pupil instead of a
monolithic one, we can show that $I=|T_{12}|^2$ and $I=|\rho_{12}|^2$ are also
following a Rician distribution. The parameters $a^2$ and $\sigma^2$
relative to both instantaneous interferometric Strehl ratio can be
directly derived from the formalism of previous section and their
expressions are given in Table \ref{tab_rician} of Appendix \ref{app_rician}.\\
\begin{figure*}
\begin{center}
\begin{tabular}{cc}
\includegraphics[width=0.4\textwidth]{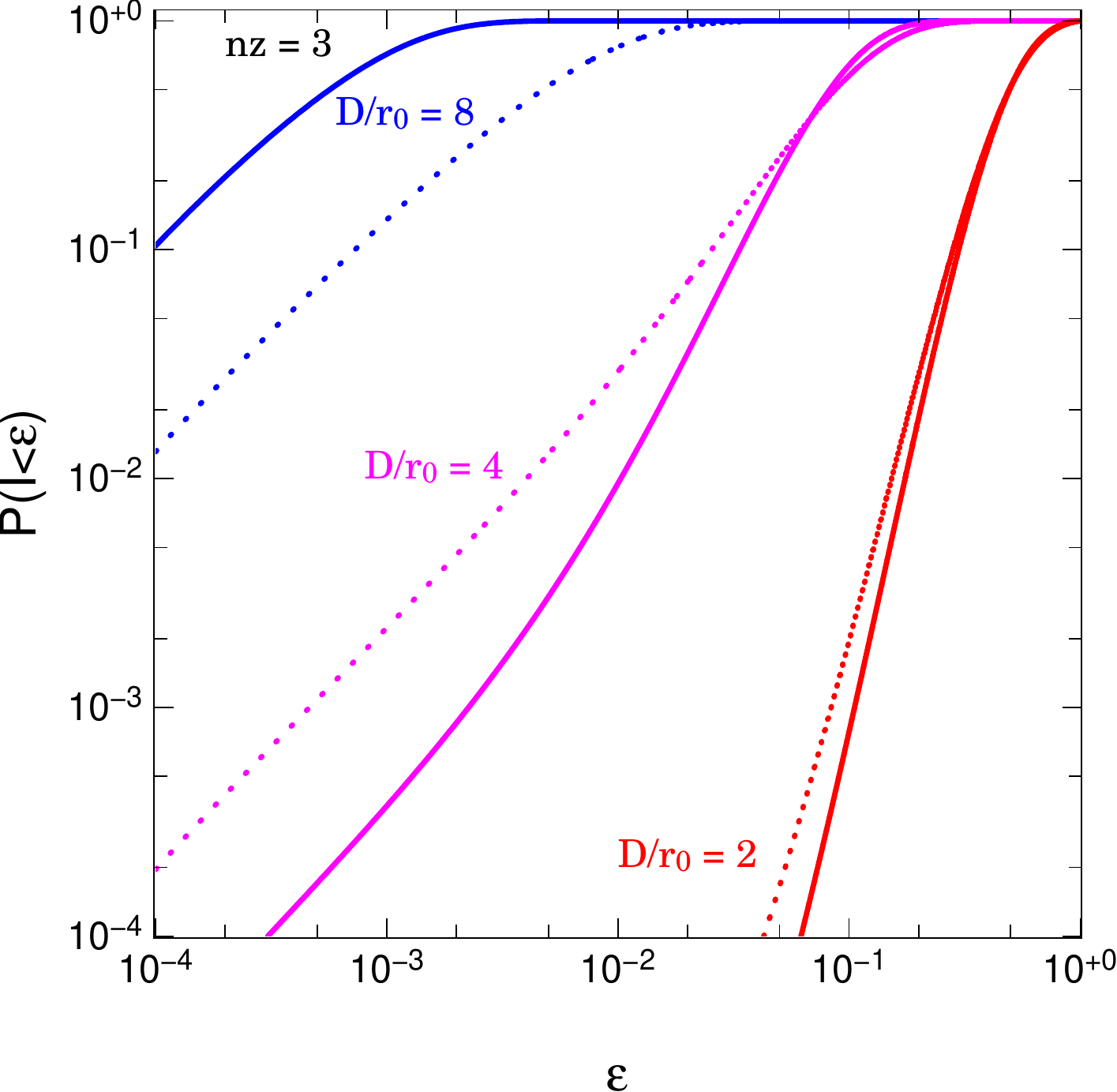}&\includegraphics[width=0.4\textwidth]{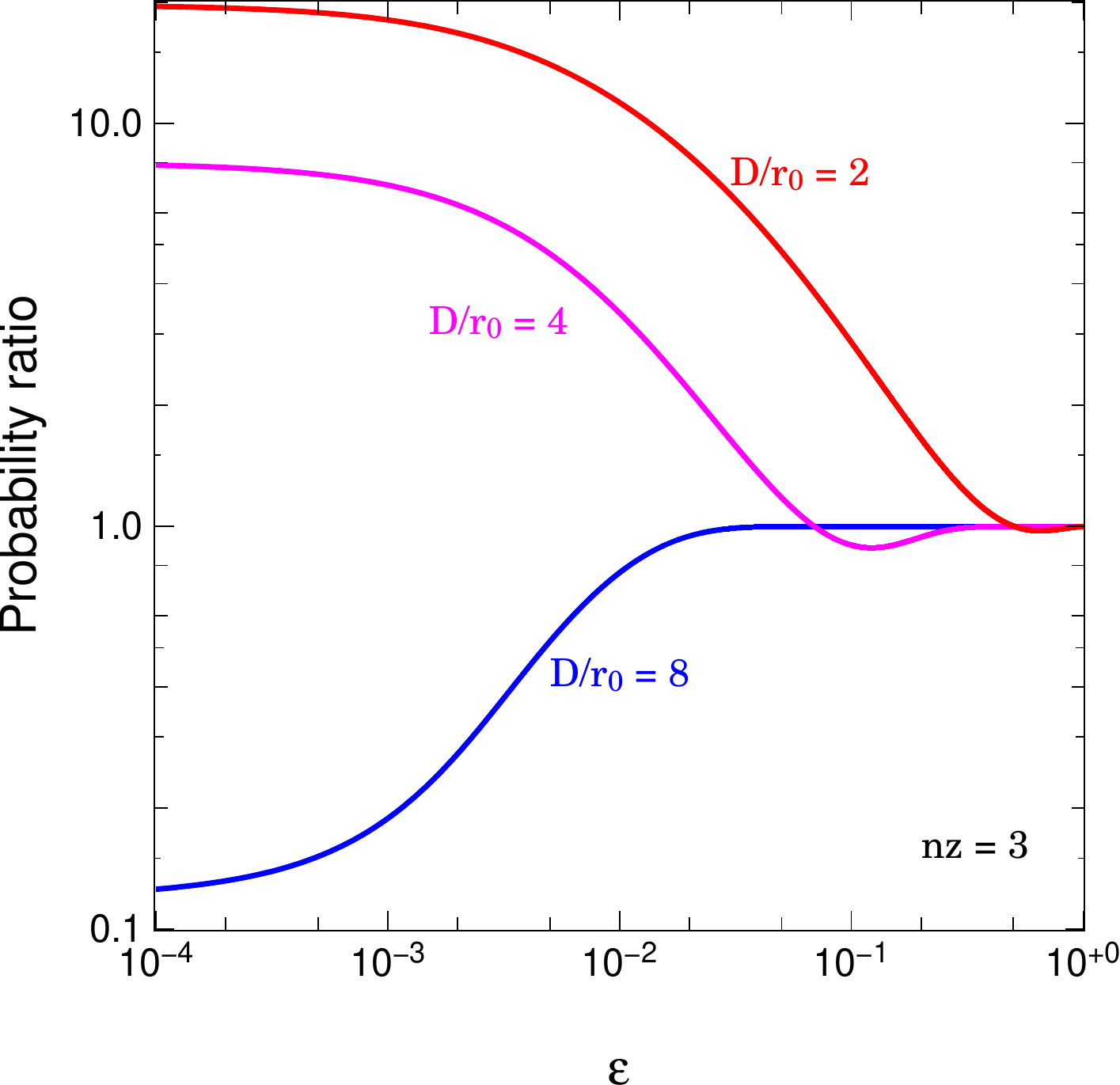}
\end{tabular}
\caption{\label{fig_rician_2} Left: probability for the instantaneous interferometric Strehl ratio I to be lower than the value $\epsilon$, in the case of tip-tilt correction only, for different strengths of turbulence $D/r_0 = 8, 4, 2$ (respectively in blue, magenta and red), for the multimode (dotted line) and single-mode (solid line) cases. Right: Multimode versus single-mode probability ratio.}
\end{center}
\end{figure*}
Figure \ref{fig_rician} shows the density probability of
$|T_{12}|^2$ and $|\rho_{12}|^2$ for various levels of AO
correction and turbulence strength. \\ \\
\textbf{High-order AO correction:} First, we can see that, as soon as
moderate AO correction is applied -- roughly a
few tens of modes or a Strehl ratio $\ga 0.3$, the probability
density displays a bell shape, all the more narrow than the AO
correction is high, and the probability
that $|T_{12}|$ or $|\rho_{12}|$ goes to very low values is null
(typically,  for $n_z = 15$ and $D/r0=8$, the probability for the
interferometric instantaneous Strehl ratio to be $< 0.01$ is $< 10^{-9}$, and  $< 0.05$ is $< 0.02\%$), emphasizing again the relevance of associating high-performance AO
systems to fringe tracking devices, the wavefront being spatially filtered or not. We note that the intensity mean value ($<I>
= 2\sigma^2 + a^2$) is systematically slightly higher in the single-mode
case and the dispersion ($\sigma^2_I = 2\sigma^2 + 4\sigma^2a^2$)
slightly smaller. The difference is however not critical. \\ \\  
\textbf{Low-order AO correction:} For low-order AO such as tip-tilt correction only,
the situation is fairly different. The
shape of the distribution has changed, peaking at zero in the case of
strong turbulence with high values of $D/r_0 \ga 5$. The probability
to have a very small value of the intensity is thus significant, and events such as phase jumps are likely to happen. In other
words, tip-tilt correction is a too low order correction to prevent
from such events. \\
Figure \ref{fig_rician_2} displays the probability $P(I)$ for the
instantaneous interferometric Strehl ratio to reach values close to
zero (i.e. $P(I<\epsilon)$  with $\epsilon$ very small) for different
strengths of the atmospheric turbulence. We can see that, for strong
turbulence ($D/r_0 \ga 5$), the probability $P(I)<\epsilon$ 
is always higher in the single-mode case. In
such cases, we expect phase jumps to occur more frequently with single-mode fringe tracker, as the likelihood that two speckles are
simultaneously entering the single-mode fibers of the two telescopes is weak.  However, this does not mean that phase jumps should not
occur often with multimode fringe tracker too. As an example the
probability to have  instantaneous interferometric Strehl lower than
$0.01$ in the multimode case is $\sim 78\%$ for $D/r_0 = 8$, whereas it is $100\%$ in the
single-mode one. In other words, regardless of the multimode versus
single-mode issue, one should never consider to perform fringe
tracking with big telescope apertures associated with the sole tip-tilt correction.\\ 
For weak turbulence (roughly $D/r_0 \la 5$, Fig. \ref{fig_rician},
bottom middle and right), the distribution starts to look like a bell shape
again, with however non null probability to have a zero intensity. 
But as shown in Fig. \ref{fig_rician_2}
this probability is now higher in the multimode case, as only a few
and big speckles are present in the images. As a matter of fact,
this case is the one treated by \citet[see Fig. 4, with $D/r_0 = 4$, tip-tilt correction]{buscher_1} in his simulations, and we come to the same
conclusion than his: in these conditions, spatial filtering enables to
decrease the number of phase jump events. We however emphasize here  that this
statement is true only for cases of moderate turbulence strength.\\ \\
\textbf{Application to the VLTI:} 
%For $D/r_0 = 4$ the probability to have the interferometric
% instantaneous Strehl lower than $0.01$ is reasonably low, being $\sim
% 1\%$ and $\sim 3\%$, for the single-mode and multimode cases
% respectively, {\bf and the same probability is in both cases $<1\promille$ whe% $D/r_0 = 2$}. 
We recall that at Paranal
the average $r_0$ is $\sim 1$m in the K-band, which gives turbulence
strength of $D/r_0 \sim 8$ and $D/r_0 \sim 1.8$ for the UTs  and ATs, respectively. UTs come with high-order Adaptive Optics systems (MACAO, \citealt{arsenault_1}) therefore one can equally choose multimode or single-mode fringe tracking schemes since (i) phase jumps are unlikely to occur in any case, and (ii) multimode and single-mode fringe trackers will provide the same robustness to eventual phase jumps. In the case of ATs where tip-tilt correction is currently provided, the likelihood to endure phase jumps is also very low. However to maximize the stability of their fringe tracking system, it still seems appropriate to accompany the ATs with single-mode spatial filtering devices as the probability to undergo such phase jumps remains roughly 10 times lower (see Fig. \ref{fig_rician_2}, right). Alternatively, providing higher order AO correction to ATs will fix the phase jump issue.

\begin{figure*}
\begin{center}
\begin{tabular}{cc}
\includegraphics[width=0.4\textwidth]{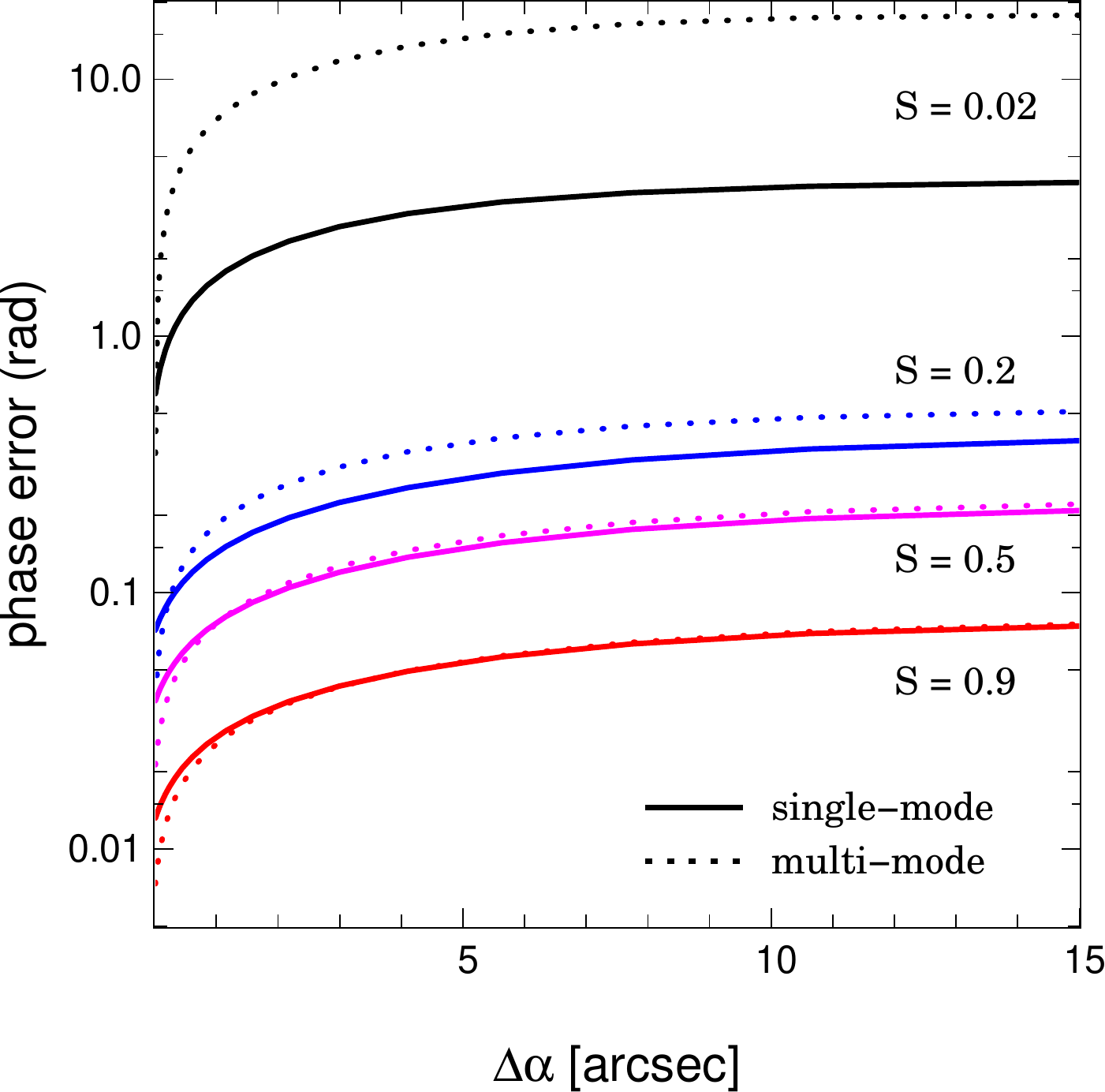}
&  \includegraphics[width=0.4\textwidth]{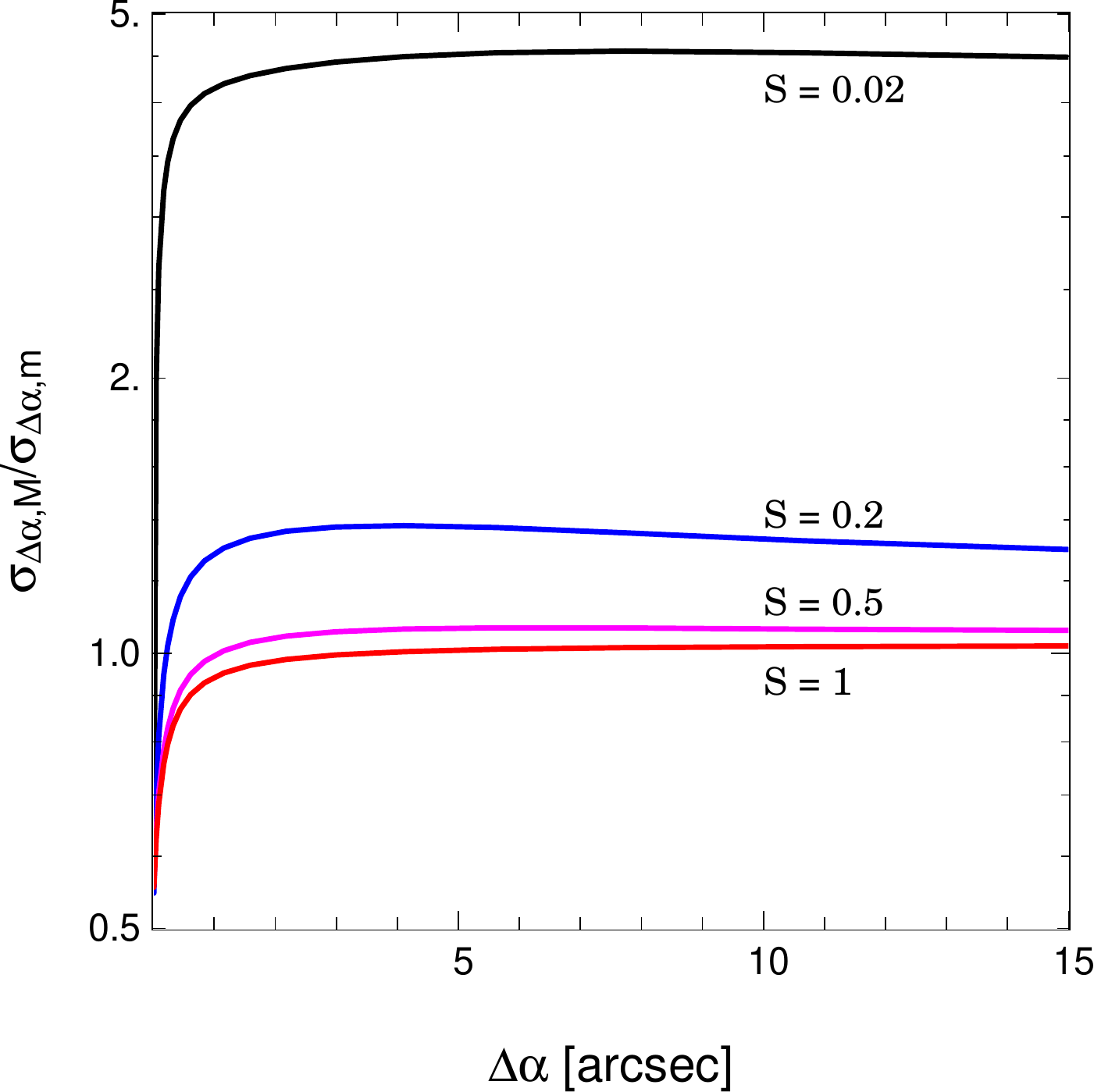}
\end{tabular}
\caption{\label{fig_astrometry}Left: astrometric phase error due to
  atmospheric turbulence, as the function of the angular
distance $\vec\Delta{\alpha}$ between the source and the reference
star, for both single-mode (solid lines) and multimode (dotted lines). 
Right: ratio of the multimode 
 versus the single-mode astrometric error. 
Results are shown for various AO correction levels as described in
previous Figures. We have 
assumed one single turbulent layer located at $h=10$km.}
\end{center}
\end{figure*}
\subsection{Phase referencing - astrometry}
The interest of phase referencing instruments such as PRIMA recently
installed on the VLTI \citep{delplancke_1}, is (i) to provide sub-microarcsecond precision astrometry, allowing e.g. to detect the presence of a faint companion (extrasolar planet) around the central star \citep{launhardt_1}, and (ii) to drastically increase the limiting magnitude of the interferometer by locking the fringes of the (possibly faint) science object on a simultaneously observed bright off-axis reference star whose phase is used as reference \citep{sahlmann_1}. 
Such method  requires that the nearby reference star is close enough -- typically in the isoplanatic patch -- in order to
assume that both the wavefront of the source and the reference are
identically perturbed by the atmospheric turbulence. Strictly
speaking, this assumption is not true as the optical paths of the two
stars are different, crossing different part of the atmosphere
in the turbulent layers. As a result, the loss of correlation between the two
wavefronts will drive to an atmospheric noise on the astrometric phase
which will lower the ultimate performance of this method.
 \\
We call respectively $\phi_s$, and $\phi_r^{\Delta{\alpha}h}$, the phase of
the astrophysical target and the phase estimated from the off-axis
reference source, located at an angular distance $\Delta\alpha$ of the science
object, and where we assume for the sake of simplicity a single turbulent layer located at an height $h$ from the
ground.  The astrometric phase is then simply defined by:
\begin{equation}
\Delta_{\phi}^{\Delta\alpha{h}} = \phi_s - \phi_r^{\Delta\alpha{h}}
\end{equation}
and its associated error writes:
 \begin{equation}
\sigma^2_{\Delta_{\phi}^{\Delta\alpha{h}}} = 2\sigma^2_{\phi} -
2\mathrm{cov}(\phi_s, \phi_r^{\Delta\alpha{h}}) \label{eq_errnarrow}
\end{equation}
Equation \ref{eq_errnarrow} tells us that the noise of
the astrometric phase increases as long as the correlation between
the two turbulent wavefronts $\mathrm{cov}(\phi_s,
\phi_r^{\Delta\alpha{h}})$ decreases, that is as long as the spatial
distance $\Delta\alpha h$ between both turbulent phase screens is
getting higher. The quantitative effect of the separation between the
star and its reference on the interferometric narrow-angle astrometric
error has been already studied by \citet{shao_1} and we refer the readers to
their paper for further details. In our analysis, we focus on the
relative performance of astrometry between multimode
and single-mode scheme, studying how spatial filtering of the wavefronts
will impact on the astrometric error, as developed in Appendix \ref{app_astrometry}.\\
For sake of simplicity we have assumed in the following one single
turbulent layer located at $h=10$km as it is the strongest layer at
the Cerro Paranal site \citep{masciadri_1}. There is also a strong layer very
close to the ground at around $20$m \citep{martin_1} but that only
marginally contributes to the phase decorrelation since the linear
distance at stake is $500$ times smaller than the one relative to the
high altitude layer. Figure \ref{fig_astrometry} (left) shows the evolution of the
astrometric error as a function of the angular distance between the
source and the reference, for both single-mode and
multimode schemes. \\ \\
{\bf Large field of view:} When the
reference source is further than roughly $\simeq 1^{''}$ from the astrophysical object of interest,  we can see that spatial filtering  allows to increase the precision of the astrometric phase. The improvement is better as the AO correction is low, reaching a factor of $\sim 4$ without AO
correction, as emphasized by Fig. \ref{fig_astrometry} (right).  This argues in favor of using spatial-filtering elements in the design of astrometric instruments  with large field of view (FOV)
as it has been chosen for PRIMA ($\mathrm{FOV} \simeq 30^{''}$). %as well as
%for GRAVITY the second generation of astrometric instruments of the VLTI
%\citep{gillessen_1}.  
Looking in further details, one can notice that
the phase tends asymptotically towards a plateau, from an angular
separation of $\sim 10^{"}$ which roughly corresponds to the
isoplanatic angle of Paranal in the K-band\footnote{considering the
measurements of \citet{martin_1}  which have found an average
isoplanatic angle of $\sim 1.9^{''}$ in the visible, and recalling
that it evolves with wavelength as $\lambda^{1.2}$ \citep{shao_1}.}. 
Indeed from this angle, wavefronts can
  be considered as  uncorrelated and the astrometric error converges
  to $\sqrt{2}\sigma_{\phi}$, which does not depend on the separation
  $\Delta\alpha$ any longer. As a consequence in this regime
  (i.e. $\Delta\alpha \gtrsim 10^{"}$), the
astrometric phase error ratio can be approximated by the speckle phase
error ratio, which has already been discussed in Sect. \ref{sec_specknoise}.\\ \\
{\bf Narrow field of view:} At the contrary, when wavefronts are still strongly correlated ($\Delta\alpha \la 1^{''}$), the
performance of single-mode astrometry is slightly lower (by a factor
$\sim 1$ to $\sim 1.5$) than that of multimode astrometry, the error
of the latter decreasing faster as the separation between the
astrophysical source and the reference shrinks to zero. 
By smoothing the turbulent wavefronts across the apertures,
spatial filtering is indeed lowering the effect of the strong
correlation between the wavefronts. Such situation 
concerns the cases for which the reference is very close to the star, like GRAVITY the second generation of astrometric instruments of the VLTI
\citep[$\mathrm{FOV} \simeq 2^{''}$]{gillessen_1}.

\section{Summary and conclusions}

In this paper, we have provided a theoretical formalism that allows to
derive the error of the interferometric phase, both in cases of
single-mode and multimode interferometry. From these derivations, we
have demonstrated that:
\begin{itemize}
\item Contrarily to a widespread idea, losing flux by injecting the
  light into single-mode spatial filters is not a performance killer
  for estimating the phase. As the matter of fact, single-mode
  interferometry provides better performance than
    that of multimode interferometry, unless when the interferometer is working in the detector noise regime (faint sources). In such cases, multimode interferometry is slightly better, providing a phase
  error smaller by a factor $\rho_0 \simeq 0.8$ which is the maximum
  fraction of flux that can be injected in single-mode devices.
\item In cases of bright sources observations, spatial filtering is
  shown to be very efficient, especially when the AO correction of
  the turbulent wavefronts is poor or absent. In such situations, the
  precision of the single-mode interferometric phase is better than
  that of the multimode one by a factor of $2$ and more when the Strehl is
  below $10\%$. 
\item Single-mode interferometry also proves to be more
  robust to the turbulence to both locking and coherently integrating the
  fringes and providing a better astrometric precision when using phase
  referencing techniques, except for narrow field of views ($\mathrm{FOV} \la 1^{''}$).
\end{itemize}
In conclusion, from a theoretical point of view and contrarily
  to a widespread opinion, single-mode fringe tracking should be seriously considered as an advantageous technical solution.
  %% especially in the case of bright targets observed
  %% with a very partial correction of the atmosphere. However, in the
  %% other case of faint sources, where the multimode approach can be
  %% better than the single-mode one, the gain of using multimode
  %% techniques is never exceeding $20\%$, a percentage corresponding
  %% to the loss of photons due to the coupling into the optical
  %% guides.
  Furthermore, the astronomers should realize that the many
  gains of single-mode interferometry (flexibility of the solutions,
  robustness to the alignment, less optical elements,...) may
  compensate significantly a modest and limited loss of performance
  in the faint sources case compared to multimode interferometry solutions. This is all the more true that the detectors should soon
  evolve to reach photon counting capability.

\Online

\appendix

\onecolumn

\section{Computation of the interferometric phase error}\label{app_sigma}
\subsection{General formalism} \label{sec_goodman}
It follows Goodman's approach \citep{goodman_1}, based on a continuous model of detection process, where the recorded signal (i.e. the interferogram) can be represented as:
\begin{equation}
d(x,y) = \sum_{k=1}^{K}\delta(x-x_k,y-y_k)
\end{equation}
and its Fourier transform: 
\begin{equation}
D({\vec f}) =  D(f_x,f_y) = \sum_{k=1}^{K}\expo^{-2i\pi(f_xx_k+f_yy_k)} \label{eq_dfourier}
\end{equation}
where the position ($x_n$) and the number of photoevents (per time unit) ($K$) are statistical processes with probability laws depending on the intensity distribution $I({\vec x})$. To take also into account the detector noise, one needs to add Gaussian additive noise ($\epsilon$) to the previous equation \citep{tatulli_1, tatulli_3}:
\begin{equation}
 Q({\vec f}) = D({\vec f}) + \epsilon({\vec f}) = \sum_{k=1}^{K}\expo^{-2i\pi(f_xx_k+f_yy_k)}  + \epsilon({\vec f}) \label{eq_noisig}
\end{equation} 
Note that this formalism is also valid for a temporal coding of the interferogram, that is replacing the spatial position ($x_k, y_k$) by the (1-D) temporal sampling $t$, and similarly the spatial frequency ${\vec f}$ by the temporal one $\nu$.\\ As Q represents the spectrum of the interferogram, the phase of the
spectrum is merely the argument of this estimator:
\begin{equation}
\phi =\mathrm{atan}\left(\mathrm{Im}(Q),\mathrm{Re}(Q)\right)\end{equation}
In this framework, \citet{chelli_1} has shown that in first approximation (i.e. small noise error), the
variance of the phase can be expressed as following:
\begin{equation}
\sigma^2_{\phi} = \frac{1}{2}\frac{\mathrm{E}(|Q|^2)-\mathrm{Re}[\mathrm{E}(Q^2)]}{[\mathrm{E}(Q)]^2}
\end{equation}  \\
then using Goodman's formalism described above, one can show that \citep{chelli_1, tatulli_1}:
\begin{eqnarray}
\mathrm{E}(|Q|^2) &=& <\overline{K}^2|i({\vec f})|^2 +
\overline{K} + N_{pix}\sigma^2_{det}>_{\Phi} \label{eq_goodman_1}\\
\mathrm{E}(Q^2) &=& <\overline{K}^2 i^2({\vec f})>_{\Phi} \label{eq_goodman_2}\\
\mathrm{E}(Q) &=& <\overline{K}i({\vec f})>_{\Phi} \label{eq_goodman_3}
\end{eqnarray}
where $\overline{K}$ is the average number of photoevents per time unit, 
$\sigma_{det}$ is the detector noise and
$N_{pix}$ is the number of pixel to sample the signal. $i({\vec f})$ is the
normalized spectrum of the interferogram (i.e. $i(\vec{0})=1$).\\
$<>_{\Phi}$ denotes the expected value with respect to the
atmosphere. How the previous expressions are unfolding depends on
whether we deal with multimode or single-mode interferometry.

\subsection{Useful definitions}
In this section we introduce the concepts and notations that we will use in our formalism to derive the expression of the noise of the interferometric phase.
Let us consider (see also Fig. \ref{fig_turb} in the paper):
\begin{itemize}
\item  an interferometer made of 2 telescopes described by their pupil 
functions $P_1({\vec r})$, $P_2({\vec r})$ and their associated transmission $\sqrt{t_1}$, $\sqrt{t_2}$. Note that with such a definition $P_1$ and $P_2$ are respectively centered at the position ${\vec r_1} =\lambda{\vec f_1}$ and ${\vec r_2}=\lambda{\vec f_2}$, and we
introduce the quantity ${\vec f_{12}} = {\vec f_2} - {\vec f_1}$ as the baseline frequency of the interferometer.
\item $\phi^r_1({\vec r})$ and $\phi^r_2({\vec r})$ the partially AO-corrected atmospheric turbulent phase screens. We also introduce  $\mathcal{D}_{\phi_r}$ the structure function of the residual phases (assuming the same level of AO correction for each telescope) together with $\sigma^2_{\phi_r}$, its associated residual phase variance (see also Appendix \ref{app_ao}).
\item the incoming wavefront of the source $\Psi({\vec r})$. By definition we have  $|\Psi(\vec{0})|^2 = N$, $N$ being the number of photons per surface unit and per time unit emitted by the source, and $[\Psi({\vec r})\Psi^{\ast}({\vec r}+\lambda{\vec f})]/|\Psi(\vec{0})|^2 = V({\vec f})$ the visibility of the source at the spatial frequency ${\vec f}$.   
\end{itemize}
We emphasized that the following derivations will be done assuming monochromatic interferograms at the wavelength $\lambda$. Indeed, taking into account a non-null spectral bandwidth $\delta\lambda$ is equivalent to consider a monochromatic interferogram at the effective wavelength $\lambda_0$ of the filter, modulated in amplitude by an envelope, which width is fixed by the spectral coherence length $L_c = \lambda_0^2/\delta\lambda$. Moreover, such effect is independent of the multimode or single-mode nature of the interferogram.

\subsection{Phase noise in multimode interferometry} \label{subapp_multimode_noise}
In the following, we derive the estimator of the coherent flux of the interferogram, and subsequently the error of the interferometric phase in the multimode case. We provide two formalisms that respectively describe both beam recombination schemes: in the image plane and in the pupil plane.

\subsubsection{Image plane recombination} \label{app_image_plane}
Image plane recombination is well suited to describe multiaxial interferometric schemes where the fringe pattern is spatially coded on the detector, such as for the GI2T and VEGA/CHARA instruments. Note that co-axial (i.e. temporal) coding can also be performed in the image plane but in practice -- to our knowledge -- no interferometric instruments are using this technique. \\
In such case, the spatial distribution of the complex amplitude of the radiation field $E({\vec r})$ can be written as following:
\begin{equation}
E({\vec r}) = \sqrt{t_1}P_1({\vec r})\exp^{i\phi^r_1({\vec r})}\Psi({\vec r}) +  \sqrt{t_2}P_2({\vec r})\exp^{i\phi^r_2({\vec r})}\Psi({\vec r})
\end{equation}
By rules of diffraction theory, the Fourier Transform of the interference pattern formed in the image plane is the autocorrelation of the complex amplitude $E({\vec r})$, namely:
\begin{eqnarray}
I({\vec f}) &=& \int E({\vec r})E^{\ast}({\vec r}+\lambda{\vec f}) \mathrm{d}{\vec r} \nonumber \\
&=&  t_1 \int P_1({\vec r}) P_1({\vec r}+\lambda{\vec f})\exp^{i(\phi^r_1({\vec r})-\phi^r_1({\vec r}+\lambda{\vec f}))} \Psi({\vec r})\Psi^{\ast}({\vec r}+\lambda{\vec f}) \mathrm{d}{\vec r} \nonumber \\ 
&+&  t_2 \int P_2({\vec r}) P_2({\vec r}+\lambda{\vec f})\exp^{i(\phi^r_2({\vec r})-\phi^r_2({\vec r}+\lambda{\vec f}))} \Psi({\vec r})\Psi^{\ast}({\vec r}+\lambda{\vec f}) \mathrm{d}{\vec r} \nonumber \\ 
&+& \sqrt{t_1t_2} \int P_1({\vec r}) P_2({\vec r}+\lambda{\vec f})\exp^{i(\phi^r_1({\vec r})-\phi^r_2({\vec r}+\lambda{\vec f}))} \Psi({\vec r})\Psi^{\ast}({\vec r}+\lambda{\vec f}) \mathrm{d}{\vec r} \nonumber \\ 
&+& \sqrt{t_1t_2} \int P_1({\vec r}+\lambda{\vec f}) P_2({\vec r})\exp^{i(\phi^r_2({\vec r})-\phi^r_1({\vec r}+\lambda{\vec f}))} \Psi^{\ast}({\vec r}+\lambda{\vec f})\Psi({\vec r}) \mathrm{d}{\vec r}
\end{eqnarray}
Introducing $|\Psi(\vec{0})|^2 = N$  in previous equation enables to make the complex visibility function appear:
\begin{equation}
I({\vec f}) = N t_1 V({\vec f}) S_1({\vec f}) + N t_2 V({\vec f}) S_2({\vec f}) + N \sqrt{t_1t_2} V({\vec f})
S_{12}({\vec f}) + N \sqrt{t_1t_2} V({\vec f}) S^{\ast}_{12}(-{\vec f}) \label{eq_multi_image}
\end{equation} 
with $S_1({\vec f})$ and $S_2({\vec f})$ the so-called photometric peaks resulting on the autocorrelation of each corrugated pupil, and $S_{12}({\vec f})$ the interferometric peak
arising from the cross-correlation between both pupils:
\begin{eqnarray}
S_{i}(f) &=& \int
  P_i({\vec r})P_i({\vec r}+\lambda{\vec f})\exp^{i(\phi^r_i({\vec r})-\phi^r_i({\vec r}+\lambda{\vec f}))}\mathrm{d}{\vec r},~~\mbox{(i=[1,2])}\\
S_{12}(f) &=& \int P_1({\vec r})P_2({\vec r}+\lambda{\vec f})\exp^{i(\phi^r_1({\vec r})-\phi^r_2({\vec r}+\lambda{\vec f}))}\mathrm{d}{\vec r}
\end{eqnarray}
From  equation \ref{eq_multi_image}, we can straightforward derive the number of photoevents:
\begin{equation}
\overline{K} = I({\vec 0}) = \Sigma_P N  (t_1 + t_2) \label{eq_fK}
\end{equation}
which is turbulence independent (neglecting scintillation). $\Sigma_P$ is the collecting area of a single telescope:
\begin{equation}
\Sigma_P = \int  [P_1({\vec r})]^2 \mathrm{d}{\vec r} = \int  [P_2({\vec r})]^2 \mathrm{d}{\vec r}
\end{equation}
assuming for sake of simplicity that both pupils are identical.\\ Eq. (\ref{eq_multi_image}) also shows that multimode interferometry with image plane recombination continuously transmits the whole spatial frequencies ${\vec f}$. The complex visibility $V_{12}=V({\vec f_{12}})$  is then derived from the estimation of the coherent flux at the baseline frequency of the interferometer  ${\vec f_{12}}$, that is:
\begin{equation}
F^c_{12} = N\sqrt{t_1t_2}V_{12}S_{12}({\vec f_{12}}) \label{eq_fc}
\end{equation}
and the normalized spectrum of the interferogram takes the form:
\begin{equation}
i({\vec f_{12}}) = \frac{\sqrt{t_1t_2}V_{12}}{(t_1+t_2)\Sigma_P}S_{12}({\vec f_{12}}) = \frac{\sqrt{t_1t_2}V_{12}}{(t_1+t_2)}T_{12}({\vec f_{12}})\label{eq_norm_if12}
\end{equation}
where $T_{12}({\vec f}) = S_{12}({\vec f})/\Sigma_P$ is the normalized interferometric transfer function, such as $|T_{12}({\vec f_{12}})|=1$. As a consequence, by analogy with the definition of the Strehl ratio for a single telescope, $|T_{12}({\vec f_{12}})|^2$ can be seen as the instantaneous multimode {\it interferometric} Strehl ratio. 

\subsubsection{Pupil plane recombination} \label{app_pupil_plane}
In such a scheme, fringes are formed in the pupil plane by means of geometrical 
operations of the entrance pupil \citep{chelli_2}. This formalism is usually well suited to describe co-axial recombination with temporal coding where the pupils are superimposed at the beam splitter, hence the fringes formed in the pupil plane, as in the IOTA instrument. In practice, the fringes are form by translating one pupil over the other (translation of vector ${\vec f_{12}}$), and by introducing an optical path delay ($\theta_1(t)$ and $\theta_2(t)$) on each beam (see e.g. \citealt[Appendix A2]{buscher_2}) to modulate the fringe pattern. In this case, the complex amplitude of the superimposed electric fields writes:
\begin{equation}
E({\vec r}) = \sqrt{t_1}P_1({\vec r})\exp^{i\phi^r_1({\vec r})}\Psi({\vec r})\exp^{i\theta_1(t)} +  \sqrt{t_2}P_2({\vec r}+\lambda{{\vec f_{12}}})\exp^{i\phi^r_2({\vec r}+\lambda{{\vec f_{12}}})}\Psi({\vec r}+\lambda{{\vec f_{12}}})\exp^{i\theta_2(t)}
\end{equation}
The fringe pattern $I(t)$ is then formed by focusing the light on one single pixel of the detector. Again, by rules of image formation, it writes as following:
\begin{equation}
I(t) =  \int |E({\vec r})|^2  \mathrm{d}{\vec r}
\end{equation}
which, by making use of the expression of $E({\vec r})$ rewrites:
\begin{eqnarray}
I(t) &=& \Sigma_P N (t_1 + t_2)  + N \sqrt{t_1t_2} \exp^{i\Delta^{12}\theta(t)} \int P_1({\vec r})P_2({\vec r}+\lambda{{\vec f_{12}}})\exp^{i(\phi^r_1({\vec r})-\phi^r_2({\vec r}+\lambda{{\vec f_{12}}}))} \Psi({\vec r})\Psi^{\ast}({\vec r}+\lambda{{\vec f_{12}}}) \mathrm{d}{\vec r} + CC \nonumber \\
&=& \Sigma_P N (t_1 + t_2) + N \sqrt{t_1t_2} V({\vec f_{12}})S_{12}({\vec f_{12}})\exp^{i\Delta^{12}\theta(t)} +  N \sqrt{t_1t_2} V(-{\vec f_{12}})S^{\ast}_{12}(-{\vec f_{12}})\exp^{-i\Delta^{12}\theta(t)}
\end{eqnarray}
where $\Delta^{12}\theta(t) = \theta_1(t) - \theta_2(t)$ is the phase delay between the two beams which samples the fringe pattern. In the Fourier plane, the interferometric equation takes the form:
\begin{equation}
I(\nu) = \Sigma_P N (t_1 + t_2)  + N \sqrt{t_1t_2} V({\vec f_{12}})S_{12}({\vec f_{12}}) \delta_{\nu_{12}} + N \sqrt{t_1t_2} V^{\ast}({\vec f_{12}})S^{\ast}_{12}({\vec f_{12}}) \delta_{-\nu_{12}}
\end{equation}
$\delta_{\nu}$ is the Dirac function, and $\nu_{12}$ is the temporal frequency at which the interferogram is modulated, that is the frequency of the moving piezoelectric mirror. \\
Note that, at the difference of the image plane recombination, pupil plane interferometry transmits only the frequency baseline ${\vec f_{12}}$.   Nonetheless, the coherent flux, estimated in the Fourier plane at the frequency mirror ($\nu = \nu_{12}$) takes the same form as in the image plane case, namely:
\begin{equation}
F^c_{12} = N \sqrt{t_1t_2} V_{12}S_{12}({\vec f_{12}})
\end{equation}
and similarly the total number of detected photoevents is:
\begin{equation}
\overline{K} = I(\vec{0}) = \Sigma_P N (t_1 + t_2)
\end{equation}
and the normalized spectrum of the interferogram at the frequency $\nu_{12}$ is:
\begin{equation}
i(\nu_{12}) = \frac{\sqrt{t_1t_2} V_{12}}{(t_1 + t_2)\Sigma_P}S_{12}({\vec f_{12}})
\end{equation}

\subsubsection{The interferometric phase error}
In previous sections, we have shown that independently of the chosen recombination scheme, the total flux, the coherent flux and consequently the  normalized spectrum of the interferogram can take the form of Eqs. (\ref{eq_fK},\,\ref{eq_fc},\,\ref{eq_norm_if12}), respectively. Using these expressions, Eqs. (\ref{eq_goodman_1},\,\ref{eq_goodman_2},\,\ref{eq_goodman_3}) can be rewritten as following:
\begin{eqnarray}
\mathrm{E}(|Q|^2) &=& N^2 t_1t_2 |V_{12}|^2 <|S_{12}({\vec f_{12}})|^2>_{\Phi} + \Sigma_P N (t_1 + t_2) +  N_{pix}\sigma_{det}^2\label{eq_absq2_multi}\\
\mathrm{E}(Q^2) &=& N^2 t_1t_2
V^2_{12} <S^2_{12}({\vec f_{12}})>_{\Phi} \label{eq_q2_multi} \\
\mathrm{E}(Q) &=& N \sqrt{t_1t_2}
V_{12} <S_{12}({\vec f_{12}})>_{\phi} \label{eq_q_multi}
\end{eqnarray}
{\it NB: \citet{chelli_1} has shown that the error of the phase does not depends
on the object phase, hence one can consider in the following that the
object is centro-symmetric, that is $V_{12} = |V_{12}| = \mathrm{Re}[V_{12}]$.}\\
It remains to derive in previous equations the moments of the
interferometric transfer function $S_{12}({\vec f_{12}})$:
\begin{eqnarray}
<S_{12}({\vec f_{12}})>_{\Phi} &=& \int P_1({\vec r})P_2({\vec r}+\lambda{{\vec f_{12}}})<\exp^{i(\phi^r_1({\vec r})-\phi^r_2({\vec r}+\lambda{{\vec f_{12}}}))}>_{\Phi}\mathrm{d}{\vec r}\nonumber \\
&=&  \int
  P_1({\vec r})P_2({\vec r}+\lambda{{\vec f_{12}}})\exp^{-\frac{1}{2}<|(\phi^r_1({\vec r})-\phi^r_2({\vec r}+\lambda{{\vec f_{12}}}))|^2>_{\Phi}}\mathrm{d}{\vec r} \nonumber \\
&=& \Sigma_P \exp^{-\sigma^2_{\phi_r}}\label{eq_i_multi}
\end{eqnarray}
\begin{eqnarray}
<S^2_{12}({\vec f_{12}})>_{\Phi} &=& \int
  P_1({\vec r})P_2({\vec r}+\lambda{{\vec f_{12}}})P_1({\vec r}^{\prime})P_2({\vec r}^{\prime}+\lambda{{\vec f_{12}}})<\exp^{i(\phi^r_1({\vec r})-\phi^r_2({\vec r}+\lambda{{\vec f_{12}}})+\phi^r_1({\vec r}^{\prime})-\phi^r_2({\vec r}^{\prime}+\lambda{{\vec f_{12}}}))}>_{\Phi}\mathrm{d}{\vec r}\mathrm{d}{{\vec r^{\prime}}}\nonumber \\
&=& 
  P_1({\vec r})P_2({\vec r}+\lambda{{\vec f_{12}}})P_1({\vec r}^{\prime})P_2({\vec r}^{\prime}+\lambda{{\vec f_{12}}})\exp^{-\frac{1}{2}<|(\phi^r_1({\vec r})-\phi^r_2({\vec r}+\lambda{{\vec f_{12}}})+\phi^r_1({\vec r}^{\prime})-\phi^r_2({\vec r}^{\prime}+\lambda{{\vec f_{12}}}))|^2>_{\Phi}}\mathrm{d}{\vec r}\mathrm{d}{{\vec r^{\prime}}} \nonumber \\
&& \mbox{knowing that}~<\phi^r_1({\vec r})\phi^r_1({\vec r}^{\prime})>_{\Phi} = 
\sigma^2_{\phi_r} - \frac{1}{2}\mathcal{D}_{\phi_r}({\vec r},{\vec r^{\prime}}),~\mbox{it comes}, \nonumber \\
&=& \exp^{-4\sigma^2_{\phi_r}} \int
  P_1({\vec r})P_2({\vec r}+\lambda{{\vec f_{12}}})P_1({\vec r}^{\prime})P_2({\vec r}^{\prime}+\lambda{{\vec f_{12}}})\exp^{\frac{1}{2}\mathcal{D}_{\phi_r}({\vec r},{\vec r^{\prime}})+\frac{1}{2}\mathcal{D}_{\phi_r}({\vec r}+\lambda{{\vec f_{12}}},{\vec r^{\prime}}+\lambda{{\vec f_{12}}})}\mathrm{d}{\vec r}\mathrm{d}{{\vec r^{\prime}}} \nonumber \\
&& \mbox{changing axes reference:}~P_1({\vec r}),~P_2({\vec r})~\rightarrow~P({\vec r})~\mbox{centered on 0},\nonumber \\
&=& \exp^{-4\sigma^2_{\phi_r}}
\int \left[
  P({\vec r})P({\vec r}^{\prime})\exp^{\frac{1}{2}\mathcal{D}_{\phi_r}({\vec r},{\vec r^{\prime}})}\right]\left[P({\vec r})P({\vec r}^{\prime})\exp^{\frac{1}{2}\mathcal{D}_{\phi_r}({\vec r},{\vec r^{\prime}})}\right]\mathrm{d}{\vec r}\mathrm{d}{{\vec r^{\prime}}}\label{eq_i2_multi}
\end{eqnarray}
\begin{eqnarray}
<|S_{12}({\vec f_{12}})|^2>_{\Phi} &=& \int
  P_1({\vec r})P_2({\vec r}+\lambda{{\vec f_{12}}})P_1({\vec r}^{\prime})P_2({\vec r}^{\prime}+\lambda{{\vec f_{12}}})<\exp^{i(\phi^r_1({\vec r})-\phi^r_2({\vec r}+\lambda{{\vec f_{12}}})-\phi^r_1({\vec r}^{\prime})+\phi^r_2({\vec r}^{\prime}+\lambda{{\vec f_{12}}}))}>_{\Phi}\mathrm{d}{\vec r}\mathrm{d}{{\vec r^{\prime}}} \nonumber \\
&=&  \int
  P_1({\vec r})P_2({\vec r}+\lambda{{\vec f_{12}}})P_1({\vec r}^{\prime})P_2({\vec r}^{\prime}+\lambda{{\vec f_{12}}})\exp^{-\frac{1}{2}<|(\phi^r_1({\vec r})-\phi^r_2({\vec r}+\lambda{{\vec f_{12}}})-\phi^r_1({\vec r}^{\prime})+\phi^r_2({\vec r}^{\prime}+\lambda{{\vec f_{12}}}))|^2>_{\Phi}}\mathrm{d}{\vec r}\mathrm{d}{{\vec r^{\prime}}} \nonumber \\
&=& \int
  P_1({\vec r})P_2({\vec r}+\lambda{{\vec f_{12}}})P_1({\vec r}^{\prime})P_2({\vec r}^{\prime}+\lambda{{\vec f_{12}}})\exp^{-\frac{1}{2}\mathcal{D}_{\phi_r}({\vec r},{\vec r^{\prime}})-\frac{1}{2}\mathcal{D}_{\phi_r}({\vec r}+\lambda{{\vec f_{12}}},{\vec r^{\prime}}+\lambda{{\vec f_{12}}})}\mathrm{d}{\vec r}\mathrm{d}{{\vec r^{\prime}}}\nonumber \\
&& \mbox{changing axes reference:} \nonumber \\
&=& \int \left[
  P({\vec r})P({\vec r}^{\prime})\exp^{-\frac{1}{2}\mathcal{D}_{\phi_r}({\vec r},{\vec r^{\prime}})}\right]\left[P({\vec r})P({\vec r}^{\prime})\exp^{-\frac{1}{2}\mathcal{D}_{\phi_r}({\vec r},{\vec r^{\prime}})}\right]\mathrm{d}{\vec r}\mathrm{d}{{\vec r^{\prime}}} \label{eq_absi2_multi}
\end{eqnarray}
Putting Eqs. (\ref{eq_i_multi},\,\ref{eq_i2_multi},\,\ref{eq_absi2_multi}) into  Eqs. (\ref{eq_absq2_multi},\,\ref{eq_q2_multi},\,\ref{eq_q_multi}), we
finally find that the variance of the phase can
be written as the quadratic sum of 3 terms, corresponding 
respectively to the detector, photon and atmospheric regimes $\sigma^2_{\phi} = \sigma^2_{det_{\phi}} + \sigma^2_{phot_{\phi}} +  \sigma^2_{atm_{\phi}}$,
with:
\begin{eqnarray}
\sigma^2_{det_{\phi}}&=&\frac{1}{2}\frac{
  N_{pix}\sigma^2_{det}}{\Sigma_P^2N^2t_1t_2|V_{12}|^2\mathrm{e}^{-2\sigma^2_{\phi_r}}}\\
\sigma^2_{phot_{\phi}}&=&
\frac{1}{2}\frac{t_1+t_2}{\Sigma_PNt_1t_2|V_{12}|^2\mathrm{e}^{-2\sigma^2_{\phi_r}}}\\
\sigma^2_{atm_{\phi}}&=&\frac{1}{2}\frac{\int \left[P({\vec r})P({\vec r}^{\prime})\mathrm{e}^{-\frac{1}{2}\mathcal{D}_{\phi_r}({\vec r},{\vec r^{\prime}})}\right]^2\mathrm{d}{\vec r}\mathrm{d}{{\vec r^{\prime}}}-\mathrm{e}^{-4\sigma^2_{\phi_r}}\int \left[P({\vec r})P({\vec r}^{\prime})\mathrm{e}^{\frac{1}{2}\mathcal{D}_{\phi_r}({\vec r},{\vec r^{\prime}})}\right]^2\mathrm{d}{\vec r}\mathrm{d}{{\vec r^{\prime}}}}{\left[\int
      P^2({\vec r})\mathrm{d}{\vec r}\right]^2\mathrm{e}^{-2\sigma^2_{\phi_r}}}
\end{eqnarray}
We note that the expression of the variance in the photon noise case
($\sigma^2_{phot_{\phi}}$) is the generalization of
the formula found by \cite{vannier_1} (Eq. (14)), when the visibility
 is corrupted by the partially AO-corrected turbulent phase. 

\subsection{Phase noise in single-mode interferometry}\label{app_singlemode_noise}
Because of the spatial filtering specific properties of single-mode
devices, the single-mode interferometric equation changes slightly
with respect to that of the multimode as we have to take into account
the coupling efficiency which extenuates the number of coherent and
incoherent photoevents. The coupling coefficients, respectively for the
photometric ($\rho_1(V)$, $\rho_2(V)$) and interferometric ($\rho^{12}(V)$) 
channels write \citep{mege_2, tatulli_1}: 
\begin{eqnarray}
\rho_i(V)&=&\rho_0\frac{(V \ast S_{i})(\vec{0})}{\int
    S^0_i({\vec f})\mathrm{d}{\vec f}},~~\mbox{(i=[1,2])} \nonumber \\
&=& \rho_0\frac{\int V(f)
  P_i({\vec r})P_i({\vec r}+\lambda{\vec f})\exp^{i(\phi^r_i({\vec r})-\phi^r_i({\vec r}+\lambda{\vec f}))}\mathrm{d}{\vec r}\mathrm{d}{\vec f}}{\left[\int
  P_i({\vec r})\mathrm{d}{\vec r}\right]^2} \label{app_eq_rhoi}\\
\rho^{12}(V) &=& \rho_0\frac{(V \ast S_{12})({\vec f_{12}})}{\sqrt{\int
    S^0_1({\vec f}) \mathrm{d}{\vec f}\int S^0_2({\vec f}) \mathrm{d}{\vec f}}}\nonumber \\
&=& \rho_0\frac{\int V(f-{\vec f_{12}})
  P_1({\vec r})P_2({\vec r}+\lambda{\vec f})\exp^{i(\phi^r_1({\vec r})-\phi^r_2({\vec r}+\lambda{\vec f}))}\mathrm{d}{\vec r}\mathrm{d}{\vec f}}{\int
  P_1({\vec r})\mathrm{d}{\vec r} \int P_2({\vec r})\mathrm{d}{\vec r}} \label{app_eq_rhoij}\\
\end{eqnarray}
where $S^0_1({\vec f})$, $S^0_2({\vec f})$ are the photometric transfer functions of a perfect (atmosphere-free) interferometer, and $\rho_0$  the maximum coupling efficiency, shown to be $\sim 80\%$ \citep{shaklan_1}.\\
And the single-mode interferometric equation writes:
\begin{equation}
I({\vec f}) = \Sigma_P N t_1 \rho_1(V) H_1({\vec f}) + \Sigma_P N t_2 \rho_2(V) H_2({\vec f}) + \Sigma_P N \sqrt{t_1t_2} \rho^{12}(V)
H_{12}({\vec f}) + \Sigma_P N \sqrt{t_1t_2} (\rho^{12}(V))^{\ast} H^{\ast}_{12}(-{\vec f})
\end{equation}
where $H_1({\vec f})$, $H_2({\vec f})$ and $H_{12}({\vec f})$ are the normalized (i.e. $H_1(\vec{0})=H_2(\vec{0})=1$, and $H_{12}({\vec f_{12}})=1$) single-mode photometric and interferometric transfer
functions, so-called the {\it carrying wave} \citep{mege_3} which are fixed by the geometry of the single-mode device and therefore independent of the atmosphere. As a result, the single-mode coherent flux is defined by:
\begin{equation}
F^c_{12} = \Sigma_P N \sqrt{t_1t_2} \rho^{12}(V) \label{eq_singlemode_fc}
\end{equation}
\subsubsection{Compact sources case}
In the case of compact sources -- that is unresolved by the telescope, which is the most common case in interferometry -- Equations \ref{app_eq_rhoi} and \ref{app_eq_rhoij} are taking the simplified following form:
\begin{eqnarray}
\rho_i(V)&=& \rho_0\frac{\int P_i({\vec r})P_i({\vec r}+\lambda{\vec f})\exp^{i(\phi^r_i({\vec r})-\phi^r_i({\vec r}+\lambda{\vec f}))}\mathrm{d}{\vec r}\mathrm{d}{\vec f}}{\left[\int
  P_i({\vec r})\mathrm{d}{\vec r}\right]^2} = \rho_0 \mathcal{S}_i\\
\rho^{12}(V) &=&  \rho_0 V_{12}  \frac{\int 
  P_1({\vec r})P_2({\vec r}+\lambda{\vec f})\exp^{i(\phi^r_1({\vec r})-\phi^r_2({\vec r}+\lambda{\vec f}))}\mathrm{d}{\vec r}\mathrm{d}{\vec f}}{\int
  P_1({\vec r})\mathrm{d}{\vec r} \int P_2({\vec r})\mathrm{d}{\vec r}} =  \rho_0 V_{12} \rho_{12}\\
\end{eqnarray}
where  $\mathcal{S}_i$ is by definition the instantaneous Strehl ratio relative to the $i^{th}$ telescope, and with:
\begin{eqnarray}
\rho_{12} &=& \frac{\int 
  P_1({\vec r})P_2({\vec r}+\lambda{\vec f})\exp^{i(\phi^r_1({\vec r})-\phi^r_2({\vec r}+\lambda{\vec f}))}\mathrm{d}{\vec r}\mathrm{d}{\vec f}}{\int
  P_1({\vec r})\mathrm{d}{\vec r} \int P_2({\vec r})\mathrm{d}{\vec r}}
\end{eqnarray}
$|\rho_{12}|^2$ being by  analogy the instantaneous single-mode {\it interferometric} Strehl ratio. The interferometric equation finally rewrites:
\begin{equation}
I({\vec f}) = \Sigma_P  N t_1 \rho_0 \mathcal{S}_1  H_1({\vec f}) + \Sigma_P N t_2 \rho_0 \mathcal{S}_2 H_2({\vec f}) + \Sigma_P N \sqrt{t_1t_2} V_{12}\rho_{12}
H_{12}({\vec f}) + \Sigma_P N \sqrt{t_1t_2} \rho_0 V^{\ast}_{12} \rho^{\ast}_{12} H^{\ast}_{12}(-{\vec f})
\end{equation}
And the single-mode coherent flux, at the spatial frequency ${\vec f_{12}}$ takes the form:
\begin{equation}
F^c_{12} = \Sigma_P N \sqrt{t_1t_2}V_{12}\rho_{12}
\end{equation}
whereas the number of photoevents writes:
\begin{equation}
\overline{K} = I({\vec 0}) = \Sigma_P N \rho_0 \mathcal{S}_1  + \Sigma_P N t_2 \rho_0 \mathcal{S}_2
\end{equation}
which depends on the turbulent atmosphere through the fluctuations of
the instantaneous Strehl ratio $\mathcal{S}_1$ and $\mathcal{S}_2$. And the normalized spectrum of the interferogram is:
\begin{equation}
i({\vec f_{12}}) = \frac{\sqrt{t_1t_2}\rho_{12}V_{12}}{t_1 \mathcal{S}_1 +t_2 \mathcal{S}_2} 
\end{equation}
The generic equations of the phase noise thus rewrites:
\begin{eqnarray}
\mathrm{E}(|Q|^2) &=& \Sigma_P^2 N^2 t_1t_2 \rho_0^2  |V_{12}|^2 <|\rho_{12}|^2>_{\Phi} +
\Sigma_P N t_1 \rho_0 <\mathcal{S}_1>_{\Phi} + \Sigma_P N t_2 \rho_0 <\mathcal{S}_2>_{\Phi}  +  N_{pix}\sigma_{det}^2 \label{eq_absq2_mono}\\
\mathrm{E}(Q^2) &=& \Sigma_P^2 N^2 t_1t_2  \rho_0^2 V_{12}^2 <\rho_{12}^2>_{\Phi} \label{eq_q2_mono}\\
\mathrm{E}(Q) &=& \Sigma_P N \sqrt{t_1t_2} \rho_0 V_{12} <\rho_{12}>_{\Phi}\label{eq_q_mono}
\end{eqnarray}
It now remains to compute the moments of the coupling coefficients:
\begin{eqnarray}
<\mathcal{S}_i>_{\Phi} &=&  \overline{\mathcal{S}_i},~~\mbox{(i=[1,2])} \label{eq_rho_mono}
\end{eqnarray}
where $\overline{\mathcal{S}_i}$ is the long exposure Strehl ratio of
the $i^{th}$ telescope. Assuming equivalent AO correction for both
telescopes, we have $\overline{\mathcal{S}_1} = \overline{\mathcal{S}_2} = \overline{\mathcal{S}}$ 
\begin{eqnarray}
<\rho_{12}>_{\Phi} &=& \frac{\int 
  P_1({\vec r})P_2({\vec r}+\lambda{\vec f})<\exp^{i(\phi^r_1({\vec r})-\phi^r_2({\vec r}+\lambda{\vec f}))}>_{\Phi}\mathrm{d}{\vec r}\mathrm{d}{\vec f}}{\int
  P_1({\vec r})\mathrm{d}{\vec r} \int P_2({\vec r})\mathrm{d}{\vec r}} = \mathrm{e}^{-\sigma^2_{\phi_r}}\frac{\int
  P_1({\vec r})P_2({\vec r}+\lambda{\vec f})\mathrm{d}{\vec r}\mathrm{d}{\vec f}}{\int
  P_1({\vec r})\mathrm{d}{\vec r} \int P_2({\vec r})\mathrm{d}{\vec r}} = \mathrm{e}^{-\sigma^2_{\phi_r}}  \label{eq_rhoij_mono}
\end{eqnarray}
which shows that the long exposure {\it interferometric} Strehl ratio is equal to the coherent energy $\mathrm{e}^{-\sigma^2_{\phi_r}}$.
\begin{eqnarray}
<\rho_{12}^2>_{\Phi}&=&\frac{\int P_1({\vec r})P_2({\vec r}+\lambda{\vec f})P_1({\vec r}^{\prime})P_2({\vec r}^{\prime}+\lambda{{\vec f^{\prime}}})<\exp^{i(\phi^r_1({\vec r})+\phi^r_1({\vec r}^{\prime})-\phi^r_2({\vec r}+\lambda{\vec f})-\phi^r_2({\vec r}^{\prime}+\lambda{{\vec f^{\prime}}}))}>_{\Phi}\mathrm{d}{\vec r}\mathrm{d}{\vec r^{\prime}}\mathrm{d}{\vec f}\mathrm{d}{\vec f}^{\prime}}{\left[\int P_1({\vec r})\mathrm{d}{\vec r} \int P_2({\vec r})\mathrm{d}{\vec r}\right]^2}\nonumber\\
&=&\mathrm{e}^{-4\sigma^2_{\phi_r}}\frac{\int P_1({\vec r})P_2({\vec r}+\lambda{\vec f})P_1({\vec r}^{\prime})P_2({\vec r}^{\prime}+\lambda{{\vec f^{\prime}}})\exp^{\frac{1}{2}\mathcal{D}_{\phi_r}({\vec r},{\vec r^{\prime}})+\frac{1}{2}\mathcal{D}_{\phi_r}({\vec r}+\lambda{\vec f},{\vec r^{\prime}}+\lambda{{\vec f^{\prime}}})}\mathrm{d}{\vec r}\mathrm{d}{\vec r^{\prime}}\mathrm{d}{\vec f}\mathrm{d}{\vec f}^{\prime}}{\left[\int
  P_1({\vec r})\mathrm{d}{\vec r} \int P_2({\vec r})\mathrm{d}{\vec r}\right]^2}\nonumber\\
&& \mbox{change of variables}~{\vec r}+\lambda{\vec f} \rightarrow {\vec u}, {\vec r^{\prime}}+\lambda{{\vec f^{\prime}}} \rightarrow {\vec u^{\prime}}:\nonumber\\
&=&\mathrm{e}^{-4\sigma^2_{\phi_r}}\frac{\int
  P_1({\vec r})P_1({\vec r}^{\prime})\exp^{\frac{1}{2}\mathcal{D}_{\phi_r}({\vec r},{\vec r^{\prime}})}\mathrm{d}{\vec r}\mathrm{d}{\vec r^{\prime}}\int
  P_2({\vec r})P_2({\vec r}^{\prime})\exp^{\frac{1}{2}\mathcal{D}_{\phi_r}({\vec r},{\vec r^{\prime}})}\mathrm{d}{\vec r}\mathrm{d}{\vec r^{\prime}}}{\left[\int
    P_1({\vec r})\mathrm{d}{\vec r} \int P_2({\vec r})\mathrm{d}{\vec r}\right]^2} \label{eq_rhoij2_mono}
\end{eqnarray}
\begin{eqnarray}
<|\rho_{12}|^2>_{\Phi}&=&\frac{\int P_1({\vec r})P_2({\vec r}+\lambda{\vec f})P_1({\vec r}^{\prime})P_2({\vec r}^{\prime}+\lambda{{\vec f^{\prime}}})<\exp^{i(\phi^r_1({\vec r})-\phi^r_1({\vec r}^{\prime})-\phi^r_2({\vec r}+\lambda{\vec f})+\phi^r_2({\vec r}^{\prime}+\lambda{{\vec f^{\prime}}}))}>_{\Phi}\mathrm{d}{\vec r}\mathrm{d}{\vec r^{\prime}}\mathrm{d}{\vec f}\mathrm{d}{\vec f}^{\prime}}{\left[\int
  P_1({\vec r})\mathrm{d}{\vec r} \int P_2({\vec r})\mathrm{d}{\vec r}\right]^2}\nonumber\\
&=&\frac{\int P_1({\vec r})P_2({\vec r}+\lambda{\vec f})P_1({\vec r}^{\prime})P_2({\vec r}^{\prime}+\lambda{{\vec f^{\prime}}})\exp^{-\frac{1}{2}\mathcal{D}_{\phi_r}({\vec r},{\vec r^{\prime}})-\frac{1}{2}\mathcal{D}_{\phi_r}({\vec r}+\lambda{\vec f},{\vec r^{\prime}}+\lambda{{\vec f^{\prime}}})}\mathrm{d}{\vec r}\mathrm{d}{\vec r^{\prime}}\mathrm{d}{\vec f}\mathrm{d}{\vec f}^{\prime}}{\left[\int
  P_1({\vec r})\mathrm{d}{\vec r} \int P_2({\vec r})\mathrm{d}{\vec r}\right]^2}\nonumber\\
&& \mbox{change of variables}~{\vec r}+\lambda{\vec f} \rightarrow {\vec u}, {\vec r^{\prime}}+\lambda{{\vec f^{\prime}}} \rightarrow {\vec u^{\prime}}:\nonumber\\
&=&\frac{\int
  P_1({\vec r})P_1({\vec r}^{\prime})\exp^{-\frac{1}{2}\mathcal{D}_{\phi_r}({\vec r},{\vec r^{\prime}})}\mathrm{d}{\vec r}\mathrm{d}{\vec r^{\prime}}\int
  P_2({\vec r})P_2({\vec r}^{\prime})\exp^{-\frac{1}{2}\mathcal{D}_{\phi_r}({\vec r},{\vec r^{\prime}})}\mathrm{d}{\vec r}\mathrm{d}{\vec r^{\prime}}}{\left[\int
    P_1({\vec r})\mathrm{d}{\vec r} \int P_2({\vec r})\mathrm{d}{\vec r}\right]^2} \label{eq_absrhoij2_mono}
\end{eqnarray}
Putting Eqs. (\ref{eq_rho_mono},\,\ref{eq_rhoij_mono},\,\ref{eq_rhoij2_mono},\,\ref{eq_absrhoij2_mono}) into  Eqs. (\ref{eq_absq2_mono},\,\ref{eq_q2_mono},\,\ref{eq_q_mono}),  we find again that the variance of the phase can be written as the quadratic sum of 3 terms,
corresponding respectively to the detector, photon and
atmospheric regimes $\sigma^2_{\phi} = \sigma^2_{det_{\phi}} +
\sigma^2_{phot_{\phi}} +  \sigma^2_{atm_{\phi}}$, with:
\begin{eqnarray}
\sigma^2_{det_{\phi}}&=&\frac{1}{2} \frac{
  N_{pix}\sigma^2_{det}}{\rho_0^2 \Sigma^2_P N^2 t_1t_2|V_{12}|^2\mathrm{e}^{-2\sigma^2_{\phi_r}}}\\
\sigma^2_{phot_{\phi}}&=&\frac{1}{2}
\frac{(t_1 + t_2) \overline{\mathcal{S}}}{\rho_0 \Sigma_P N t_1t_2|V_{12}|^2\mathrm{e}^{-2\sigma^2_{\phi_r}}}\\
\sigma^2_{atm_{\phi}}&=&\frac{1}{2}\frac{\left[\int P({\vec r})P({\vec r}^{\prime})\mathrm{e}^{-\frac{1}{2}\mathcal{D}_{\phi_r}({\vec r},{\vec r^{\prime}})}\mathrm{d}{\vec r}\mathrm{d}{{\vec r^{\prime}}}\right]^2-\mathrm{e}^{-4\sigma^2_{\phi_r}}\left[\int P({\vec r})P({\vec r}^{\prime})\mathrm{e}^{\frac{1}{2}\mathcal{D}_{\phi_r}({\vec r},{\vec r^{\prime}})}\mathrm{d}{\vec r}\mathrm{d}{{\vec r^{\prime}}}\right]^2}{\left[\int
      P({\vec r})\mathrm{d}{\vec r}\right]^4\mathrm{e}^{-2\sigma^2_{\phi_r}}}
\end{eqnarray}

\section{Coherent flux dropouts - Rician density probability} \label{app_rician}
By analogy with the study of \citet{canales_1} on the
statistics of the instantaneous Strehl ratio in partial AO correction,
the instantaneous {\it interferometric} Strehl ratio, namely
$I=|T_{12}|^2$ and $I=|\rho_{12}|^2$ for multimode and single-mode
interferometers respectively are shown to follow Rician density
probability statistics of the form:
\begin{equation}
dp(I) = \frac{1}{2\sigma^2}\exp\left(-\frac{I+a^2}{2\sigma^2}\right)I_0\left(-\frac{a\sqrt{I}}{\sigma^2}\right)
\end{equation}
In this appendix, we aim to derive the theoretical expression of the
moments of the Rician distribution (namely $a$ and $\sigma$), in both multimode
and single-mode cases. Following \citet{canales_1}, the quantity  $a$
and $\sigma$ can be expressed from the moments of the real ($A_r$) and
imaginary ($A_i$) parts of the complex instantaneous interferometric Strehl ratio
(i.e. such as $I = A_r^2 + A_i^2$):
\begin{eqnarray}
a^4 &=& <A_r>^4_{\Phi} + 2<A_r>^2_{\Phi}(\sigma^2_i - \sigma^2_r) -
(\sigma^2_i - \sigma^2_r)^2  \label{eq_a} \\
2\sigma^2 &=& \sigma^2_r + \sigma^2_i + <A_r>^2_{\Phi} - a^2  \label{eq_sig}
\end{eqnarray} 
where $\sigma^2_r = <A^2_r>_{\Phi} - <A_r>_{\Phi}^2$ and $\sigma^2_i = <A^2_i>_{\Phi} - <A_i>_{\Phi}^2$.

\subsection{Multimode case}
In multimode interferometry we have $I=|T_{12}|^2$, hence:
\begin{eqnarray}
A_r &=& \mathrm{Re}(T_{12}) = \frac{\int P_1({\vec r})P_2({\vec r}+\lambda{\vec f})\cos(\phi^r_1({\vec r})-\phi^r_2({\vec r}+\lambda{\vec f}))\mathrm{d}{\vec r}}{\Sigma_P}\\
A_i &=& \mathrm{Im}(T_{12}) = \frac{\int P_1({\vec r})P_2({\vec r}+\lambda{\vec f})\sin(\phi^r_1({\vec r})-\phi^r_2({\vec r}+\lambda{\vec f}))\mathrm{d}{\vec r}}{\Sigma_P}
\end{eqnarray}
Following the same approach as in appendix \ref{subapp_multimode_noise}, we compute the first and second order moments of these quantities.
\begin{eqnarray}
<A_r>_{\Phi} &=& \mathrm{Re}(T_{12}) = \frac{\int P_1({\vec r})P_2({\vec r}+\lambda{\vec f})<\cos(\phi^r_1({\vec r})-\phi^r_2({\vec r}+\lambda{\vec f}))>_{\Phi}\mathrm{d}{\vec r}}{\Sigma_P} =  \exp^{-\sigma^2_{\phi_r}} \label{eq_ar_multi}\\
<A_i>_{\Phi} &=& \mathrm{Im}(T_{12}) = \frac{\int P_1({\vec r})P_2({\vec r}+\lambda{\vec f})<\sin(\phi^r_1({\vec r})-\phi^r_2({\vec r}+\lambda{\vec f}))>_{\Phi}\mathrm{d}{\vec r}}{\Sigma_P} = 0 \label{eq_ai_multi}
\end{eqnarray}
\begin{eqnarray}
<A_r^2>_{\Phi} &=& \frac{\int
  P_1({\vec r})P_2({\vec r}+\lambda{{\vec f_{12}}})P_1({\vec r}^{\prime})P_2({\vec r}^{\prime}+\lambda{{\vec f_{12}}})<\cos(\phi^r_1({\vec r})-\phi^r_2({\vec r}+\lambda{{\vec f_{12}}}))\cos(\phi^r_1({\vec r}^{\prime})-\phi^r_2({\vec r}^{\prime}+\lambda{{\vec f_{12}}}))>_{\Phi}\mathrm{d}{\vec r}\mathrm{d}{{\vec r^{\prime}}}}{\Sigma_P^2} \nonumber \\
&& \mbox{using}~\cos(a)\cos(b) = (\cos(a-b) + \cos(a+b))/2 \nonumber\\
&=&  \frac{1}{2}\frac{\int
  P_1({\vec r})P_2({\vec r}+\lambda{{\vec f_{12}}})P_1({\vec r}^{\prime})P_2({\vec r}^{\prime}+\lambda{{\vec f_{12}}})<\cos(\phi^r_1({\vec r})-\phi^r_2({\vec r}+\lambda{{\vec f_{12}}})-\phi^r_1({\vec r}^{\prime})+\phi^r_2({\vec r}^{\prime}+\lambda{{\vec f_{12}}}))>_{\Phi}\mathrm{d}{\vec r}\mathrm{d}{{\vec r^{\prime}}}}{\Sigma_P^2}  \nonumber \\
&& + \frac{1}{2}\frac{\int
  P_1({\vec r})P_2({\vec r}+\lambda{{\vec f_{12}}})P_1({\vec r}^{\prime})P_2({\vec r}^{\prime}+\lambda{{\vec f_{12}}})<\cos(\phi^r_1({\vec r})-\phi^r_2({\vec r}+\lambda{{\vec f_{12}}})+\phi^r_1({\vec r}^{\prime})-\phi^r_2({\vec r}^{\prime}+\lambda{{\vec f_{12}}}))>_{\Phi}\mathrm{d}{\vec r}\mathrm{d}{{\vec r^{\prime}}}}{\Sigma_P^2} \nonumber \\
&& \mbox{changing axes reference:}~P_1,~P_2~\mbox{centered on 0},\nonumber \\
&=& \frac{1}{2} \frac{\int \left[
  P({\vec r})P({\vec r}^{\prime})\exp^{-\frac{1}{2}\mathcal{D}_{\phi_r}({\vec r},{\vec r^{\prime}})}\right]\left[P({\vec r})P({\vec r}^{\prime})\exp^{-\frac{1}{2}\mathcal{D}_{\phi_r}({\vec r},{\vec r^{\prime}})}\right]\mathrm{d}{\vec r}\mathrm{d}{{\vec r^{\prime}}}}{\left[\int
    P^2({\vec r})\mathrm{d}{\vec r}\right]^2}  \nonumber \\
&& + \frac{1}{2}  \exp^{-4\sigma^2_{\phi_r}}
\frac{\int \left[
  P({\vec r})P({\vec r}^{\prime})\exp^{\frac{1}{2}\mathcal{D}_{\phi_r}({\vec r},{\vec r^{\prime}})}\right]\left[P({\vec r})P({\vec r}^{\prime})\exp^{\frac{1}{2}\mathcal{D}_{\phi_r}({\vec r},{\vec r^{\prime}})}\right]\mathrm{d}{\vec r}\mathrm{d}{{\vec r^{\prime}}}}{\left[\int
    P^2({\vec r})\mathrm{d}{\vec r}\right]^2} \label{eq_ar2_multi}
\end{eqnarray}
\begin{eqnarray}
<A_i^2>_{\Phi} &=& \frac{\int
  P_1({\vec r})P_2({\vec r}+\lambda{{\vec f_{12}}})P_1({\vec r}^{\prime})P_2({\vec r}^{\prime}+\lambda{{\vec f_{12}}})<\sin(\phi^r_1({\vec r})-\phi^r_2({\vec r}+\lambda{{\vec f_{12}}}))\sin(\phi^r_1({\vec r}^{\prime})-\phi^r_2({\vec r}^{\prime}+\lambda{{\vec f_{12}}}))>_{\Phi}\mathrm{d}{\vec r}\mathrm{d}{{\vec r^{\prime}}}}{\Sigma_P^2} \nonumber \\
&& \mbox{using}~\sin(a)\sin(b) = (\cos(a-b) - \cos(a+b))/2 \nonumber\\
&=&  \frac{1}{2}\frac{\int
  P_1({\vec r})P_2({\vec r}+\lambda{{\vec f_{12}}})P_1({\vec r}^{\prime})P_2({\vec r}^{\prime}+\lambda{{\vec f_{12}}})<\cos(\phi^r_1({\vec r})-\phi^r_2({\vec r}+\lambda{{\vec f_{12}}})-\phi^r_1({\vec r}^{\prime})+\phi^r_2({\vec r}^{\prime}+\lambda{{\vec f_{12}}}))>_{\Phi}\mathrm{d}{\vec r}\mathrm{d}{{\vec r^{\prime}}}}{\Sigma_P^2}  \nonumber \\
&& - \frac{1}{2}\frac{\int
  P_1({\vec r})P_2({\vec r}+\lambda{{\vec f_{12}}})P_1({\vec r}^{\prime})P_2({\vec r}^{\prime}+\lambda{{\vec f_{12}}})<\cos(\phi^r_1({\vec r})-\phi^r_2({\vec r}+\lambda{{\vec f_{12}}})+\phi^r_1({\vec r}^{\prime})-\phi^r_2({\vec r}^{\prime}+\lambda{{\vec f_{12}}}))>_{\Phi}\mathrm{d}{\vec r}\mathrm{d}{{\vec r^{\prime}}}}{\Sigma_P^2} \nonumber \\
&& \mbox{changing axes reference:}~P_1,~P_2~\mbox{centered on 0},\nonumber \\
&=& \frac{1}{2} \frac{\int \left[
  P({\vec r})P({\vec r}^{\prime})\exp^{-\frac{1}{2}\mathcal{D}_{\phi_r}({\vec r},{\vec r^{\prime}})}\right]\left[P({\vec r})P({\vec r}^{\prime})\exp^{-\frac{1}{2}\mathcal{D}_{\phi_r}({\vec r},{\vec r^{\prime}})}\right]\mathrm{d}{\vec r}\mathrm{d}{{\vec r^{\prime}}}}{\left[\int
    P^2({\vec r})\mathrm{d}{\vec r}\right]^2}  \nonumber \\
&& - \frac{1}{2}  \exp^{-4\sigma^2_{\phi_r}}
\frac{\int \left[
  P({\vec r})P({\vec r}^{\prime})\exp^{\frac{1}{2}\mathcal{D}_{\phi_r}({\vec r},{\vec r^{\prime}})}\right]\left[P({\vec r})P({\vec r}^{\prime})\exp^{\frac{1}{2}\mathcal{D}_{\phi_r}({\vec r},{\vec r^{\prime}})}\right]\mathrm{d}{\vec r}\mathrm{d}{{\vec r^{\prime}}}}{\left[\int
    P^2({\vec r})\mathrm{d}{\vec r}\right]^2} \label{eq_ai2_multi}
\end{eqnarray}
Using Eqs. (\ref{eq_ar_multi},\,\ref{eq_ai_multi},\,\ref{eq_ar2_multi},\,\ref{eq_ai2_multi}) in Eqs. (\ref{eq_a},\,\ref{eq_sig}), we obtain the theoretical expressions of the parameters $a$ and $\sigma$ of the Rician distribution is the multimode case, as summarized in Table \ref{tab_rician} of the appendix. 

\subsection{Single-mode case}
In single-mode interferometry we have $I=|\rho_{12}|^2$, hence:
\begin{eqnarray}
A_r &=& \mathrm{Re}(\rho_{12}) = \frac{\int P_1({\vec r})P_2({\vec r}+\lambda{\vec f})\cos(\phi^r_1({\vec r})-\phi^r_2({\vec r}+\lambda{\vec f}))\mathrm{d}{\vec r}\mathrm{d}{\vec f}}{\int
  P_1({\vec r})\mathrm{d}{\vec r} \int P_2({\vec r})\mathrm{d}{\vec r}}\\
A_i &=& \mathrm{Im}(\rho_{12}) =  \frac{\int P_1({\vec r})P_2({\vec r}+\lambda{\vec f})\sin(\phi^r_1({\vec r})-\phi^r_2({\vec r}+\lambda{\vec f}))\mathrm{d}{\vec r}\mathrm{d}{\vec f}}{\int
  P_1({\vec r})\mathrm{d}{\vec r} \int P_2({\vec r})\mathrm{d}{\vec r}}
\end{eqnarray}
Following the same formalism as previously, we compute the first and second order moments of these quantities.
\begin{eqnarray}
<A_r>_{\Phi} &=& \frac{\int P_1({\vec r})P_2({\vec r}+\lambda{\vec f})<\cos(\phi^r_1({\vec r})-\phi^r_2({\vec r}+\lambda{\vec f}))>_{\Phi}\mathrm{d}{\vec r}\mathrm{d}{\vec f}}{\int
  P_1({\vec r})\mathrm{d}{\vec r} \int P_2({\vec r})\mathrm{d}{\vec r}} = \exp^{-\sigma^2_{\phi_r}}\label{eq_ar_mono}\\
<A_i>_{\Phi} &=& \frac{\int P_1({\vec r})P_2({\vec r}+\lambda{\vec f})<\sin(\phi^r_1({\vec r})-\phi^r_2({\vec r}+\lambda{\vec f}))>_{\Phi}\mathrm{d}{\vec r}\mathrm{d}{\vec f}}{\int
  P_1({\vec r})\mathrm{d}{\vec r} \int P_2({\vec r})\mathrm{d}{\vec r}} = 0 \label{eq_ai_mono}
\end{eqnarray}
\begin{eqnarray}
<A_r^2>_{\Phi} &=& \frac{\int P_1({\vec r})P_2({\vec r}+\lambda{\vec f})P_1({\vec r}^{\prime})P_2({\vec r}^{\prime}+\lambda{{\vec f^{\prime}}})<\cos(\phi^r_1({\vec r})-\phi^r_2({\vec r}+\lambda{\vec f}))\cos(\phi^r_1({\vec r}^{\prime})-\phi^r_2({\vec r}^{\prime}+\lambda{{\vec f^{\prime}}}))>_{\Phi}\mathrm{d}{\vec r}\mathrm{d}{\vec r^{\prime}}\mathrm{d}{\vec f}\mathrm{d}{\vec f}^{\prime}}{\left[\int
  P_1({\vec r})\mathrm{d}{\vec r} \int P_2({\vec r})\mathrm{d}{\vec r}\right]^2} \nonumber \\
&& \mbox{using}~\cos(a)\cos(b) = (\cos(a-b) + \cos(a+b))/2 \nonumber\\
&=& \frac{1}{2}\frac{\int P_1({\vec r})P_2({\vec r}+\lambda{\vec f})P_1({\vec r}^{\prime})P_2({\vec r}^{\prime}+\lambda{{\vec f^{\prime}}})<\cos(\phi^r_1({\vec r})-\phi^r_2({\vec r}+\lambda{\vec f})-\phi^r_1({\vec r}^{\prime})+\phi^r_2({\vec r}^{\prime}+\lambda{{\vec f^{\prime}}})>_{\Phi}\mathrm{d}{\vec r}\mathrm{d}{\vec r^{\prime}}\mathrm{d}{\vec f}\mathrm{d}{\vec f}^{\prime}}{\left[\int
  P_1({\vec r})\mathrm{d}{\vec r} \int P_2({\vec r})\mathrm{d}{\vec r}\right]^2}  \nonumber\\
&& + \frac{1}{2}\frac{\int P_1({\vec r})P_2({\vec r}+\lambda{\vec f})P_1({\vec r}^{\prime})P_2({\vec r}^{\prime}+\lambda{{\vec f^{\prime}}})<\cos(\phi^r_1({\vec r})-\phi^r_2({\vec r}+\lambda{\vec f})+\phi^r_1({\vec r}^{\prime})-\phi^r_2({\vec r}^{\prime}+\lambda{{\vec f^{\prime}}}))>_{\Phi}\mathrm{d}{\vec r}\mathrm{d}{\vec r^{\prime}}\mathrm{d}{\vec f}\mathrm{d}{\vec f}^{\prime}}{\left[\int
  P_1({\vec r})\mathrm{d}{\vec r} \int P_2({\vec r})\mathrm{d}{\vec r}\right]^2}  \nonumber \\
&& \mbox{change of variables}~{\vec r}+\lambda{\vec f} \rightarrow u, {\vec r^{\prime}}+\lambda{{\vec f^{\prime}}} \rightarrow {\vec u^{\prime}}:\nonumber\\
&=&\frac{1}{2}\frac{\int
  P_1({\vec r})P_1({\vec r}^{\prime})\exp^{-\frac{1}{2}\mathcal{D}_{\phi_r}({\vec r},{\vec r^{\prime}})}\mathrm{d}{\vec r}\mathrm{d}{\vec r^{\prime}}\int
  P_2({\vec r})P_2({\vec r}^{\prime})\exp^{-\frac{1}{2}\mathcal{D}_{\phi_r}({\vec r},{\vec r^{\prime}})}\mathrm{d}{\vec r}\mathrm{d}{\vec r^{\prime}}}{\left[\int
    P_1({\vec r})\mathrm{d}{\vec r} \int P_2({\vec r})\mathrm{d}{\vec r}\right]^2}   \nonumber\\
&& + \mathrm{e}^{-4\sigma^2_{\phi_r}}\frac{\int
  P_1({\vec r})P_1({\vec r}^{\prime})\exp^{\frac{1}{2}\mathcal{D}_{\phi_r}({\vec r},{\vec r^{\prime}})}\mathrm{d}{\vec r}\mathrm{d}{\vec r^{\prime}}\int
  P_2({\vec r})P_2({\vec r}^{\prime})\exp^{\frac{1}{2}\mathcal{D}_{\phi_r}({\vec r},{\vec r^{\prime}})}\mathrm{d}{\vec r}\mathrm{d}{\vec r^{\prime}}}{\left[\int
    P_1({\vec r})\mathrm{d}{\vec r} \int P_2({\vec r})\mathrm{d}{\vec r}\right]^2} \label{eq_ar2_mono}
\end{eqnarray}
\begin{eqnarray}
<A_i^2>_{\Phi} &=& \frac{\int P_1({\vec r})P_2({\vec r}+\lambda{\vec f})P_1({\vec r}^{\prime})P_2({\vec r}^{\prime}+\lambda{{\vec f^{\prime}}})<\sin(\phi^r_1({\vec r})-\phi^r_2({\vec r}+\lambda{\vec f}))\sin(\phi^r_1({\vec r}^{\prime})-\phi^r_2({\vec r}^{\prime}+\lambda{{\vec f^{\prime}}}))>_{\Phi}\mathrm{d}{\vec r}\mathrm{d}{\vec r^{\prime}}\mathrm{d}{\vec f}\mathrm{d}{\vec f}^{\prime}}{\left[\int
  P_1({\vec r})\mathrm{d}{\vec r} \int P_2({\vec r})\mathrm{d}{\vec r}\right]^2} \nonumber \\
&& \mbox{using}~\sin(a)\sin(b) = (\cos(a-b) - \cos(a+b))/2 \nonumber\\
&=& \frac{1}{2}\frac{\int P_1({\vec r})P_2({\vec r}+\lambda{\vec f})P_1({\vec r}^{\prime})P_2({\vec r}^{\prime}+\lambda{{\vec f^{\prime}}})<\cos(\phi^r_1({\vec r})-\phi^r_2({\vec r}+\lambda{\vec f})-\phi^r_1({\vec r}^{\prime})+\phi^r_2({\vec r}^{\prime}+\lambda{{\vec f^{\prime}}})>_{\Phi}\mathrm{d}{\vec r}\mathrm{d}{\vec r^{\prime}}\mathrm{d}{\vec f}\mathrm{d}{\vec f}^{\prime}}{\left[\int
  P_1({\vec r})\mathrm{d}{\vec r} \int P_2({\vec r})\mathrm{d}{\vec r}\right]^2}  \nonumber\\
&& - \frac{1}{2}\frac{\int P_1({\vec r})P_2({\vec r}+\lambda{\vec f})P_1({\vec r}^{\prime})P_2({\vec r}^{\prime}+\lambda{{\vec f^{\prime}}})<\cos(\phi^r_1({\vec r})-\phi^r_2({\vec r}+\lambda{\vec f})+\phi^r_1({\vec r}^{\prime})-\phi^r_2({\vec r}^{\prime}+\lambda{{\vec f^{\prime}}})>_{\Phi}\mathrm{d}{\vec r}\mathrm{d}{\vec r^{\prime}}\mathrm{d}{\vec f}\mathrm{d}{\vec f}^{\prime}}{\left[\int
  P_1({\vec r})\mathrm{d}{\vec r} \int P_2({\vec r})\mathrm{d}{\vec r}\right]^2}  \nonumber \\
&& \mbox{change of variables}~{\vec r}+\lambda{\vec f} \rightarrow {\vec u}, {\vec r^{\prime}}+\lambda{{\vec f^{\prime}}} \rightarrow {\vec u^{\prime}}:\nonumber\\
&=&\frac{1}{2}\frac{\int
  P_1({\vec r})P_1({\vec r}^{\prime})\exp^{-\frac{1}{2}\mathcal{D}_{\phi_r}({\vec r},{\vec r^{\prime}})}\mathrm{d}{\vec r}\mathrm{d}{\vec r^{\prime}}\int
  P_2({\vec r})P_2({\vec r}^{\prime})\exp^{-\frac{1}{2}\mathcal{D}_{\phi_r}({\vec r},{\vec r^{\prime}})}\mathrm{d}{\vec r}\mathrm{d}{\vec r^{\prime}}}{\left[\int
    P_1({\vec r})\mathrm{d}{\vec r} \int P_2({\vec r})\mathrm{d}{\vec r}\right]^2}   \nonumber\\
&& - \mathrm{e}^{-4\sigma^2_{\phi_r}}\frac{\int
  P_1({\vec r})P_1({\vec r}^{\prime})\exp^{\frac{1}{2}\mathcal{D}_{\phi_r}({\vec r},{\vec r^{\prime}})}\mathrm{d}{\vec r}\mathrm{d}{\vec r^{\prime}}\int
  P_2({\vec r})P_2({\vec r}^{\prime})\exp^{\frac{1}{2}\mathcal{D}_{\phi_r}({\vec r},{\vec r^{\prime}})}\mathrm{d}{\vec r}\mathrm{d}{\vec r^{\prime}}}{\left[\int
    P_1({\vec r})\mathrm{d}{\vec r} \int P_2({\vec r})\mathrm{d}{\vec r}\right]^2}  \label{eq_ai2_mono}
\end{eqnarray}
Putting  Eqs. (\ref{eq_ar_mono},\,\ref{eq_ai_mono},\,\ref{eq_ar2_mono},\,\ref{eq_ai2_mono}) in Eqs. (\ref{eq_a},\,\ref{eq_sig}), we obtain the theoretical expressions of the parameters $a$ and $\sigma$ of the Rician distribution is the single-mode case, as summarized in Table \ref{tab_rician} of the appendix. 

\begin{table}
\caption{Parameters $a$ and $\sigma$ of the Rician probability distribution function for both multimode and single-mode cases.\label{tab_rician}}
\begin{center}
\begin{tabular}{|lll|}
\multicolumn{3}{l}{Multimode case:} \\
\hline
&&\\
$a^2$ & = & $\ds \exp^{-2\sigma^2_{\phi_r}}\sqrt{\left(2-\exp^{-4\sigma^2_{\phi_r}}\left[\frac{\int \left[P({\vec r})P({\vec r}^{\prime})\mathrm{e}^{\frac{1}{2}\mathcal{D}_{\phi_r}({\vec r},{\vec r^{\prime}})}\right]^2\mathrm{d}{\vec r}\mathrm{d}{{\vec r^{\prime}}}}{\left[\int
    P^2({\vec r})\mathrm{d}{\vec r}\right]^2}\right]^2\right)}$\\
$\sigma^2$ & = & $\ds \frac{1}{2}\left(\frac{\int
  \left[P({\vec r})P({\vec r}^{\prime})\mathrm{e}^{-\frac{1}{2}\mathcal{D}_{\phi_r}({\vec r},{\vec r^{\prime}})}\right]^2\mathrm{d}{\vec r}\mathrm{d}{{\vec r^{\prime}}}}{\left[\int
    P^2({\vec r})\mathrm{d}{\vec r}\right]^2} - a^2\right) $\\&&\\
\hline
\multicolumn{3}{l}{Single-mode case:} \\
\hline
&&\\
$a^2$ & = & $\ds \exp^{-2\sigma^2_{\phi_r}}\sqrt{\left(2-\exp^{-4\sigma^2_{\phi_r}}\left[\frac{\left[\int P({\vec r})P({\vec r}^{\prime})\mathrm{e}^{\frac{1}{2}\mathcal{D}_{\phi_r}({\vec r},{\vec r^{\prime}})}\mathrm{d}{\vec r}\mathrm{d}{{\vec r^{\prime}}}\right]^2}{\left[\int
      P({\vec r})\mathrm{d}{\vec r}\right]^4}\right]^2\right)}$\\
$\sigma^2$ & = & $\ds \frac{1}{2}\left(\frac{\left[\int P({\vec r})P({\vec r}^{\prime})\mathrm{e}^{-\frac{1}{2}\mathcal{D}_{\phi_r}({\vec r},{\vec r^{\prime}})}\mathrm{d}{\vec r}\mathrm{d}{{\vec r^{\prime}}}\right]^2}{\left[\int
      P({\vec r})\mathrm{d}{\vec r}\right]^4}- a^2\right) $\\&&\\
\hline
\end{tabular}
\end{center}
\end{table}

\section{Astrometric phase error} \label{app_astrometry}
The astrometric phase error writes:
\begin{equation}
\sigma^2_{\Delta_{\phi}^{\Delta\alpha{h}}} = 2\sigma^2_{\phi} -
2\mathrm{cov}(\phi_s, \phi_r^{\Delta\alpha{h}}) \label{app_eq_errnarrow}
\end{equation}
where $\phi_s$, and $\phi_r^{\Delta\alpha{h}}$ are respectively the partially AO-corrected phase of the astrophysical target and the phase estimated from the off-axis reference source. $\vec{\Delta \alpha}$ is the angular separation between both sources, and h is the altitude of the turbulent layer. \\
Since $\sigma^2_{\phi}$ has been computed in previous sections, it remains to compute the covariance part of Eq. (\ref{app_eq_errnarrow}). The formal expression of the covariance is given by \citet{chelli_1}:
\begin{equation}
\mathrm{cov}(\phi_s, \phi_r^{\Delta\alpha{h}}) = \frac{1}{2}\frac{\mathrm{Re}[\mathrm{E}(QQ^{\ast}_{\Delta\alpha{h}})]-\mathrm{Re}[\mathrm{E}(QQ_{\Delta\alpha{h}})]}{\mathrm{E}(Q)\mathrm{E}(Q_{\Delta\alpha{h}})}
\end{equation}
where $Q_{\Delta\alpha{h}}$ is the estimator of the coherent flux for the off-axis reference source. 

\subsection{Multimode case}
Using formalism of Sect. \ref{app_sigma}, one can derive the moments of the multimode estimator:
\begin{eqnarray}
\mathrm{E}(Q) &=&  N \sqrt{t_1t_2} V_{12} <S_{12}({\vec f_{12}})>_{\phi} \label{eq_qcov_multi}\\
\mathrm{E}(Q_{\Delta\alpha{h}}) &=&  N^{\Delta\alpha{h}} \sqrt{t_1t_2} V^{\Delta\alpha{h}}_{12} <S^{\Delta\alpha{h}}_{12}({\vec f_{12}})>_{\phi} \label{eq_qcovh_multi}\\
\mathrm{E}(QQ^{\ast}_{\Delta\alpha{h}}) &=& N N^{\Delta\alpha{h}} t_1t_2  V_{12} (V^{\Delta\alpha{h}}_{12})^{\ast} <S_{12}({\vec f_{12}})(S^{\Delta\alpha{h}}_{12})^{\ast}({\vec f_{12}})>_{\phi}\label{eq_absq2cov_multi}\\
\mathrm{E}(QQ_{\Delta\alpha{h}}) &=&  N N^{\Delta\alpha{h}} t_1t_2  V_{12} V^{\Delta\alpha{h}}_{12}  <S_{12}({\vec f_{12}})S^{\Delta\alpha{h}}_{12}({\vec f_{12}})>_{\phi}\label{eq_q2cov_multi}
\end{eqnarray}
where $N^{\Delta\alpha{h}}$ and $V^{\Delta\alpha{h}}_{12}$ are respectively the number of photons (per surface unit and per time unit) and the visibility of the reference source, and with:
\begin{eqnarray}
<S^{\Delta\alpha{h}}_{12}({\vec f_{12}})>_{\Phi} &=& \int
  P_1({\vec r})P_2({\vec r}+\lambda{{\vec f_{12}}})<\exp^{i(\phi^{\Delta\alpha{h}}_1({\vec r})-\phi^{\Delta\alpha{h}}_2({\vec r}+\lambda{{\vec f_{12}}}))}>_{\Phi}\mathrm{d}{\vec r} = \Sigma_P  \exp^{-\sigma^2_{\phi_r}} \label{eq_icov_multi}
\end{eqnarray}
Since we assume one turbulent layer located at the altitude $h$, we have by definition $\phi^{\Delta\alpha{h}}_1({\vec r}) = \phi_1({\vec r}+\vec{\Delta\alpha}{h})$ and $\phi^{\Delta\alpha{h}}_2({\vec r}) = \phi_2({\vec r}+\vec{\Delta\alpha}{h})$, hence:
\begin{eqnarray}
<S^{\Delta\alpha{h}}_{12}({\vec f_{12}})>_{\Phi} &=& \int
  P_1({\vec r})P_2({\vec r}+\lambda{{\vec f_{12}}})<\exp^{i(\phi_1({\vec r}+\vec{\Delta\alpha}{h})-\phi_2({\vec r}+\lambda{{\vec f_{12}}}+\vec{\Delta\alpha}{h}))}>_{\Phi}\mathrm{d}{\vec r} = \Sigma_P \exp^{-\sigma^2_{\phi_r}}\label{eq_icovh_multi}
\end{eqnarray}
Furthermore, we have:
\begin{eqnarray}
&& <S_{12}({\vec f_{12}})S^{\Delta\alpha{h}}_{12}({\vec f_{12}})>_{\Phi}  \nonumber \\
&&= \int
  P_1({\vec r})P_2({\vec r}+\lambda{{\vec f_{12}}})P_1({\vec r}^{\prime})P_2({\vec r}^{\prime}+\lambda{{\vec f_{12}}})<\exp^{i(\phi_1({\vec r})-\phi_2({\vec r}+\lambda{{\vec f_{12}}})+\phi_1({\vec r}^{\prime}+\vec{\Delta\alpha}{h})-\phi_2({\vec r}^{\prime}+\lambda{{\vec f_{12}}+\vec{\Delta\alpha}{h}}))}>_{\Phi}\mathrm{d}{\vec r}\mathrm{d}{{\vec r^{\prime}}} \nonumber \\
&&= \int
  P_1({\vec r})P_2({\vec r}+\lambda{{\vec f_{12}}})P_1({\vec r}^{\prime})P_2({\vec r}^{\prime}+\lambda{{\vec f_{12}}})\exp^{-\frac{1}{2}<|(\phi_1({\vec r})-\phi_2({\vec r}+\lambda{{\vec f_{12}}})+\phi_1({\vec r}^{\prime}+\vec{\Delta\alpha}{h})-\phi_2({\vec r}^{\prime}+\lambda{{\vec f_{12}}}+\vec{\Delta\alpha}{h}))|^2>_{\Phi}}\mathrm{d}{\vec r}\mathrm{d}{{\vec r^{\prime}}}  \nonumber \\
&& \mbox{knowing that}~<\phi_1({\vec r})\phi_1({\vec r}^{\prime}+\vec{\Delta\alpha}{h})>_{\Phi} = 
\sigma^2_{\phi_r} - \frac{1}{2}\mathcal{D}_{\phi_r}({\vec r},{\vec r^{\prime}}+\vec{\Delta\alpha}{h}),~\mbox{it comes}, \nonumber \\
&&= \exp^{-4\sigma^2_{\phi_r}} \int
  P_1({\vec r})P_2({\vec r}+\lambda{{\vec f_{12}}})P_1({\vec r}^{\prime})P_2({\vec r}^{\prime}+\lambda{{\vec f_{12}}})\exp^{\frac{1}{2}\mathcal{D}_{\phi_r}({\vec r},{\vec r^{\prime}}+\vec{\Delta\alpha}{h})+\frac{1}{2}\mathcal{D}_{\phi_r}({\vec r}+\lambda{{\vec f_{12}}},{\vec r^{\prime}}+\lambda{{\vec f_{12}}}+\vec{\Delta\alpha}{h})}\mathrm{d}{\vec r}\mathrm{d}{{\vec r^{\prime}}} \nonumber \\
&& \mbox{changing axes reference:}~P_1,~P_2~\mbox{centered on 0},\nonumber \\
&&= \exp^{-4\sigma^2_{\phi_r}} \int \left[
  P({\vec r})P({\vec r}^{\prime})\exp^{\frac{1}{2}\mathcal{D}_{\phi_r}({\vec r},{\vec r^{\prime}}+\vec{\Delta\alpha}{h})}\right]\left[P({\vec r})P({\vec r}^{\prime})\exp^{\frac{1}{2}\mathcal{D}_{\phi_r}({\vec r},{\vec r^{\prime}}+\vec{\Delta\alpha}{h})}\right]\mathrm{d}{\vec r}\mathrm{d}{{\vec r^{\prime}}} \label{eq_i2cov_multi}
\end{eqnarray}
\begin{eqnarray}
&&<S_{12}({\vec f_{12}})(S^{\Delta\alpha{h}}_{12})^{\ast}({\vec f_{12}})>_{\Phi} \\
&&= \int
  P_1({\vec r})P_2({\vec r}+\lambda{{\vec f_{12}}})P_1({\vec r}^{\prime})P_2({\vec r}^{\prime}+\lambda{{\vec f_{12}}})<\exp^{i(\phi_1({\vec r})-\phi_2({\vec r}+\lambda{{\vec f_{12}}})-\phi_1({\vec r}^{\prime}+\vec{\Delta\alpha}{h})+\phi_2({\vec r}^{\prime}+\lambda{{\vec f_{12}}}+\vec{\Delta\alpha}{h}))}>_{\Phi}\mathrm{d}{\vec r}\mathrm{d}{{\vec r^{\prime}}} \nonumber \\
&&=  \int
  P_1({\vec r})P_2({\vec r}+\lambda{{\vec f_{12}}})P_1({\vec r}^{\prime})P_2({\vec r}^{\prime}+\lambda{{\vec f_{12}}})\exp^{-\frac{1}{2}<|(\phi_1({\vec r})-\phi_2({\vec r}+\lambda{{\vec f_{12}}})-\phi_1({\vec r}^{\prime}+\vec{\Delta\alpha}{h})+\phi_2({\vec r}^{\prime}+\lambda{{\vec f_{12}}}+\vec{\Delta\alpha}{h}))|^2>_{\Phi}}\mathrm{d}{\vec r}\mathrm{d}{{\vec r^{\prime}}} \nonumber \\
&&= \int
  P_1({\vec r})P_2({\vec r}+\lambda{{\vec f_{12}}})P_1({\vec r}^{\prime})P_2({\vec r}^{\prime}+\lambda{{\vec f_{12}}})\exp^{-\frac{1}{2}\mathcal{D}_{\phi_r}({\vec r},{\vec r^{\prime}}+\vec{\Delta\alpha}{h})-\frac{1}{2}\mathcal{D}_{\phi_r}({\vec r}+\lambda{{\vec f_{12}}},{\vec r^{\prime}}+\lambda{{\vec f_{12}}}+\vec{\Delta\alpha}{h})}\mathrm{d}{\vec r}\mathrm{d}{{\vec r^{\prime}}} \nonumber \\
&& \mbox{changing axes reference:}~P_1,~P_2~\mbox{centered on 0},\nonumber \\
&&=  \int \left[
  P({\vec r})P({\vec r}^{\prime}))\exp^{-\frac{1}{2}\mathcal{D}_{\phi_r}({\vec r},{\vec r^{\prime}}+\vec{\Delta\alpha}{h})}\right]\left[P({\vec r}P({\vec r}^{\prime})\exp^{-\frac{1}{2}\mathcal{D}_{\phi_r}({\vec r},{\vec r^{\prime}}+\vec{\Delta\alpha}{h})}\right]\mathrm{d}{\vec r}\mathrm{d}{{\vec r^{\prime}}} \label{eq_absi2cov_multi}
\end{eqnarray}
Putting Eqs. (\ref{eq_icov_multi},\,\ref{eq_icovh_multi},\,\ref{eq_i2cov_multi},\,\ref{eq_absi2cov_multi}) into  Eqs. (\ref{eq_absq2cov_multi},\,\ref{eq_q2cov_multi},\,\ref{eq_qcov_multi},\,\ref{eq_qcovh_multi}), we
finally find that the covariance part of the multimode astrometric phase writes:
\begin{equation}
\mathrm{cov}(\phi_s, \phi_r^{\Delta\alpha{h}}) = \frac{1}{2}\frac{\int \left[P({\vec r})P({\vec r}^{\prime})\mathrm{e}^{-\frac{1}{2}\mathcal{D}_{\phi_r}({\vec r},{\vec r^{\prime}}+\vec{\Delta\alpha}{h})}\right]^2\mathrm{d}{\vec r}\mathrm{d}{{\vec r^{\prime}}}-\mathrm{e}^{-4\sigma^2_{\phi_r}}\int \left[P({\vec r})P({\vec r}^{\prime})\mathrm{e}^{\frac{1}{2}\mathcal{D}_{\phi_r}({\vec r},{\vec r^{\prime}}+\vec{\Delta\alpha}{h})}\right]^2\mathrm{d}{\vec r}\mathrm{d}{{\vec r^{\prime}}}}{\left[\int
      P^2({\vec r})\mathrm{d}{\vec r}\right]^2\mathrm{e}^{-2\sigma^2_{\phi_r}}}
\end{equation}

\subsection{Single-mode case}
We once again use formalism of Sect. \ref{app_sigma} to compute the moments of the single-mode estimator:
\begin{eqnarray}
\mathrm{E}(Q) &=&  \Sigma_P N \sqrt{t_1t_2} \rho_0 V_{12} <\rho_{12}>_{\phi} \label{eq_qcov_mono}\\
\mathrm{E}(Q_{\Delta\alpha{h}}) &=&  \Sigma_P N^{\Delta\alpha{h}} \sqrt{t_1t_2} \rho_0 V^{\Delta\alpha{h}}_{12} <\rho^{\Delta\alpha{h}}_{12}>_{\phi} \label{eq_qcovh_mono}\\
\mathrm{E}(QQ^{\ast}_{\Delta\alpha{h}}) &=& \Sigma_P^2 N N^{\Delta\alpha{h}} t_1t_2 \rho^2_0 V_{12} (V^{\Delta\alpha{h}}_{12})^{\ast} <\rho_{12}(\rho^{\Delta\alpha{h}}_{12})^{\ast}>_{\phi}\label{eq_absq2cov_mono}\\
\mathrm{E}(QQ_{\Delta\alpha{h}}) &=&  \Sigma_P^2 N N^{\Delta\alpha{h}} t_1t_2  \rho^2_0 V_{12} V^{\Delta\alpha{h}}_{12} <\rho_{12}\rho^{\Delta\alpha{h}}_{12}>_{\phi}\label{eq_q2cov_mono}
\end{eqnarray}
with
\begin{eqnarray}
<\rho_{12}>_{\Phi} &=&  \mathrm{e}^{-\sigma^2_{\phi_r}}  \label{eq_rhoijcov_mono} \\
<\rho^{\Delta\alpha{h}}_{12}>_{\Phi} &=&  \mathrm{e}^{-\sigma^2_{\phi_r}} \label{eq_rhoijcovh_mono}
\end{eqnarray}
We also have:
\begin{eqnarray}
<\rho_{12}\rho^{\Delta\alpha{h}}_{12}>_{\Phi}&=& \frac{
  P_1({\vec r})P_2({\vec r}+\lambda{\vec f})P_1({\vec r}^{\prime})P_2({\vec r}^{\prime}+\lambda{{\vec f^{\prime}}})<\exp^{i(\phi_1({\vec r})+\phi_1({\vec r}^{\prime}+\vec{\Delta\alpha}{h})-\phi_2({\vec r}+\lambda{\vec f})-\phi_2({\vec r}^{\prime}+\lambda{{\vec f^{\prime}}}+\vec{\Delta\alpha}{h}))}>_{\Phi}\mathrm{d}{\vec r}\mathrm{d}{\vec r^{\prime}}\mathrm{d}{\vec f}\mathrm{d}{\vec f}^{\prime}}{\left[\int
  P_1({\vec r})\mathrm{d}{\vec r} \int P_2({\vec r})\mathrm{d}{\vec r}\right]^2}\nonumber\\
&=&\mathrm{e}^{-4\sigma^2_{\phi_r}}\frac{\int P_1({\vec r})P_2({\vec r}+\lambda{\vec f})P_1({\vec r}^{\prime})P_2({\vec r}^{\prime}+\lambda{{\vec f^{\prime}}})\exp^{\frac{1}{2}\mathcal{D}_{\phi_r}({\vec r},{\vec r^{\prime}}+\vec{\Delta\alpha}{h})+\frac{1}{2}\mathcal{D}_{\phi_r}({\vec r}+\lambda{\vec f},{\vec r^{\prime}}+\lambda{{\vec f^{\prime}}}+\vec{\Delta\alpha}{h})}\mathrm{d}{\vec r}\mathrm{d}{\vec r^{\prime}}\mathrm{d}{\vec f}\mathrm{d}{\vec f}^{\prime}}{\left[\int
  P_1({\vec r})\mathrm{d}{\vec r} \int P_2({\vec r})\mathrm{d}{\vec r}\right]^2}\nonumber\\
&& \mbox{change of variables}~{\vec r}+\lambda{\vec f} \rightarrow {\vec u}, {\vec r^{\prime}}+\lambda{{\vec f^{\prime}}} \rightarrow {\vec u^{\prime}}:\nonumber\\
&=&\mathrm{e}^{-4\sigma^2_{\phi_r}}\frac{\int
  P_1({\vec r})P_1({\vec r}^{\prime})\exp^{\frac{1}{2}\mathcal{D}_{\phi_r}({\vec r},{\vec r^{\prime}}+\vec{\Delta\alpha}{h})}\mathrm{d}{\vec r}\mathrm{d}{\vec r^{\prime}}\int
  P_2({\vec r})P_2({\vec r}^{\prime})\exp^{\frac{1}{2}\mathcal{D}_{\phi_r}({\vec r},{\vec r^{\prime}}+\vec{\Delta\alpha}{h})}\mathrm{d}{\vec r}\mathrm{d}{\vec r^{\prime}}}{\left[\int
    P_1({\vec r})\mathrm{d}{\vec r} \int P_2({\vec r})\mathrm{d}{\vec r}\right]^2} \label{eq_rhoij2cov_mono}
\end{eqnarray}
\begin{eqnarray}
<\rho_{12}(\rho^{\vec{\Delta\alpha}{h}}_{12})^{\ast}>_{\Phi}&=&\frac{P_1({\vec r})P_2({\vec r}+\lambda{\vec f})P_1({\vec r}^{\prime})P_2({\vec r}^{\prime}+\lambda{{\vec f^{\prime}}})<\exp^{i(\phi_1({\vec r})-\phi_1({\vec r}^{\prime}+\vec{\Delta\alpha}{h})-\phi_2({\vec r}+\lambda{\vec f})+\phi_2({\vec r}^{\prime}+\lambda{{\vec f^{\prime}}}+\vec{\Delta\alpha}{h}))}>_{\Phi}\mathrm{d}{\vec r}\mathrm{d}{\vec r^{\prime}}\mathrm{d}{\vec f}\mathrm{d}{\vec f}^{\prime}}{\left[\int
  P_1({\vec r})\mathrm{d}{\vec r} \int P_2({\vec r})\mathrm{d}{\vec r}\right]^2}\nonumber\\
&=&\frac{\int P_1({\vec r})P_2({\vec r}+\lambda{\vec f})P_1({\vec r}^{\prime})P_2({\vec r}^{\prime}+\lambda{{\vec f^{\prime}}})\exp^{-\frac{1}{2}\mathcal{D}_{\phi_r}({\vec r},{\vec r^{\prime}}+\vec{\Delta\alpha}{h})-\frac{1}{2}\mathcal{D}_{\phi_r}({\vec r}+\lambda{\vec f},{\vec r^{\prime}}+\lambda{{\vec f^{\prime}}}+\vec{\Delta\alpha}{h})}\mathrm{d}{\vec r}\mathrm{d}{\vec r^{\prime}}\mathrm{d}{\vec f}\mathrm{d}{\vec f}^{\prime}}{\left[\int
  P_1({\vec r})\mathrm{d}{\vec r} \int P_2({\vec r})\mathrm{d}{\vec r}\right]^2}\nonumber\\
&& \mbox{change of variables}~{\vec r}+\lambda{\vec f} \rightarrow {\vec u}, {\vec r^{\prime}}+\lambda{{\vec f^{\prime}}} \rightarrow {\vec u^{\prime}}:\nonumber\\
&=&\frac{\int
  P_1({\vec r})P_1({\vec r}^{\prime})\exp^{-\frac{1}{2}\mathcal{D}_{\phi_r}({\vec r},{\vec r^{\prime}}+\vec{\Delta\alpha}{h})}\mathrm{d}{\vec r}\mathrm{d}{\vec r^{\prime}}\int
  P_2({\vec r})P_2({\vec r}^{\prime})\exp^{-\frac{1}{2}\mathcal{D}_{\phi_r}({\vec r},{\vec r^{\prime}}+\vec{\Delta\alpha}{h})}\mathrm{d}{\vec r}\mathrm{d}{\vec r^{\prime}}}{\left[\int
    P_1({\vec r})\mathrm{d}{\vec r} \int P_2({\vec r})\mathrm{d}{\vec r}\right]^2} \label{eq_absrhoij2cov_mono}
\end{eqnarray}
Putting Eqs. (\ref{eq_rhoijcov_mono},\,\ref{eq_rhoijcovh_mono},\,\ref{eq_rhoij2cov_mono},\,\ref{eq_absrhoij2cov_mono}) into  Eqs. (\ref{eq_absq2cov_mono},\,\ref{eq_q2cov_mono},\,\ref{eq_qcov_mono},\,\ref{eq_qcovh_mono}), we
find that the covariance part of the single-mode astrometric phase writes:
\begin{equation}
\mathrm{cov}(\phi_s, \phi_r^{\Delta\alpha{h}}) = \frac{1}{2}\frac{\left[\int P({\vec r})P({\vec r}^{\prime})\mathrm{e}^{-\frac{1}{2}\mathcal{D}_{\phi_r}({\vec r},{\vec r^{\prime}}+\vec{\Delta\alpha}{h})}\mathrm{d}{\vec r}\mathrm{d}{{\vec r^{\prime}}}\right]^2-\mathrm{e}^{-4\sigma^2_{\phi_r}}\left[\int P({\vec r})P({\vec r}^{\prime})\mathrm{e}^{\frac{1}{2}\mathcal{D}_{\phi_r}({\vec r},{\vec r^{\prime}}+\vec{\Delta\alpha}{h})}\mathrm{d}{\vec r}\mathrm{d}{{\vec r^{\prime}}}\right]^2}{\left[\int
      P({\vec r})\mathrm{d}{\vec r}\right]^4\mathrm{e}^{-2\sigma^2_{\phi_r}}}
\end{equation}

\section{Phase spatial fluctuations in partial AO correction}\label{app_ao} 
The spatial fluctuations of the atmospheric phase can be characterized by the computation of the so-called spatial structure function $\mathcal{D}_{\phi}({\vec r},{\vec r^{\prime}})$ defined as the variance of the phase difference taken at two radii ${\vec r}$ and ${\vec r^{\prime}}$:
\begin{equation}
\mathcal{D}_{\phi}({\vec r},{\vec r^{\prime}}) = \mathcal{D}_{\phi}({\vec r}-{\vec r^{\prime}}) = <|\phi({\vec r}) - \phi({\vec r}^{\prime})|^2>
\end{equation}
In the fully turbulent case, and assuming Kolmogorov's description of
the atmosphere, the phase structure function can be written \citep{roddier_2}:
\begin{equation}
\mathcal{D}_{\phi}({\vec \rho}) = 6.88\left(\frac{{\vec \rho}}{r_0}\right)^{5/3}
\end{equation}
$r_0$ being the fried parameter, that gives the ultimate spatial
resolution achievable in presence of atmospheric turbulence, that is
$\theta_{turb} = \lambda/r_0$.\\
When partial AO correction is applied, low (spatial) orders of the
corrugated wavefront are real-time compensated thanks to a deformable
mirror. The variance $\sigma^2_{\phi_r}$ of the residual turbulent phase $\phi^r({\vec r})$ not corrected by the system, can be approximated as following, providing a full
correction of the first $J \ll 1$ Zernike modes of the turbulence \cite{noll_1}:
\begin{equation}
\sigma^2_{\phi_r} \simeq 0.2944 J^{-\sqrt{3/2}}\left(\frac{D}{r_0}\right)^{5/3}
\end{equation}
Furthermore, the phase structure function of the {\it residual} phase can be written:
\begin{equation}
\mathcal{D}_{\phi_r}({\vec \rho}) =
2\sigma^2_{\phi_r}\left[1-\frac{<\phi^r({\vec r})-\phi^r({\vec r}+{\vec \rho})>}{\sigma^2_{\phi_r}}\right] \label{eq_aocorrel}
\end{equation} 
For $\rho$ bigger than typically the distance between the mirror's
actuators, the phases are uncorrelated, that is $<\phi^r({\vec r})-\phi^r({\vec r}+\rho)> = 0$, and consequently $\mathcal{D}_{\phi_r}$ saturates at $2\sigma^2_{\phi_r}$. \\
The long exposure AO corrected
transfer function  of the atmosphere is by definition:
\begin{equation}
B({\vec f}) = \exp\left[-\frac{1}{2}\mathcal{D}_{\phi_r}(\lambda{\vec f})\right]
\end{equation} 
Taking into account the properties of the phase structure function of
the residual phase, it comes \citep{conan_2} that $B({\vec f})$ can be
decomposed in two parts: one low frequency term -- namely the  halo
($FTO_{halo}$) --
and one high frequency term saturating at
$e^{-\sigma^2_{\phi_r}}$, namely the coherent energy. It comes that the long exposure transfer
function of the telescope $T({\vec f})=T_0({\vec f}).B({\vec f})$, where $T_0({\vec f})$ is the
transfer function of a perfect telescope, can be in first approximation
modelled by a sum of two components:
\begin{eqnarray}
T(f) &=& T_0(f).[e^{-\sigma^2_{\phi_r}} + FTO_{halo}(f)] \nonumber
\\
&\simeq& T_0(f).[e^{-\sigma^2_{\phi_r}}+ (1-\exp^{-\sigma^2_{\phi_r}})\exp^{-\frac{1}{2}\mathcal{D}_{\phi}(\lambda{\vec f})}]
\end{eqnarray} 
Eq. (\ref{eq_aocorrel}) also demonstrates that for interferometers
equipped with AO systems, the phases $\phi^r_1$ and $\phi^r_2$ of telescopes $1$ and $2$ are always uncorrelated. As a consequence, the phase structure
function of the residual piston phase $\phi_{12} = \phi^r_1 - \phi^r_2$ is twice
the phase structure function of the residual phase over one single
telescope (assuming equivalent AO systems for both telescopes), that is $\mathcal{D}_{\phi_{12}}({\vec \rho}) =
2\mathcal{D}_{\phi_r}({\vec \rho})$ . As a matter of fact, even if no AO correction is
applied, the phases $\phi_1$ and $\phi_2$ can still be consider
uncorrelated, as soon as the baseline is longer than the outer scale
of the atmosphere $\mathcal{L}_0$. This is for
  example the case for the VLTI at Paranal, where the outer scale has
  been estimated to be   $\mathcal{L}_0 \simeq 22$m \citep{martin_1}. 

% LocalWords:  interferogram photoevents multimode multiaxial autocorrelation
% LocalWords:  Strehl

\end{document}